\newif\iffigs
  \figsfalse
\documentclass[12pt]{article}
\usepackage{latexsym,amssymb,lscape}

\iffigs
    \usepackage{graphicx}
    \input{epsf}
\else
               \usepackage[all]{xy}
\fi
\newcommand{\ft}[2]{{\textstyle\frac{#1}{#2}}}
\def\tilde{\widetilde}

\newsavebox{\uuunit}
\sbox{\uuunit}
                             {\setlength{\unitlength}{0.825em}
                                      \begin{picture}(0.6,0.7)
                                                                  \thinlines
                                                                  \put(0,0){\line(1,0){0.5}}
                                                                  \put(0.15,0){\line(0,1){0.7}}
                                                                  \put(0.35,0){\line(0,1){0.8}}
                                                                 \multiput(0.3,0.8)(-0.04,-0.02){10}{\rule{0.5pt}{0.5pt}}
                                      \end {picture}}
\newcommand {\unity}{\mathord{\!\usebox{\uuunit}}}
\makeatletter \@addtoreset{equation}{section} \makeatother

\def\Solv{\mathop{\rm Solv}\nolimits}
\def\dim{\mathop{\rm dim}\nolimits}

\def\rmd{{\mathrm d}}
\def\rme{{\mathrm e}}
\def\Riem{\mathop{\rm Riem}\nolimits}
\newcommand{\SU}{\mathop{\rm SU}}
\newcommand{\SO}{\mathop{\rm SO}}
\newcommand{\U}{\mathop{\rm {}U}}
\newcommand{\USp}{\mathop{\rm {}USp}}
\newcommand{\Sp}{\mathop{\rm {}Sp}}

\newcommand{\Sl}{\mathop{\rm {}S}\ell }

\newcommand{\Spin}{\mathop{\rm {}Spin}}
\newcommand{\uxi}{\underline{\xi}}
\newcommand{\ulambda}{\underline{\lambda}}
\newcommand{\uzeta}{\underline{\zeta}}

\newcommand{\ualpha}{\underline{\alpha}}
\newcommand{\ua}{\underline{a}}
\newcommand{\ub}{\underline{b}}
\newcommand{\uepsilon}{\underline{\epsilon}}

\newcommand{\so}{\mathfrak{so}}
\newcommand{\su}{\mathfrak{su}}
\newcommand{\usp}{\mathfrak{usp}}
\newcommand{\uu}{\mathfrak{u}}

\usepackage{ifpdf}
\ifpdf
   \pdfoutput=1
   \pdftrue
  \usepackage[pdftex]{hyperref}
\else
   \pdffalse
   \usepackage{cite}
\fi
\begin{document}
\begin{titlepage}
\begin{flushright}
DFTT 09/2006\\
KUL-TF-06/21\\
hep-th/yymmnnn
\end{flushright}
\vskip 1.5cm
\begin{center}
{\LARGE \bf  Tits-Satake projections of \\ \vskip 0.2cm homogeneous  special geometries} \\
 \vfill
{\large Pietro Fr{\'e}$^1$, Floriana Gargiulo$^2$, Jan Rosseel$^3$ \\ \vskip
0.2cm
Ksenya Rulik$^1$, Mario Trigiante$^2$ and Antoine Van Proeyen$^3$} \\
\vfill {
$^1$ Dipartimento di Fisica Teorica, Universit{\`a} di Torino, \\
$\&$ INFN -
Sezione di Torino\\
via P. Giuria 1, I-10125 Torino, Italy\\
$^2$ Dipartimento di Fisica Politecnico di Torino,\\ C.so Duca degli
Abruzzi, 24,
I-10129 Torino, Italy, \\
$^3$ Institute for theoretical physics, K.U. Leuven \\ Celestijnenlaan
200D, B-3001 Leuven, Belgium}
\end{center}
\vfill
\begin{abstract}
We organize the homogeneous special geometries, describing as well the
couplings of $D=6$, 5, 4 and 3 supergravities with 8 supercharges, in a
small number of universality classes. This relates manifold on which
similar types of dynamical solutions can exist. The mathematical
ingredient is the Tits-Satake projection of real simple Lie algebras,
which we extend to all solvable Lie algebras occurring in these
homogeneous special geometries. Apart from some exotic cases all the
other, 'very special', homogeneous manifolds can be grouped in seven
universality classes. The organization of these classes, which capture
the essential features of their basic dynamics, commutes with the $r$-
and $c$-map. Different members are distinguished by different choices of
the paint group, a notion discovered in the context of cosmic billiard
dynamics of non maximally supersymmetric supergravities. We comment on
the usefulness of this organization in universality class both in
relation with cosmic billiard dynamics and with configurations of branes
and orbifolds defining special geometry backgrounds.
\end{abstract}
\vspace{2mm} \vfill \hrule width 3.cm
{\footnotesize \noindent
e-mails: \{rulik, fre, gargiulo, mario.trigiante\}@to.infn.it,\\
\phantom{e-mails: } \{jan.rosseel, antoine.vanproeyen\}@fys.kuleuven.be}
\end{titlepage}
%%%%%%%%%%%%%%%%%%%%%%%%%%%%%%%%%%%%%%%%%%%%%%%%%%%%%%%%%%%%%%%%%%%%%%%%%%%
\addtocounter{page}{1}
 \tableofcontents{}
\newpage
%%%%%%%%%%%%%%%%%%%%%%%%%%%%%%%%%%%%%%%%%%%%%%%%%%%%%%%%%%%%%%%%%%%%%%%%%%%
\section{Introduction}
 \label{introibo}
The recent interest for applications of string theory to cosmology has
stimulated the investigation of supergravity solutions where all the
fields depend only on time \cite{Damour:2002cu,Damour:2002et}. As it was
pointed out in \cite{Fre:2003ep}, the search and classification of such
backgrounds amounts, via dimensional reduction and oxidation, to a
systematic study of supergravity dimensionally reduced to $1+0$
dimensions, where all degrees of freedom are represented by scalars and
where the duality symmetries of string/supergravity theory are maximally
enlarged and made manifest. Actually the reduction of all degrees of
freedom to scalars occurs already at the level of $D=3$ dimensions
\cite{Cremmer:1999du,Keurentjes:2002xc} and, for this reason, a
convenient view point is provided by regarding the effective
one-dimensional $\sigma$--model:
\begin{equation}
  S=\int \sqrt{\gamma(t)}\, \rmd t\left[ - \gamma^{00}(t)\partial_t \phi^I\partial_t \phi^J \,
  g_{IJ}(\phi) \right]
 \label{1dim}
\end{equation}
as embedded in the $D=3$ $\sigma$--model:
\begin{equation}
  S=\int \sqrt{\gamma(x)}\, \rmd^3x \,\left[ - \gamma^{\mu\nu}(x)\partial_\mu \phi^I\partial_\nu \phi^J \,
  g_{IJ}(\phi) \right],
 \label{3dim}
\end{equation}
where $g_{IJ}(\phi)$ is the  metric on the $D=3$ scalar manifold
$\mathcal{M}_{\rm scalar}$, whose geometry and structure are dictated,
for various values of $\mathcal{N}_Q=\# \mbox{ of supercharges} $, by
supersymmetry and whose isometries encode the string duality algebras.
For instance, when $\mathcal{N}_Q \ge 12$ the manifold $\mathcal{M}_{\rm
scalar}$ is necessarily a homogeneous symmetric space within certain
classes ($16 \ge \mathcal{N}_Q \ge 12$) or just a completely fixed coset
manifold $\mathrm{G/H}$ for $32 \ge \mathcal{N}_Q > 16$.
\par
For $12 > \mathcal{N}_Q \ge 8$ the scalar manifold $\mathcal{M}_{\rm
scalar}$ is not necessarily a homogeneous space $\mathrm{G/H}$, yet its
geometry is severely restricted to fall into a hierarchy of special
classes  that have received the, by now well established, name of
\textit{$\mathcal{N}=2$ special geometries}. These geometries are mainly
defined by a constraint of restricted holonomy and by the existence of
certain characteristic bundles dictating the structure of their metric.
$\mathcal{N}=2$ special geometries include:\footnote{A six-dimensional
counterpart was discussed in \cite{deWit:1991nm,Andrianopoli:2004xu} and
will be reviewed in section \ref{ss:6dorigin}.}
\begin{enumerate}
  \item The very special real geometry described by the vector multiplet scalars
  in $D=5$ dimensions  \cite{Gunaydin:1984bi,deWit:1992cr}.
  \item The special K{\"a}hler geometry of vector multiplet scalars in
  $D=4$ space-time dimensions \cite{deWit:1984pk,Castellani:1990zd,D'Auria:1991fj}.
  \item The quaternionic-K{\"a}hler geometry of hypermultiplet scalars
  in $D=4$ or of just all the available scalars in the  sigma
  model to which any $\mathcal{N}_Q = 8$ supergravity reduces in
  $D=3$ \cite{Bagger:1983tt}.
\end{enumerate}
Dimensional reduction on a circle provides a well understood and
extremely powerful relation of inclusion between the above mentioned
three classes of special geometries, usually named the $\mathbf{c}$-map
\cite{Cecotti:1989qn} and the $\mathbf{r}$-map \cite{deWit:1992nm}:
\begin{equation}
  \mbox{Very Special real} \, \stackrel{\mbox{$\mathbf{r}$-map}}{\Longrightarrow}
  \, \mbox{ Special K{\"a}hler} \, \stackrel{\mbox{$\mathbf{c}$-map}}{\Longrightarrow}
  \, \mbox{ Quaternionic-K{\"a}hler}
\label{cmappa}
\end{equation}
The $\mathbf{r}$-map defines for any very special real manifold, say of
dimension $n-1$, a special K{\"a}hler manifold of real dimension $2n$, and
the $\mathbf{c}$-map defines for any special K{\"a}hler manifold a
quaternionic-K{\"a}hler manifold of dimension $4(n+1)$. To clarify names, the
quaternionic spaces that are in the image of the $\mathbf{c}$--map are
called \textit{special quaternionic} manifolds, and those that are in the
image of the composed $\mathbf{r}\circ\mathbf{c}$--map are called
\textit{very special quaternionic}. The similar convention is used also
for special K{\"a}hler spaces, that are called \textit{very special K{\"a}hler}
if they are in the image of the $\mathbf{r}$--map.
 \par
Homogeneous quaternionic-K{\"a}hler manifolds have rank at most~4. Since
under reduction on a circle the rank of the space is increased by
precisely one unit, all quaternionic manifolds of rank $r\geq 3$ have
nonempty $\mathbf{c}$ and $\mathbf{r}\circ\mathbf{c}$ predecessors, and
are thus `very special'. It happens that for these manifolds the mapping
could be extended further to a supergravity theory for $D=6$
\cite{deWit:1991nm,Andrianopoli:2004xu} and for the case $r=4$ this also
includes a scalar manifold. These steps allow us to understand the higher
dimensional origin of some important compact symmetries acting on the
scalar manifolds of $D=3,4,5$ supergravities, as we will discuss in
section \ref{ss:6dorigin}.
 \par
Hence in the context of the construction and classification of
time-dependent supergravity solutions, the reasoning recalled above shows
that for theories with $\mathcal{N}_Q = 8$, the relevant special geometry
is that popping out in $D=3$, namely quaternionic geometry. Yet the
inclusion relations of the $\mathbf{r}$- and $\mathbf{c}$-maps are just
of vital use in dimensional oxidation, namely they are essential to
reinterpret the sigma model solutions as full-fledged supergravity
backgrounds in higher dimensions.
\par
As stressed above, special geometry manifolds are not necessarily
homogeneous spaces: for instance they include the moduli spaces of
complex structures or of K{\"a}hler classes for Calabi-Yau threefolds.
Yet a large and rich subclass of special manifolds is provided by the
\textit{special homogeneous manifolds} $\mathcal{SH}$ which have all been
classified and studied systematically \cite{Cremmer:1985hc,deWit:1992nm}.
They describe a large class of supergravity models associated for
instance with orbifold or orientifold compactification of superstrings
and also with a variety of brane constructions
\cite{D'Auria:2004kx,D'Auria:2004cu,Smet:2004da}. They occur in
$\mathrm{T^2 \times K3}$ compactifications and typically they describe
the large radius limit of Calabi-Yau moduli spaces.
\par
The main point of this paper deals with the \textbf{Tits-Satake (TS)
projection} of these \textit{special homogeneous ($\mathcal{SH}$)
manifolds}:
\begin{equation}
  \mathcal{SH} \, \stackrel{\mbox{Tits-Satake}}{\Longrightarrow} \,
  \mathcal{SH}_{\rm TS}
\label{titssatake}
\end{equation}
and with the related concept of \textbf{Paint Group} which was introduced
in \cite{Fre':2005sr}.
\par
Let us explain what this means and let us summarize our main result.
\par
It turns out that one can define an algorithm, the Tits-Satake projection
$\pi_{\rm TS}$ which works on the space of homogeneous manifolds with a
solvable transitive group of motions $\mathcal{G}_M$ and associates to
any such manifold another one.
This map has a series of very strong distinctive features:
\begin{enumerate}
  \item $\pi_{\rm TS}$ is a projection operator, so that several different
  manifolds $\mathcal{SH}_i$ ($i=1,\dots ,k$) have the same image $\pi_{\rm TS}\left(\mathcal{SH}_i \right)
  $.
  \item $\pi_{\rm TS}$ preserves the rank of $\mathcal{G}_M$ namely the
  dimension of the maximal Abelian subalgebra with semisimple elements (Cartan
  subalgebra or CSA) of $\mathcal{G}_M$.
  \item \label{Pipreserves} $\pi_{\rm TS}$ maps special homogeneous into special homogeneous
  manifolds, and preserves the three classes of special
  manifolds discussed above, except for the quaternionic-K{\"a}hler manifolds that are not very special.
  Thus, apart from these exceptions, it maps \textit{very special real} into
  \textit{very special real}, maps \textit{special K{\"a}hler} into \textit{special
  K{\"a}hler} and maps \textit{quaternionic-K{\"a}hler} into
  \textit{quaternionic-K{\"a}hler}.
  \item{ $\pi_{\rm TS}$ commutes with the $\mathbf{r}$- and $\mathbf{c}$-map, so that we obtain the
  following commutative diagram:}
\begin{equation}
  \begin{array}{ccccc}
    \mbox{Very Special real} & \stackrel{\mbox{$\mathbf{r}$-map}}{\Longrightarrow} & \mbox{Special K{\"a}hler} &
    \stackrel{\mbox{$\mathbf{c}$-map}}{\Longrightarrow} & \mbox{Quaternionic-K{\"a}hler} \\
    \begin{array}{cc}
      \pi_{\rm TS} & \Downarrow \\
    \end{array} & \null & \begin{array}{cc}
      \pi_{\rm TS} & \Downarrow \\
    \end{array} & \null & \begin{array}{cc}
      \pi_{\rm TS} & \Downarrow \\
    \end{array} \\
    \left( \mbox{Very Special real}\right) _{\rm TS} & \stackrel{\mbox{$\mathbf{r}$-map}}{\Longrightarrow} &
    \left( \mbox{Special K{\"a}hler}\right) _{\rm TS} &
    \stackrel{\mbox{$\mathbf{c}$-map}}{\Longrightarrow} & \left( \mbox{Quaternionic-K{\"a}hler}\right) _{\rm TS} \
  \end{array}
\label{diagrammo}
\end{equation}
The exception mentioned under \ref{Pipreserves} means that the
(Quaternionic-K{\"a}hler)$_{\rm TS}$ is itself not quaternionic-K{\"a}hler for
the cases in which the original manifold does not have a very special
real predecessor. For Special K{\"a}hler manifolds a similar conclusion holds
: (Special K{\"a}hler)$_{\rm TS}$ is itself not necessarily special K{\"a}hler
when the original manifold does not have a very special real predecessor.
\end{enumerate}
The main consequence of the above features is that the whole set of
special homogeneous manifolds and hence of associated supergravity models
is distributed into a set of \textit{universality classes} which turns
out to be composed of extremely few elements, altogether 7 very special
classes, 3 further special classes, and 2 classes of quaternionic
projective spaces. By definition we put into the same class all those
manifolds ($i.e.$ supergravity models) that have the same Tits-Satake
projection or are connected by a $\mathbf{c}$-- or $\mathbf{r}$--map.
\par
Such an organization of supergravity theories into universality classes
might seem at first sight quite arbitrary, but there are instead strong
indications that just the opposite is true. Indeed the Tits-Satake
projection seems to capture for each model the few truly essential
degrees of freedom in whose dynamics is encoded the qualitative behaviour
of the solutions for the whole class. Specifically all the classical
solutions of the Tits-Satake projected theory are \textit{bona fide}
solutions for all the theories in its universality class, although not
all solutions of each element of the class can be obtained in this way.
The additional solutions which are specific to each element of the same
class seem to be finer modulations of the solutions contained in the
Tits-Satake universal model. In certain string solutions, which we shall
discuss later, the Tits-Satake universal model describes the bulk
excitations and positions of branes, while the related unprojected models
describe configurations with multiple overlapping branes.
\par
That such behaviour is the case became clear in the study of cosmic
billiards \cite{Damour:2002et}, which is  the context where the very
notion of Tits-Satake projection of a supergravity model first emerged
\cite{Henneaux:2003kk,Fre':2005sr}, and where the key notion of
\textit{paint group} was discovered\footnote{In the context of symmetric
spaces this corresponds to the group $G_c$ in \cite{Keurentjes:2002rc}.}.
We shall now briefly review these facts in order to illustrate our
previous statements and introduce the concept of paint group. Yet we
immediately want to stress that cosmic billiards are just only one
example of possible applications of the organization of supergravity
models into Tits-Satake universality classes. Many other applications are
likely to be possible and relevant (black hole solutions just to mention
some) since the projection map is a map on the entire theory and not just
on the billiard degrees of freedom.
\par
The solvable Lie algebra parametrization of noncompact homogeneous
manifolds, based on classical theorems of differential geometry
\cite{Helgason}, was introduced into supergravity literature in
\cite{Andrianopoli:1996bq} and proved to be an extremely useful tool to
address a variety of supergravity/superstring problems
\cite{Andrianopoli:1996zg,Trigiante:1998vu}. Its essential virtue is that
it provides an algebraic characterization of the scalar fields by
identifying the scalar manifold $\mathcal{M}_{\rm scalar}$ with a group
manifold of a solvable Lie algebra $\rm{Solv}_{\mathcal{M}}$ on which one
can introduce an invariant non-degenerate symmetric form $< \, , \,>$,
obviously different from the Cartan Killing form, but diffeomorphic at
each point of the manifold $\mathcal{M}$ to its Riemannian metric.
\par
A general mathematical theory \cite{Trigiante:1998vu,Fre:2001jd} allows
us to construct $\Solv_{\mathrm{G_R/H}}$ whenever
$\mathcal{M}=\mathrm{G_R/H}$ is a noncompact symmetric space. In this
case, $\mathrm{G_R} = \exp [\mathbb{G_R}]$ is the exponential of a
noncompact real form $\mathbb{G_R}$ of a semisimple Lie algebra
$\mathbb{G}$ and $\mathrm{H}=\exp [\mathbb{H}]$ is the exponential
\footnote{Note that here and in the rest of this paper, we will talk
about the exponential of a Lie algebra as a local statement.} of its
maximal compact subalgebra $\mathbb{H}\subset \mathbb{G_R}$. The algebra
$\Solv_{\mathrm{G_R/H}}$ is constructed in terms of Cartan generators and
step operators associated with positive roots. There exist also
homogeneous manifolds that are not symmetric spaces. In this case
$\Solv_\mathcal{M}$ is also defined, but is not derived from the real
form of any semisimple Lie algebra $\mathbb{G_R}$. In this class fall
several of the homogeneous special manifolds classified in
\cite{Alekseevsky1975,deWit:1991nm} and considered in the present paper.
A related analysis for symmetric spaces is contained in the oxidation
results of \cite{Keurentjes:2002rc}.
\par
Irrespectively whether $\mathcal{M}$ is a symmetric space or not, the
structure of $\Solv_\mathcal{M}$ is always of the following type. There
are $r$ semisimple generators $\mathcal{H}_i$, named the \textit{Cartan
generators}, and $r$ is called the \textit{(real) rank} of
$\Solv_\mathcal{M}$. Furthermore there are nilpotent generators
$\mathcal{W}^{\ell(\alpha)}_\alpha$. Here, $\alpha $ labels
$r$--dimensional vectors $\vec{\alpha }$, with components $\alpha_i$,
which we call `\textit{restricted roots}' analogous to the terminology
used for symmetric spaces. The set of these vectors is denoted by
$\overline{\Delta}$. These vectors have in general multiplicities
$n(\vec{\alpha })$. This leads to the notation $\ell(\alpha )$, where for
fixed $\alpha $, the number $\ell$ labels the multiplicities. The general
structure is as follows:
\begin{eqnarray}
 \Solv_{\mathcal{M}}  & = & \left\{ \mathcal{H}_i, \mathcal{W}_\alpha^{\ell(\alpha )} \right\} ,
   \qquad  i=1,\ldots ,r,\qquad
   \forall\vec\alpha\in\overline{\Delta}\,:\,\ell(\alpha)=1,\ldots, n(\vec{\alpha }),
 \nonumber\\
   &   & r+\sum_{\alpha \in\overline{\Delta} } n(\vec{\alpha })= \dim\mathcal{M},\nonumber\\
&& \left \{
\begin{array}{lcl}
 \left[\mathcal{H}_i \, , \, \mathcal{H}_j\right]  & = & 0  \\
\left[\mathcal{H}_i \, , \, \mathcal{W}_\alpha^{\ell(\alpha )}\right] & =
&
\alpha_i \, \mathcal{W}_\alpha^{\ell(\alpha )}  \\
\left[\mathcal{W}_\alpha^{\ell(\alpha )} \, , \,
\mathcal{W}_\beta^{m(\beta )}\right] & = &
 N_{\ell(\alpha ) m(\beta ) }^{r(\gamma)}
\mathcal{W}_\gamma^{r(\gamma)} \, \qquad N_{\ell(\alpha ) m(\beta )
}^{r(\gamma)}\neq 0 \ \Rightarrow \ \vec{\alpha }+\vec{\beta }=
\vec{\gamma }\in \overline{\Delta}.
\end{array}
\right. \label{genestrucca}
\end{eqnarray}
\par
If we consider the $1$-dimensional $\sigma$-model (\ref{1dim}) on a
solvable target manifold that is the group manifold of
(\ref{genestrucca}), the action takes the general form:
\begin{equation}
  S=\int \rmd t\left[  \dot{ \vec{h}}\cdot\dot{\vec{h}} +\sum_{\alpha \in\overline{\Delta} } \,
  \rme^{-2\vec{\alpha } \cdot \vec{h}(t)
  }N_\alpha \left(\phi,\dot\phi\right)\right],
 \label{ShN}
\end{equation}
where $\vec{h}$ is the vector of scalar fields $h_i$ associated with the
Cartan generators, while $N_\alpha (\phi,\dot\phi)$ are a set of
polynomial functions depending on the scalar fields $\phi_{\ell(\alpha)}$
associated with the nilpotent generators $\mathcal{W}_\alpha^{\ell(\alpha
)}$ and on their time derivatives $\dot\phi_{\ell(\alpha)}$. The fields
$h_i$ evolve within the domain where $\vec{\alpha }\cdot \vec{h}>0$, for
all $\vec{\alpha }\in\overline{\Delta} $ and they experience bounces when
approaching the walls of an ideal \textit{Weyl chamber}, defined by
$\vec{\alpha } \cdot \vec{h}=0$ for any $\vec{\alpha }$. The name Weyl
chamber is justified by the following consideration: as already stated,
the scalar fields $\phi_{\ell (\alpha) }$ correspond to the nilpotent
generators $\mathcal{W}_\alpha^{\ell(\alpha )}$, which have gradings with
respect to the $\mathcal{H}_i$ encoded in the restricted root vectors
$\vec{\alpha}$. These determine the position of the walls and this
explains the adopted nomenclature \textit{Weyl chamber}. Recall that the
walls are only determined by different restricted root vectors,
irrespective of the multiplicity with which these occur. This is the
essence of the Tits-Satake projection, where we restrict the generators
to one for each different root vector.
\par
This type of scalar field evolution leads to the cosmic billiard
paradigm. Indeed the fields ${\rm e}^{h_i(t)}$ can be identified with the
cosmic scale factors relative to the various compact and noncompact
dimensions of the $10$ or $11$--dimensional spacetime manifold and the
bouncing phenomena correspond to inversions in the expansion/contraction
development of these scale factors. In order to define the main features
of billiard dynamics only a subset of the scalar fields
$\phi_{\ell(\alpha)}$ is needed. This subset contains \textit{one
nilpotent scalar} for each root entering the definition of the
"restricted root system" $\overline{\Delta}$. This is sufficient to
define the positions of all the walls that cause the inversions in the
motion of a billiard ball with coordinates $h_i(t)$.
 \par
The scalar manifolds of maximal supergravity theories are symmetric
spaces. Moreover, these symmetric spaces have isometry groups that are
maximally noncompact (`split real forms'). In such cases, the dynamical
equations are integrable \cite{Fre:2003ep,Fre:2005bs}. We are, however,
interested also in non-maximal supergravity theories. In these cases the
isometry groups are not maximally noncompact. When we step down to
$\mathcal{N}_Q =8$ real supersymmetries or less, the isometry group is
not even necessarily a semi-simple group.
\par
As we shall recall and summarize in section \ref{Titsatesection} the
procedure leading to the identification of the subalgebra
$\mathbb{G}_{\mathrm{TS}} \subset \mathbb{G}_\mathrm{R}$ is known in the
mathematical literature as the Tits-Satake projection \cite{Helgason},
since it is due to the named authors, see for example \cite{BorelTits}.
It corresponds to a geometrical projection of the root system
$\Delta_\mathrm{G}$ of the original algebra onto a restricted root system
$\Delta_{\rm TS}$, which entirely lies in the noncompact subspace of the
Cartan subalgebra, the compact part being projected out. This projection
mechanism was applied in \cite{Henneaux:2003kk} to discuss  the
asymptotic features of billiard dynamics and was demonstrated in
\cite{Fre':2005sr} to generate exact analytic billiard solutions of the
complete $\mathrm{G_R/H}$ supergravity theory. In the latter paper, the
notion of paint group was introduced, which is a pivotal tool to master
the mechanism of the reduction to the TS subalgebra and to classify the
different theories belonging to the same universality class, namely
having the same Tits-Satake projection.
\par
In short the notion of paint group is described as follows. As mentioned
above, and illustrated in (\ref{genestrucca}), it happens that there are
several roots of $\Delta_\mathrm{G}$ (i.e. generators of
$\Solv_{\mathrm{G/H}}$) that have the same nonvanishing components along
the noncompact directions. Then, the Tits-Satake projection $\pi_{\rm
TS}$ projects them onto the same restricted root of $\Delta_{\rm TS}$. It
is a fact that these copies of each restricted root arrange into
representations $\left\{ \mathcal{D_\gamma}\right\} $ of a compact group
$\mathrm{G}_{\rm paint}$ named in \cite{Fre':2005sr} the \textit{paint
group}. This latter acts as an external automorphism group of the
solvable Lie algebra $\Solv_{\mathrm{G/H}}$. Altogether the set of
representations $\left\{ \mathcal{D_\gamma}\right\} $ have a little group
of stability called the subpaint group $\mathrm{G}_{\rm subpaint} \subset
\mathrm{G}_{\rm paint}$ and decomposing with respect to $\mathrm{G}_{\rm
subpaint}$ we find:
\begin{eqnarray}
 \mathcal{D_\gamma} & \stackrel{\mathrm{G}_{\rm subpaint}}{\Longrightarrow}
 & \underbrace{{\bf 1}}_{\mathrm{singlet}} \, \oplus \, \mathcal{}D^\prime_\gamma \quad ; \quad \left (\gamma = 1, \dots
 , \# \mbox{ of representations} \right )
\label{decompiDell}
\end{eqnarray}
The singlets in these decompositions span, together with the Cartan
generators, the Tits-Satake projected solvable algebra $\Solv_{\rm TS}$
that defines a smaller homogeneous manifold $\mathcal{M}_{TS}$. It often
happens that this projected manifold is itself a symmetric space
$\mathcal{M}_{\rm TS} = \mathrm{G_{\rm TS}}/H_{\rm TS}$, having a
semisimple group of isometries $\mathrm{G_{\rm TS}}$.
\par
Until now, the billiard mechanism was studied only for pure
supergravities or for supergravity theories with more than 8
supersymmetries, which implies that the scalar manifolds were all
symmetric spaces. In this paper we will consider supergravity theories
with 8 real supercharges, leading to the aforementioned special
geometries that include scalar manifolds that are homogeneous but not
symmetric manifolds. Our goal is the extension of the notion of TS
projection to all the solvable algebras underlying such manifolds. As
anticipated, we will show that this is indeed possible and that all
models distribute into a very small set of Tits-Satake universality
classes. Furthermore, we shall study the systematics of paint groups and
subpaint groups.
\par
As repeatedly emphasized, the scope of this organization of models into
Tits-Satake classes goes much beyond the context of cosmic billiards
wherefrom it originated.

Before embarking on the projections, we recall the structure of
homogeneous special geometries in section \ref{structhom}. This starts
from the connection made between these manifolds and the solvable
algebras as exploited first in \cite{Alekseevsky1975} for
quaternionic-K{\"a}hler manifolds. Then it recalls the structure of the
isometry algebras of homogeneous special manifolds, reviewing the results
of \cite{deWit:1992wf}. Finally we review the 6-dimensional origin of the
very special geometries \cite{deWit:1991nm,Andrianopoli:2004xu} and
explain how the paint group is realized in this context.

The Tits-Satake projection is introduced in section \ref{Titsatesection}.
We start by recalling these notions as they have been defined for
symmetric spaces in \cite{Fre':2005sr}. We go in detail through one
example, $E_8/(E_7\times \SU(2))$, paying special attention to the
notions of paint and subpaint group. Generalizing this allows us to
define the TS projection for homogeneous special geometries in section
\ref{Tsforsolvable}. We finish this section with giving already the
structure of the results for the TS projection for all homogeneous
special geometries.

In section \ref{pitturagruppo}, we obtain these results from the
systematic procedure outlined in section \ref{Tsforsolvable}. Again we
first discuss the paint group for the symmetric spaces and then for
homogeneous special geometries. Finally, we obtain the subpaint groups
for the latter.

This finally allows us to discuss in section \ref{universaclassa} the
universality classes of homogeneous special geometries. We end in section
\ref{ss:summary} with a summary of the results and identify outstanding
problems for which these results are useful. One of these is the
application to a system of D3/D7 branes, which is discussed in more
detail as appendix \ref{d3d7}. The appendix \ref{realcliffalg} contains
an important prerequisite for this work: a summary of properties of real
Clifford algebras.

%%%%%%%%%%%%%%%%%%%%%%%%%%%%%%%%%%%%%%%%%%%%%%%%%%%%%
% Classification
%%%%%%%%%%%%%%%%%%%%%%%%%%%%%%%%%%%%%%%%%%%%%%%%%%%

\section{Homogeneous special manifolds}
\label{structhom}

In the present section our aim is that of reviewing the structure of
homogeneous special manifolds as it emerged from the classification of
\cite{Alekseevsky1975,Cortes,deWit:1992wf}.

\subsection{Classification of homogeneous quaternionic-K{\"a}hler manifolds}
\label{ss:classhomquatK}

In general, quaternionic-K{\"a}hler manifolds can be homogeneous or non
homogeneous, compact or noncompact, and when they are homogeneous they
can be symmetric or not. The homogeneous ones are of the form
$\mathrm{G/H}$, where $\mathrm{G}$ is the group of isometries, which is
not necessarily a semi-simple group, and $\mathrm{H}$ is its isotropy
group. If $\mathrm{H}$ is a symmetric subgroup\footnote{$\mathrm{H}$ is a
symmetric subgroup of $\mathrm{G}$ if their corresponding Lie algebras
verify $\mathbb{G}=\mathbb{H}+\mathbb{K}$ with
$[\mathbb{H},\mathbb{K}]\subset\mathbb{K}$ and
$[\mathbb{K},\mathbb{K}]\subset\mathbb{H}$.} of $\mathrm{G}$, then the
space $\mathrm{G/H}$ is symmetric.
\par
In this paper we are interested in \textbf{homogeneous noncompact
quaternionic-K{\"a}hler manifolds}. The noncompactness is due to the fact
that in supergravity the Ricci curvature of the manifold should be
negative. Alekseevsky conjectured in \cite{Alekseevsky1975} that all such
spaces are exhausted by the so-called \textit{normal quaternionic
manifolds}. A quaternionic space is normal if it admits a completely
solvable\footnote{A solvable Lie algebra $s$ is completely solvable if
the adjoint operation $\mathrm{ad}_X$ for all generators $X \in s$ has
only real eigenvalues. The nomenclature of the Lie algebra is carried
over to the corresponding Lie group in general in this paper.} Lie group
$\exp [\Solv_{\mathcal{M}}]$ of isometries that acts on the manifold in a
simply transitive manner (i.e. for every 2 points in the manifold there
is one and only one group element connecting them). The group
$\Solv_{\mathcal{M}}$ is then generated by a so-called \textit{normal
metric Lie algebra}, that is a completely solvable Lie algebra endowed
with a Euclidean metric. The main tool to classify and study the normal
homogeneous quaternionic spaces is provided by the theorem that states
that if a Riemannian manifold $(\mathcal{M},g)$ admits a transitive
normal solvable group of isometries $\exp\Solv_{\mathcal{M}}$, then it is
metrically equivalent to this solvable group manifold
\begin{eqnarray}
\mathcal{M} & \simeq & \exp \left[ \Solv_{\mathcal{M}}\,\right]\nonumber  \\
g\mid_{e \in \mathcal{M}} & = & <,> \,,
\end{eqnarray}
where $<,>$ is a Euclidean metric defined on the normal solvable Lie
algebra $\Solv_{\mathcal{M}}$.
\par
The conjecture of Alekseevsky thus implies that for every homogeneous
quaternionic-K{\"a}hler space of negative Ricci curvature $\mathcal{M}$ there
exists a transitive solvable group of isometries, $\exp \left
[\Solv_{\mathcal{M}}\right ]$, which can be identified with the manifold
itself. Classifying these manifolds is then achieved by classifying the
corresponding solvable algebras.
\par
As for any other solvable group manifold with a non degenerate invariant
metric\footnote{See \cite{Andrianopoli:1996bq},
\cite{Andrianopoli:1996zg}, \cite{Fre:2001jd} for reviews on the solvable
Lie algebra approach to supergravity scalar manifolds.} the differential
geometry of the manifold is completely rephrased in algebraic language
through the relation of the Levi-Civita connection and the \textit{Nomizu
operator} acting on  the solvable Lie algebra.  The latter is defined as
\begin{eqnarray}
 \mathbb{L}& : & \Solv_\mathcal{M}\, \otimes\,
\Solv_\mathcal{M}\rightarrow \Solv_{\mathcal{M}},
\\
 \forall X, Y, Z \in
\Solv_{\mathcal{M}} & : & 2<\mathbb{L}_XY,Z> = <[X,Y],Z> - <X,[Y,Z]> -
<Y,[X,Z]>.\nonumber
\end{eqnarray}
The \textit{Riemann curvature operator} on this group manifold can be
expressed as
\begin{equation}
\Riem(X,Y) = [\mathbb{L}_X,\mathbb{L}_Y] - \mathbb{L}_{[X,Y]}.
\end{equation}
The \textit{holonomy algebra} $\Gamma$ of $\Solv_{\mathcal{M}}$ is
defined as the Lie algebra generated by the curvature operator $\Riem$
and all of its commutators with Nomizu operators
\begin{equation}
[\mathbb{L}_{X_1},\dots,[\mathbb{L}_{X_k},\Riem(X_{k+1},X_{k+2})]\dots],
\quad X_k\in \Solv_\mathcal{M}.
\end{equation}
 \par
A quaternionic structure on the Euclidean space spanned by
$\Solv_{\mathcal{M}}$ is the linear Lie algebra $Q$ generated by three
anticommuting complex structures $J_1$, $J_2$, $J_3 = J_1\, J_2$
satisfying $J_x^2 = -1$, where $x = 1,2,3$. The centralizer and the
normalizer of $Q$ in the set of all antisymmetric endomorphisms of
$\Solv_{Q}=\Solv_\mathcal{M}$ are respectively denoted $C(Q)$ and $N(Q)$.
A metric Lie algebra is called quaternionic if there exists a
quaternionic structure $Q$ such that the normalizer $N(Q)$ contains the
holonomy algebra $\Gamma$ \footnote{For manifolds of quaternionic
dimension 1 this statement is trivial, since $N(Q) =
\mathrm{USp}(2)\times \mathrm{USp}(2) = \SO(4)$ in that case. Therefore,
one usually supplements the definition of a quaternionic manifold with
the restriction that the Riemann tensor be annihilated by the three
complex structures :
\[ (J^\alpha \cdot R)_{XYWZ} \equiv J^\alpha_ X{}^V
R_{VYWZ}+ J^\alpha_Y{}^V R_{XVWZ}+ J^\alpha_Z{}^V R_{XYWV}+
J^\alpha_W{}^V R_{XYVZ} = 0 \,.
\]
This condition is trivially satisfied for manifolds of quaternionic
dimension bigger than one, but gives a non-trivial restriction for
quaternionic dimension 1, which is the restriction necessary for this
manifold to encode supersymmetric hypermultiplet couplings.}
\begin{equation}
\Gamma \subset N(Q) \\
%&& A.Riem = 0, \quad \forall A\in Q
\end{equation}
The normalizer has the structure $N(Q) = Q + C(Q)$ and the quaternionic
structure $Q$ and its centralizer $C(Q)$ are the algebras of the Lie
groups $\mathrm{USp}(2)$ and $\mathrm{USp}(2n)$, respectively. This is
the algebraic counterpart of those items in the definition of a
quaternionic-K{\"a}hler manifold,\footnote{see for instance
\cite{D'Auria:1991fj,Fre:2001jd,Bergshoeff:2002qk} for the definition of
quaternionic-K{\"a}hler manifolds.} which require its holonomy algebra to be
given by a subgroup of $\mathrm{USp}(2)\times \mathrm{USp}(2n)$. The
curvature tensor $\Riem$ will then indeed take the general form as
required by the definition of quaternionic-K{\"a}hler manifolds:
\begin{equation}
\Riem(X,Y) = R^0_x(X,Y)\, J_x + \bar{R}(X,Y) \label{RiemQuat}
\end{equation}
where $\bar{R} \in \USp(2n)$ is the curvature tensor of type $C(Q)$ and
the tensor $R^0 \in \USp(2)$ is the curvature of an $\USp(2)$ bundle
constructed over the considered manifold. The statement in the definition
of a quaternionic-K{\"a}hler manifold that the curvature of this $\USp(2)$
bundle should be proportional to the hyper-K{\"a}hler $2$--forms induced by
the three complex structures $J_{1,2,3}$, is then given in algebraic
terms as:
\begin{equation}\label{structeqn}
R^0_x(X,Y) = -\ft12\nu<J_xX,Y>.
\end{equation}
The number $\nu $ is in supergravity $-\kappa ^2$, where $\kappa $ is the
gravitational coupling constant. The complex structures $J_x$ should also
satisfy ``integrability conditions", expressed in terms of the Nijenhuis
tensor \cite{Alekseevsky1975}.
 \par
The complete solvability of the algebra $\Solv_{\mathcal{M}}$ and the
structure equation (\ref{structeqn}) imply that any normal quaternionic
algebra contains a subalgebra of quaternionic dimension one, which is
called the \textit{canonical quaternionic subalgebra} $E$. There are two
possibilities for the canonical quaternionic subalgebra: it can either be
$\Solv\left(\mathrm{SU}(2,1)\right)$ or
$\Solv\left(\mathrm{USp}(2,2)\right)$. It was found by Alekseevsky that
for any quaternionic dimension $n$ there is a unique (up to scaling)
normal quaternionic Lie algebra admitting
$\Solv\left(\mathrm{USp}(2,2)\right)$ as the canonical quaternionic
subalgebra. The corresponding quaternionic manifolds are the hyperbolic
spaces $H_{\mathbf{H}}^n$ that are the symmetric cosets $\exp
\Solv_{\mathcal{M}} \simeq
\frac{\mathrm{USp}(2,2n)}{\mathrm{USp}(2)\times \mathrm{USp}(2n)}$.
\par
On the other hand, when $E = \Solv\left(\mathrm{SU}(2,1)\right)$, the
corresponding normal quaternionic algebras have the following structure
 \begin{eqnarray}\label{genstr}
{\cal M}&=&\exp\left[ \Solv_{Q}\right] ,\qquad  \Solv_{Q}=U+\tilde{U},\nonumber\\
\left[U,U\right]&\subseteq& U,\qquad
\left[U,\tilde{U}\right]\subseteq\tilde{U},\qquad
\left[\tilde{U},\tilde{U}\right]\subseteq U,
 \end{eqnarray}
where $U\subset \Solv_{Q}$ is a subalgebra that is stable with respect to
the action of a complex structure $J_1$: $J_1\, U=U$. By restriction of
the structural equation (\ref{structeqn}) on the subspace $U$ it can be
proved that $U$ is a K{\"a}hler algebra\footnote{We define K{\"a}hler algebra as
the solvable algebra of motions, generating a (homogeneous) K{\"a}hler
manifold.}. It is called the \textit{principal K{\"a}hlerian algebra}. The
subspace $\tilde{U}$ is related to $U$ by the action of a second complex
structure $J_2$: $\tilde{U}=J_2\, U$. The representation $T_U:
\tilde{U}\rightarrow \tilde{U}$, induced by the adjoint action of $U$,
has to satisfy conditions arising from the structural equation and
integrability conditions. It was called a $Q$--representation and its
main feature is that it is symplectic with respect to a suitable form
$\hat{J}$ expressed in terms of $J_1$. The structure of $U$ can be
represented as follows:
\begin{eqnarray}
U&=& \sum_{I=1}^r U_I, \qquad U_I = F_I + X_I,
\end{eqnarray}
where $U_I$ are called \textit{elementary K{\"a}hler subalgebras} while $r$
is equal to the \textit{rank} of the normal quaternionic manifold. The
two--dimensional subalgebras $F_I$ are so called \textit{key algebras}.
In general a key algebra $F$ can be described in terms of an orthonormal
basis $\{h,g\}$, where $g= J_1\, h$. The commutation relation between the
basic elements is $[h,g]=\mu g$. The number $\mu $ is called the root of
the key algebra; from the requirement that $T_U$ is a $Q$-representation,
it follows that it can only take the values
$(1,\frac{1}{\sqrt{2}},\frac{1}{\sqrt{3}})$, defining the type I, type II
and type III key algebras, respectively. Any key algebra generates a
space $\frac{\mathrm{SU}(1,1)}{\mathrm{U}(1)}$. An elementary K{\"a}hler
subalgebra $F + X$ is then defined by the following commutation relations
\cite{Cortes}
\begin{equation}\label{key}
[h,g] = \mu \, g, \qquad [h,x] = \frac{\mu}{2}x, \qquad [g,x] = 0, \qquad
[x,y] = \mu <J_1\,x,y>g,
\end{equation}
$x,y$ being elements of the space $X$. The collection of generators $h_I$
of $F_I$ generate the Cartan subalgebra of $\Solv_{Q}$.

The canonical quaternionic subalgebra has the structure $E = F_0 + J_2\,
F_0$, where $F_0$ is a key algebra of type I, which is stable under the
action of the complex structure $J_1$. Since the intersection of $U$ with
$E$ is given by $F_0$, the structure of $U$ can also be specified as
\begin{equation}\label{Ustr}
U = F_0 + \Solv_{SK}, \qquad \Solv_{SK} = \sum_{i=1}^{r-1}F_i + X_i\,.
\end{equation}
The second term, denoted as $\Solv_{SK}$, is called a \textit{normal
K{\"a}hler subalgebra}, as it generates the normal special K{\"a}hler space
related to the quaternionic-K{\"a}hler manifold by the $\mathbf{c}$--map.
This algebra has rank $r-1$.
 \par
We now give an overview of the possible quaternionic algebras according
to their rank. While going over this classification, the reader can be
guided by the last column of table \ref{allLpq}. For later convenience,
we will also indicate the weights of the different generators of the
algebra.
\begin{description}
  \item[Rank 1.]
For the rank 1 spaces two important classes can be distinguished,
according to whether the canonical quaternionic subalgebra is
$\Solv(\mathrm{USp}(2,2))$ or $\Solv(\mathrm{SU}(2,1))$. The algebras
with $\Solv(\mathrm{USp}(2,2))$ as canonical quaternionic subalgebra will
be denoted as $\Solv_Q(-3,P)$. They are the only spaces with
$\Solv(\mathrm{USp}(2,2))$ as canonical subalgebra and they are given
explicitly by the following symmetric spaces:
\begin{equation}
  \exp \left[ \Solv_Q\left( -3,P\right ) \right] \, \simeq \,
   \frac{\mathrm{USp}(2P+2,2)}{\mathrm{USp}(2P+2)\times\mathrm{SU}(2)}
\label{-3P}
\end{equation}
When $P=0$ their solvable algebra consists of four generators : one
Cartan generator and three generators of weight 1, that is a positive
root of $\mathrm{SU}(1,1)$. The weight structure for $P>0$ is different,
there is an additional set of $4\,P$ generators with weight
$\frac{1}{2}$. In this case the full set of weights of the solvable
algebra does not represent a positive root system of a Lie algebra of
simple type.

When the canonical quaternionic subalgebra is $\Solv(\mathrm{SU}(2,1))$,
the corresponding manifold can be described according to the scheme of
(\ref{genstr}),(\ref{Ustr}). The corresponding homogeneous space of rank
1 has $\Solv_{SK}=0$ and is denoted by $\mathrm{SG}_4$, since it can be
obtained as the reduction of pure 4-dimensional supergravity. It is given
by the symmetric space $\frac{\SU(1,2)}{\SU(2) \times \U(1)}$. The
solvable algebra consists of one Cartan generator $h_0$, while there are
two distinct weights associated to two spaces $g_0$ and $q$:
\begin{equation}
  \begin{array}{|lll|}\hline
 h_0\,:\,(0)  & g_0\,:\,(1) & q\,:\,(\frac{1}{2}) \\
 \hline
  \end{array}
 \label{weightsSG4}
\end{equation}
The space $g_0$ is one-dimensional whereas the space $q$ is
two-dimensional.
  \item[Rank 2.] In this case there are two distinct possibilities. The first is given by
  $\Solv_{SK} =F$, where $F$ is a key algebra of type III. This corresponds to the quaternionic manifold
$\frac{\mathrm{G}_{2(+2)}}{\mathrm{SU}(2)\times\mathrm{SU}(2)}$, whose
isometry algebra is maximally split. This case represents the degrees of
freedom of 5-dimensional pure supergravity $\mathrm{SG}_5$.
$\Solv(\mathrm{G}_{2(2)})$ has two Cartan generators $(h_0,h_1)$ and the
weights are summarized in the following scheme
\begin{equation}
  \begin{array}{|llll|}\hline
 h_0\,:\,(0,0)  &g_0\,:\,(1,0) & q_0\,:\,(1,-\frac{1}{2\sqrt{3}}) & p_0\,:\,(1,\frac{1}{2\sqrt{3}}) \\
 h_1\,:\,(0,0) & g_1\,:\,(0,\frac{1}{\sqrt{3}}) & q_1\,:\,(1,-\frac{\sqrt{3}}{2}) & p_1\,:\,(1,\frac{\sqrt{3}}{2}) \\
 \hline
  \end{array}
 \label{weightsSG5}
\end{equation}

The second possibility is represented by the series
$\frac{\mathrm{SU}(2,P+2)}{\mathrm{S}(\mathrm{U}(2)\times
\mathrm{U}(P+2))}$ . They are characterized by the following K{\"a}hler
subalgebra
\begin{equation}\label{rank2}
\Solv_{SK} = F + Y,
\end{equation}
where $F$ is a key algebra of type I. The corresponding algebras will be
denoted by $\Solv_Q(-2,P)$ and they are more explicitly characterized by
the following weights:
\begin{equation}
  \begin{array}{|llll|}\hline
 h_0\,:\,(0,0)  &g_0\,:\,(1,0) & q\,:\,(\frac{1}{2},-\frac{1}{2}) & p\,:\,(\frac{1}{2},\frac{1}{2}) \\
 h_1\,:\,(0,0) & g_1\,:\,(0,1) & Y\,:\,(0,\frac{1}{2}) & \tilde{Y}\,:\,(\frac{1}{2},0) \\
 \hline
  \end{array}
 \label{weightsm2}
\end{equation}
These weights are associated to six different subsets of generators, two
of which, namely $g_0$ and $g_1$ are one-dimensional. The spaces $q$ and
$p$ are two-dimensional while the dimension of the spaces $Y$ and
$\tilde{Y}$ depends on the value of the parameter $P$
\begin{equation}
\dim Y = \dim \tilde{Y} = 2\,P.
\end{equation}
For the case of $\Solv_Q(-2,0)$ the spaces $Y$ and $\tilde{Y}$ are absent
and the set of weights gives a positive root system of $\mathrm{SO}(3,2)$
in this case, whereas the full set of weights (\ref{weightsm2}) does not
have a simple Lie algebra description for $P \neq 0$.

  \item[Rank 3.]
The K{\"a}hler subalgebra of the quaternionic algebras of rank 3 is a sum of
two elementary K{\"a}hler algebras of types I and II, respectively, where the
second one has no $X$-part. We rename $X_1=Y$ for the systematics that
will be explained below.
\begin{equation}\label{rank3}
\Solv_{SK} = (F_1 + Y) + F_2,
\end{equation}
The space $Y$ forms a symplectic representation of the type II key
algebra $F_2$, and under this action it splits into two subspaces $Y =
Y^+ + Y^-$, with $Y^- = J_1\,Y^+ $. The quaternionic solvable algebras of
this rank are denoted by $\Solv_Q(-1,P)$. They consist of 3 Cartan
generators $h_0$, $h_1$ and $h_+$, while the weights of the other
generators are summarized in the following table:
\begin{equation}
 \label{weightsm1}
\begin{array}{|llll|}
\hline
 h_0\,:\,(0,0,0)  &g_0\,:\,(1,0,0) & q_0\,:\,(\frac{1}{2},-\frac{1}{2},-\frac{1}{\sqrt{2}}) & p_0\,:\, (\frac{1}{2},\frac{1}{2},\frac{1}{\sqrt{2}})  \\
 h_1\,:\,(0,0,0) & g_1\,:\,(0,1,0) & q_1\,:\,(\frac{1}{2},-\frac{1}{2},\frac{1}{\sqrt{2}}) & p_1\,:\,(\frac{1}{2},\frac{1}{2},-\frac{1}{\sqrt{2}})\\
 h_+\,:\,(0,0,0) & g_+\,:\,(0,0,\frac{1}{\sqrt{2}}) & q_+\,:\,(\frac{1}{2},\frac{1}{2},0) & p_+\,:\,(\frac{1}{2},-\frac{1}{2},0) \\
 Y^+\,:\,(0,\frac{1}{2},\frac{1}{2\sqrt{2}}) & Y^-\,:\,(0,\frac{1}{2},- \frac{1}{2\sqrt{2}}) & \tilde Y^+\,:\,(\frac{1}{2},0,\frac{1}{2\sqrt{2}}) &
 \tilde Y^-\,:\,(\frac{1}{2},0,-\frac{1}{2\sqrt{2}}) \\
 \hline
\end{array}
\end{equation}
The number of generators constituting the spaces of type $Y$ are related
to the parameter $P$ in the following way
\begin{equation}
\dim Y^+ = \dim Y^-  = \dim \tilde{Y}^+ = \dim \tilde{Y}^- = P\,,
\end{equation}
while the other generators are non-degenerate. We can distinguish two
cases:
\begin{itemize}
\item $\Solv_Q(-1,0)$. In this case there are no spaces of type
$Y$ and the solvable algebra is the Borel algebra of the simple Lie
algebra $\mathrm{SO}(3,4)$. The corresponding quaternionic space
$L_Q(-1,0)$ is the symmetric space
$\frac{\mathrm{SO}(3,4)}{\mathrm{SO}(3) \times \mathrm{SO}(4)}$.
\item $\Solv_Q(-1,P)$ with $P \neq 0$. These algebras are not
related to any simple Lie algebra. The corresponding quaternionic spaces
are not symmetric.
\end{itemize}

  \item[Rank 4.]
For the rank 4 quaternionic algebras the subalgebra $U$ has the following
structure
\begin{eqnarray}
U&=& F_0 + \Solv_{SK} = F_0 + (F_1 + X_1) + (F_2 + X_2) + F_3,\nonumber\\
\left[F_I,\,F_J\right]&=&0, \qquad I,J=0,1,2,3, \nonumber\\
F_I&=&\{h_I,\,g_I\}\,\,;\,\,\,\,\left[h_I,\,g_I\right]=g_I, \quad
\left[h_i,\, X_i\right]=\frac{1}{2}X_i,\qquad i,j = 1,2,3.
\end{eqnarray}
It is convenient to set $X_2=X$ and $X_1 = Y + Z$, where $[F_2,Y] = 0$
and $[F_2,Z]=Z$. Decomposing the spaces $X,Y$ into the eigenspaces with
respect to the adjoint action of $h_3$ and the space $Z$ in eigenspaces
with respect to $h_2$, the corresponding eigenspaces are denoted as $X^+$
and $X^- = J_1X^+$, etc.

The gradings of the generators with respect to the Cartan subalgebra
$(h_0,h_1,h_2,h_3)$ are summarized in the following table
\cite{Alekseevsky1975,D'Auria:2004cu}
\begin{equation}
 \label{weightsV}
\begin{array}{|llll|}
\hline
 h_0\,:\,(0,0,0,0)  &g_0\,:\,(1,0,0,0) & q_0\,:\,(\frac{1}{2},-\frac{1}{2},-\frac{1}{2},-\frac{1}{2}) & p_0\,:\,(\frac{1}{2},\frac{1}{2},\frac{1}{2},\frac{1}{2}) \\
 h_1\,:\,(0,0,0,0) & g_1\,:\,(0,1,0,0) & q_1\,:\,(\frac{1}{2},-\frac{1}{2},\frac{1}{2},\frac{1}{2}) & p_1\,:\,(\frac{1}{2},\frac{1}{2},-\frac{1}{2},-\frac{1}{2}) \\
 h_2\,:\,(0,0,0,0) & g_2\,:\,(0,0,1,0) & q_2\,:\,(\frac{1}{2},\frac{1}{2},-\frac{1}{2},\frac{1}{2}) & p_2\,:\,(\frac{1}{2},-\frac{1}{2},\frac{1}{2},-\frac{1}{2}) \\
 h_3\,:\,(0,0,0,0) & g_3\,:\,(0,0,0,1) & q_3\,:\,(\frac{1}{2},\frac{1}{2},\frac{1}{2},-\frac{1}{2}) & p_3\,:\,(\frac{1}{2},-\frac{1}{2},-\frac{1}{2},\frac{1}{2}) \\
 X^+\,:\,(0,0,\frac{1}{2},\frac{1}{2}) & X^-\,:\,(0,0,\frac{1}{2},- \frac{1}{2}) & \tilde X^+\,:\,(\frac{1}{2},\frac{1}{2},0,0) & \tilde X^-\,:\,(\frac{1}{2},- \frac{1}{2},0,0) \\
Y^+\,:\,(0,\frac{1}{2},0,\frac{1}{2}) & Y^-\,:\,(0,\frac{1}{2},0,- \frac{1}{2}) & \tilde Y^+\,:\,(\frac{1}{2},0,\frac{1}{2},0) & \tilde Y^-\,:\,(\frac{1}{2},0,-\frac{1}{2},0) \\
Z^+\,:\,(0,\frac{1}{2},\frac{1}{2},0) & Z^-\,:\,(0,\frac{1}{2},-\frac{1}{2},0) & \tilde Z^+\,:\,(\frac{1}{2},0,0,\frac{1}{2}) & \tilde Z^-\,:\,(\frac{1}{2},0,0,-\frac{1}{2}) \\
\hline
\end{array}
\end{equation}
The corresponding solvable algebras will be denoted by
$\Solv_Q(q,P,\dot{P})$. The numbers $q$, $P$ and $\dot{P}$ are related to
the dimensions of the subspaces of type $X$, $Y$ and $Z$. The parameter
$q$ gives the dimension of the spaces\footnote{Here and below, $\dim
X=\dim X^+=\dim X^-=\dim \tilde X^+=\dim\tilde X^-$, and similar for $Y$
and $Z$.} of type $X$
  \begin{equation}\label{Xdimq}
  q =\dim X.
  \end{equation}
The spaces $Y^+ \bigcup Z^+$ form a representation of the Clifford
algebra in $q+1$ dimensions with positive signature. A similar result
holds for the other spaces of types $Y$, $Z$.  This representation can in
general be reducible. When $q \neq 0$ mod 4 however, there exists only
one irreducible representation of this Clifford algebra. The
representation formed by the spaces $Y \bigcup Z$ is thus uniquely
specified once the number of irreducible representations that constitute
it is given. This number is denoted by $P$. When $q = 0$ mod 4, there
exist 2 inequivalent representations of the Clifford algebra and one
needs 2 numbers, $P$ and $\dot{P}$ to indicate the representation content
of the representation formed by $Y \bigcup Z$. The numbers $P$ and
$\dot{P}$ are thus related to the dimension of the union of spaces of
type $Y$ and $Z$:
\begin{eqnarray}\label{YZdimP}
q\neq 0&:&(P + \dot{P}) = \frac{\dim Y + \dim
Z}{\mathcal{D}_{q+1}},\qquad \dim Y=\dim Z, \nonumber\\
q=0&:&P=\dim Y  , \qquad \dot{P} = \dim Z,
\end{eqnarray}
where ${\cal D}_{q+1}$ is the dimension of the irreducible
representations of the Clifford algebra in $q+1$ dimensions with positive
signature, which can be found in table \ref{tbl:RealCliff} of appendix
\ref{realcliffalg}. In general, the quaternionic dimension $n$ is related
to the parameters $(q,P,\dot{P})$ in the following way
\begin{equation}
\dim\Solv_Q(q,P,\dot P)=4(n+1),\qquad  n=3+q+(P+\dot P){\cal D}_{q+1}.
\label{genvaluen}
\end{equation}
All quaternionic solvable algebras of rank 4 necessarily have the subset
of generators $(h_I,g_I,q_I,p_I)$, where $I=0,1,2,3$, and can have some
or all of the three spaces of types $X,Y,Z$.  So, we can distinguish the
following particular cases:
\begin{itemize}
\item $\Solv_Q(0,0,0)\equiv\Solv(\mathrm{SO}(4,4))$, where the
spaces $X,Y,Z$ are all absent. Weights of the generators
$(h_I,g_I,q_I,p_I)$ (\ref{weightsV}) correspond to the positive root
system of $\mathrm{SO}(4,4)$. The corresponding space is
$\frac{\mathrm{SO}(4,4)}{\mathrm{SO}(4) \times \mathrm{SO}(4)}$.
\item $\Solv_Q(P,0,0) =\Solv_Q(0,P,0) =
\Solv_Q(0,0,P)\equiv\Solv(\mathrm{SO}(4,4+P))$, where only one of the
groups of spaces $X$, $Y$ or $Z$ is present. The set of weights of the
generators involved in this case (for example $(h_I,g_I,q_I,p_I,
Y^\pm,\tilde{Y}^\pm)$) corresponds to the positive root system of the
simple Lie algebra $\mathrm{SO}(4,5)$. The corresponding spaces are
$\frac{\mathrm{SO}(4,4+P)}{\mathrm{SO}(4) \times \mathrm{SO}(4+P)}$.
\item $\Solv_Q(0,P,\dot{P}) =
\Solv_Q(0,\dot{P},P)$, $P\,\dot{P}\neq 0$.  The interchange of the
subspaces $Y$ and $Z$ does not change the algebra. We note also that the
set of weights in this case does not correspond to any root system of the
simple type. These algebras lead to quaternionic spaces that are
nonsymmetric.
\item $\Solv_Q(q,P,\dot{P})$, $P+\dot{P}>0$. In these cases all
three spaces $X,Y,Z$ are present and the complete set of weights given in
(\ref{weightsV}) closes the positive root system of the Lie algebra
$\mathrm{F}_4$, whereas the full solvable algebra generically does not
give rise to a symmetric space.
\end{itemize}
\end{description}
 \par
After we have displayed the root system of the quaternionic spaces of
rank 4 in (\ref{weightsV}), we can see that the other normal quaternionic
spaces are truncations of this one (apart from an exception for the case
indicated as $SG_5$). This is symbolically indicated in the second column
of the tables \ref{tbl:SATS1} and \ref{tbl:SATS2} that are at the end of
this paper, and we will now clarify this for the various cases discussed
above.

In general, the non-generic cases can be obtained by deleting some rows
of (\ref{weightsV}), and restricting the roots consequently. The full
list of rows is $(0123XYZ)$ and that is mentioned in the second column of
the last line in table (\ref{tbl:SATS2}). For $q=0$, (\ref{Xdimq})
already implies that the row $X$ is absent. If $\dot P=0$, we also do not
have the $Z$ row, according to (\ref{YZdimP}), and for $P=\dot P=0$
neither the $Y$ row. This exhaust the rank~4 cases.

The table for rank 3, i.e. (\ref{weightsm1}), can be obtained by deleting
also one of the rows that contain generators in the Cartan subalgebra.
Keeping only $h_+=\frac{1}{\sqrt{2}}(h_2+h_3)$, rather than $h_2$ and
$h_3$, we have obtained root vectors for the rank 3 spaces from the
general ones in (\ref{weightsV}). The rows $Y$ and $Z$ become identical
when restricted to the weights under $(h_0, h_1, h_+)$ such that we only
have to keep one of them. For $P=0$ this row is also absent.

For $q=-2$, the rows $2$ and $3$ are absent. As such, the root vectors
have only two components, which implies that some roots in
(\ref{weightsV}) become identical, namely those of $q_0$ and $q_1$, of
$p_0$ and $p_1$, of $Y^+$ and $Y ^-$ and of $\tilde Y^+$ and $\tilde
Y^-$. This leads to the reduced table in (\ref{weightsm2}). The $Y$
generators are absent for $P=0$.

The other rank 2 system is $SG_5$, whose root vectors are modified with
respect to the systems denoted generically as $\Solv_Q(q,P,\dot P)$, as
shown in (\ref{weightsSG5}). Only rows 0 and 1 occur, but due to the
modified weights, there is no degeneracy as it was the case for
$\Solv_Q(-2,0)$.

Finally $SG_4$ consists of only the 0 row, and as such only the first
component of the root vectors is relevant. Then $q_0$ and $p_0$ are
identical, and this leads to (\ref{weightsSG4}).

The equations (\ref{Xdimq}) and (\ref{YZdimP}), which were mentioned in
the part for rank 4, also have a general validity, except of course for
$q<0$. However, in that case the negative value indicates the number of
rows between $(0123)$ that have to be deleted such that (\ref{genvaluen})
is also generally valid.

\subsection{The inverse $\mathbf{r}$- and $\mathbf{c}$-map}
 \label{ss:invrcmap}
{}From the above construction the one-to-one correspondence between the
quaternionic algebra $ \Solv_{Q}$ and its special K{\"a}hler subalgebra
$\Solv_{SK}$ follows. This correspondence is the inverse of the
$\mathbf{c}$-map discussed in the introduction, see (\ref{cmappa}), which
maps $\exp [\Solv_{SK}] \rightarrow \exp[ \Solv_{Q}]$. A similar
discussion can be developed for the $\mathbf{r}$-map. These relations are
better understood with the help of table~\ref{tbl:genV},
\begin{table}[!htb]
  \caption{\it Generators of the solvable algebras of special manifolds.
  The generators in different rows are related by complex structures as indicated in
  the last column. The three tables indicate the generators of the quaternionic algebra $\Solv_{Q}$,
  the $\mathbf{c}$--dual K{\"a}hlerian algebra $\Solv_{SK}$ and the real special algebra $\Solv_R$.
  The last line indicates the
  multiplicity of any entry in the column, where the last two columns are merged to
  get to a common expression. If $q$ is zero or negative, the $X$ column is absent, and the
  negative number indicates the number of columns to the left of it that are
  absent too, such that the general formula (\ref{genvaluen}) always
  holds.}\label{tbl:genV}
\begin{center}
$  \begin{array}{l||c|c|c|c|c|c|c||r|} \cline{2-9}
             &  p_0 & p_1 & p_2 & p_3 & \tilde{X}^- & \tilde{Y}^- & \tilde{Z}^- & -J_3A=J_2J_1A
             \\ \cline{2-9}
             & q_0 & q_1 & q_2 & q_3 &  \tilde{X}^+ & \tilde{Y}^+ & \tilde{Z}^+ &   J_2A
             \\ \cline{2-9}
  \mbox{ \raisebox{1.5ex}[0cm][0cm]{$ \Solv_{Q} = $ }}
             & g_0 & g_1 & g_2 & g_3 & X^- & Y^- & Z^- & J_1A\\ \cline{2-9}
             &  h_0 & h_1 & h_2 & h_3 & X^+ & Y^+ & Z^+ & A \\ \cline{2-9}
 \multicolumn{9}{ c }{ }  \\ \cline{2-9}
             &      &     &     &     &         &  \phantom{{\cal D}_{q+1}}    &    &   \\ \cline{2-9}
             &      &     &     &     &         &     &     &   \\ \cline{2-9}
 \mbox{ \raisebox{1.5ex}[0cm][0cm]{$ \Solv_{SK} = $ }}
             &      & g_1 & g_2 & g_3 & X^- & Y^- & Z^- & J_1A\\ \cline{2-9}
             &      & h_1 & h_2 & h_3 & X^+ & Y^+ & Z^+ & A \\ \cline{2-9}
 \multicolumn{9}{ c }{ }  \\ \cline{2-9}
             &      &     &     &     &         &     &     &   \\ \cline{2-9}
             &      &     &     &     &         &     &     &   \\ \cline{2-9}
 \mbox{ \raisebox{1.5ex}[0cm][0cm]{$ \Solv_{R} = $ }}
             &      &     &     &     & X^- & Y^- & Z^- &  A \\ \cline{2-9}
             &      &     & h_2 & h_3 &         &       &       &  A \\ \cline{2-9}
  \multicolumn{9}{ c }{ }  \\ \cline{2-8}
\# & 1 & 1 & 1 & 1 & q & \multicolumn{2}{c|}{(P+ \dot P){\cal D}_{q+1}}\\
 \cline{2-8}
\end{array}
$
\end{center}
\end{table}
representing first the generators of the quaternionic algebra $\Solv_{Q}$
as in (\ref{weightsV}) but rotated over 90$^\circ$. The different rows
are then related by the action of the complex structures.\footnote{The
action of $J_1$ is really column by column in these tables, but applying
$J_2$ and hence also $J_3$ on generators of the lowest row leads to a
linear combination of the generators in the row indicated by $J_2A$,
resp. $J_3A$.} Furthermore it indicates how the algebras $\Solv_{SK}$ and
$\Solv_R$ are embedded in $\Solv_{Q}$. The inverse $\mathbf{c}$--map and
$\mathbf{r}$--map can then be defined by deleting generators as
indicated, even if the root system is not that of a quaternionic space as
we will encounter for the Tits-Satake projected algebras. Observe that
the rank 1 quaternionic spaces cannot be included in this scheme.
 \par
The three solvable Lie algebras $ \Solv_{Q,SK,R}
\left(q,P,\dot{P}\right)$ constitute a family sharing the same values of
the parameters $(q,P,\dot{P})$. According to \cite{deWit:1992wf} families
of the corresponding homogeneous spaces are also denoted as
$L(q,P,\dot{P})$. The list of special homogeneous spaces is given in
table~\ref{allLpq}, where for symmetric spaces we specify the explicit
coset structure and we mention the various names assigned to some of the
spaces in the literature
\cite{Alekseevsky1975,Cecotti:1988ad,deWit:1992wf}.\footnote{The
homogeneous spaces $L_{R,SK,Q}( 0,P,\dot P)$ have been called before
$Y(P,\dot P)$, $K(P,\dot P)$ and $W(P,\dot P)$, while $L_{R,SK,Q}\left(
q,P\right)$ were called $X(P,q)$, $H(P,q)$ and $V(P,q)$. As these names
are not very illuminating, we use here the new names that we propose and
that show the systematics.}
\begin{table}[!htb]
\caption{\it Homogeneous very special real, special K{\"a}hler and
quaternionic spaces. The horizontal lines separate spaces of different
rank. The first non-empty space in each column has rank~1. Going to the
right or down a line increases the rank by~1. SG denotes an empty space,
which corresponds to supergravity models without scalars. The last 4
lines are already included in $L(q,P)$, but are mentioned separately as
they have extra symmetry, promoting them to symmetric
spaces.\label{allLpq}}
\begin{center}
\begin{tabular}{|l|l|ccc|c|}\hline
$C(h)$&range&real & K{\"a}hler & quaternionic   \\
\hline&&&&\\[-3mm]
$L(-3,P)$&$P\geq 0$&&&$\frac{\USp(2P+2,2)}{
\USp(2P+2)\times \SU(2)} $ \\[2mm]
$SG_4$&&&SG&$\frac{\SU(1,2)}{\SU(2)\times \U(1)} $ \\[2mm]
\hline&&&&\\[-3mm]
$L(-2,P)$&$P\geq 0$&&$\frac{\U(P+1,1)}{\U(P+1)\times \U(1)}$&
$\frac{\U(P+2,2)}{\U(P+2)\times \U(2)} $ \\[2mm]
$SG_5$&&SG&$\frac{\SU(1,1)}{\U(1)}$&$\frac{G_{2(2)}}{\SU(2)\times \SU(2)} $ \\[2mm]
\hline&&&&\\[-3mm]
$L(-1,0)$&&$\SO(1,1)$&$\left[\frac{\SO(2,1)}{\SO(2)}\right]^2$&$\frac{\SO(3,4)}{\SO(3) \times \SO(4) } $ \\[2mm]
$L(-1,P)$&$P\geq 1$&$\frac{\SO(P+1,1)}{\SO(P+1)}$& $\Solv_{SK}(-1,P)$ & $\Solv_Q(-1,P)$ \\[2mm]
\hline&&&&\\[-3mm]
$L(0,P)$&$P\geq 0$&$\frac{\SO(P+1,1)}{\SO(P+1)}\times
\SO(1,1)$&$\frac{\SU(1,1)}{\U(1)}\times \frac{\SO(P+2,2)}{\SO(P+2)\times
\SO(2)}$&$
\frac{\SO(P+4,4)}{\SO(P+4)\times \SO(4)} $\\[2mm]
$L(0,P,\dot P)$&$P\geq \dot P\geq 1$&
 $L_R(0,P,\dot P)$ & $L_{SK}(0,P,\dot P)$& $\!L_Q(0,P,\dot P)$\\[2mm]
%$Y(P,\dot P)$&$K(P,\dot P)$&$W(P,\dot P)$ \\[2mm]
$L(q,P)$&$\cases{q\geq 1\\ P\geq 1}$&
$L_R(q,P)$ & $L_{SK}(q,P)$& $L_Q(q,P)$\\[2mm]
%$X(P,q)$&$H(P,q)$&$V(P,q)$\\[2mm]
$L(4m,P,\dot P)$&$
\cases{  m\geq 1 \\
  P\geq \dot{P}\geq 1 }
$& $L_R(4m,P,\dot P)$ & $L_{SK}(4m,P,\dot P)$& $\!\!L_Q(4m,P,\dot P)$\\[2mm]
$L(1,1)$&& $\frac{\Sl(3,\mathbb{R})}{\SO(3)}$&$\frac{\Sp(6)}{\U(3)
}$&$\frac{F_{4(4)}}{\USp(6)\times \SU(2)}$\\[2mm]
$L(2,1)$&&
$\frac{\Sl(3,\mathbb{C})}{\SU(3)}$&$\frac{\SU(3,3)}{\SU(3)\times
\SU(3)\times \U(1)}$&$\frac{E_{6(2)}}{\SU(6)\times \SU(2)}$\\[2mm]
$L(4,1)$&& $\frac{\SU^*(6)}{\Sp(6)}$&$\frac{\SO^*(12)}{\SU(6)\times
\U(1)}$&$\frac{E_{7(-5)}}{{\SO(12)}\times \SU(2)}$\\[2mm]
$L(8,1)$&&
$\frac{E_{6(-26)}}{F_{4(-52)}}$&$\frac{E_{7(-25)}}{E_{6(-78)}\times
 \U(1)}$&$\frac{E_{8(-24)}}{E_{7(-133)}\times \SU(2)}$\\[2mm]
\hline \end{tabular}
\end{center}
\end{table}

This concludes the review of the structure of quaternionic, special
K{\"a}hler and real special solvable algebras that generate families of
homogeneous special manifolds. The corresponding solvable groups provide
translational symmetries, that are only a part of the full symmetry
groups, that we consider in the next subsection.
%%%%%%%%%%%%%%%%%%%%%%%%%%%%%%%%%
% Isometries %%%%%%%%%%%%%%%%%%%%
%%%%%%%%%%%%%%%%%%%%%%%%%%%%%%%%
\subsection{Isometry groups of homogeneous special geometries}
\label{isomhomspecgeom}

In this section we will consider the isometry algebras of homogeneous
special manifolds. This discussion will enable us to identify the part of
the isometry group that acts as a group of external automorphisms on the
solvable algebra, and hence will allow us to extract general
considerations on the concept of paint group in the context of
homogeneous special geometries. Indeed, it turns out that the paint group
has a structure that is general to all homogeneous special geometries.
This discussion will also enable us to identify the different
representations in which the generators of the solvable algebra are
grouped under the paint group.
 \par
We will concentrate on the chains of special manifolds that have a
five-dimensional origin (the very special spaces). This does not exhaust
the list of homogeneous quaternionic manifolds as clearly seen in table
\ref{allLpq}. The other cases, however, are symmetric and can for our
purpose be analysed in the standard way. We will first concentrate on the
special real manifold occurring in 5 dimensions, next we will discuss the
isometry algebras that appear upon application of the $\mathbf{r}$- and
$\mathbf{c}$-map respectively. Most of this discussion can be found in
\cite{deWit:1992wf}.

\subsubsection{Isometry algebras for homogeneous very special real
spaces}\label{homspecgeom}

\par In five dimensions, very special real manifolds can be defined
by an equation of the form $C(y) = 1$, where $C(y)$ is a cubic polynomial
of the scalar fields
% I have changed the indices such that they agree with the notations used in
% 6 dimensions. Also \mu and i where already used in different meanings.
%I have done it here with a newcommand, such that if you do not
% like the changes, they can easily be undone.
\newcommand{\indt}{\alpha }  % index of tensor multiplets
\newcommand{\indvone}{\Lambda} % index 1 of vector multiplets
\newcommand{\indvtwo}{\Sigma}  % index 2 of vector multiplets
$y=(y^1, y^2, y^\indt,y^\indvone)$ where $\indt$ and $\indvone$ run over
$q+1$ and $(P+\dot P){\cal D}_{q+1}$ values, respectively:
\begin{equation} \label{cubpolyn}
C(y) = 3 \Big\{ y^1 (y^2)^2 - y^1 y^\indt y^\indt - y^2 y^\indvone
y^\indvone + \gamma_{\indt \indvone \indvtwo} y^\indt y^\indvone
y^\indvtwo \Big\} \,,
\end{equation}
where $\gamma_{\indt \indvone \indvtwo}$ are the matrix elements of
gamma-matrices, forming a real representation of the Euclidean Clifford
algebra in $q+1$ dimensions. As already mentioned in the previous
section, this representation is not necessarily irreducible. The number
of irreducible representations contained in this representation is
denoted by numbers $P$ and $\dot{P}$, depending on whether there is only
one or there are two inequivalent irreducible representations of this
Clifford algebra. Denoting by $\mathcal{D}_{q+1}$ the dimension of an
irreducible representation of this Clifford algebra, the gamma-matrices
will thus in general be $(P + \dot{P}) \mathcal{D}_{q+1}$-dimensional
real matrices. The isometry group of the corresponding spaces is given by
the linear transformations of the scalars $y$ that leave (\ref{cubpolyn})
invariant.

The structure of the isometry algebra ${\cal X}$ can be summarized by
decomposing it with respect to the adjoint action of one of the Cartan
generators $\ulambda$. One finds that ${\cal X}$ has the following
structure:
\begin{equation} \label{decompisomalg}
{\cal X} = {\cal X}_{-3/2} \oplus {\cal X}_0 \oplus {\cal X}_{3/2}\,,
\end{equation}
where the subscript denotes the grading with respect to $\ulambda$. The
space ${\cal X}_{-3/2}$ consists of generators $\uzeta_\indvone$, that
are, however, only present for symmetric spaces. The space ${\cal
X}_{3/2}$ consists of generators $\uxi^\indvone$, which are always
present. The space ${\cal X}_0$ has the following structure:
\begin{equation} \label{decompisomalg0}
{\cal X}_0 = \so(1,1) \oplus \so(q+1,1) \oplus
\mathcal{S}_{q}(P,\dot{P})\,,
\end{equation}
where the $\so(1,1)$ factor is generated by $\ulambda$. The generators of
the additional invariances of (\ref{cubpolyn}), which are denoted by
$\mathcal{S}_q(P,\dot{P})$ are given by the antisymmetric matrices $S$
that commute with the matrices $\gamma_\indt$. Further properties on real
Clifford algebras and the groups $\mathcal{S}_q(P,\dot{P})$ are given in
appendix \ref{realcliffalg}, see especially table \ref{tbl:RealCliff}.
For later purposes we mention that the generators of ${\cal X}_{3/2}$
transform as a spinor representation under the adjoint action of
$\so(q+1,1)$, while under the adjoint action of
$\mathcal{S}_{q}(P,\dot{P})$ they transform in a vector representation.
 \par
The solvable subalgebra of the isometry group $G$ is given by
\begin{eqnarray}
\Solv_R &=& \so(1,1) \oplus \Solv(\so(q+1,1))  + {\cal
X}_{3/2}\nonumber\\
&=& \{\ulambda\,,\, \Solv(\so(q+1,1)),\,
\uxi^\indvone\}\,.\label{solvalg5d}
\end{eqnarray}
The solvable algebra of $\so(q+1,1)$ consists of one Cartan generator and
$q$ nilpotent generators. In this way we can make contact with the
Alekseevsky notations. Indeed, we have 2 Cartan generators (namely the
generator of $\so(1,1)$ and the Cartan generator of $\Solv(\so(q+1,1))$).
They agree with $h_2, h_3$ (or suitable linear combinations thereof) in
the last part of table \ref{tbl:genV}. The $q$ nilpotent generators of
$\Solv(\so(q+1,1))$ constitute the space $X^-$, while $Y^-$ and $Z^-$
together form ${\cal X}_{3/2}$.

\subsubsection{Isometry algebras for homogeneous very special
K{\"a}hler spaces}

Upon dimensional reduction, the homogeneous real spaces discussed in the
previous section, are enlarged to homogeneous special K{\"a}hler spaces. The
rank of the space is now increased with one unit. Decomposing the
isometry algebra $\mathcal{W}$ into the eigenspaces with respect to the
adjoint action of one of the Cartan generators $\ulambda'$, the following
structure occurs:
\begin{eqnarray}
&&{\cal W}={\cal W}'_0 \oplus  {\cal W}'_1 \oplus  {\cal W}'_2\ ,
\nonumber\\
&&{\cal W}'_0= \underline{\lambda }'\oplus \so(q+2,2)\oplus {\cal
S}_q(P,\dot P) \ ,\nonumber\\
&&{\cal W}'_1= \underline{\xi }^\indvone + \underline{b}_\indvone =(1,
\mathrm{spinor},
\mathrm{vector}),\nonumber\\
&&{\cal W}'_2= \underline{b}_1=(2,0,0), \label{homWroots}
\end{eqnarray}
where for ${\cal W}'_1$ and ${\cal W}'_2$ we mention in which
representations (under the adjoint action of the three subalgebras of
${\cal W}'_0 $) the generators transform. Note that in general no
generators with negative gradings occur. This is different for the
symmetric spaces, where there are generators $(\uzeta_\indvone,
\ua^\indvone)$ at grading $-1$, and where there is a generators $\ua^1$
at grading $-2$. In these cases the algebra is semisimple.

The solvable subalgebra of the isometry group $G$ is given by the
following set of generators:
\begin{equation} \label{solvalg4d}
\Solv_{SK} = \{ \ulambda', \Solv(\so(q+2,2)), \uxi^\indvone,
\ub_\indvone, \ub_1 \} \,.
\end{equation}
Again it is possible to make contact with the second part of table
\ref{tbl:genV}. The solvable algebra of $\so(q+2,2)$ consists of $2q + 4$
generators. Two of these belong to the Cartan subalgebra of $\Solv_{SK}$,
$2q$ of them constitute the 2 spaces $X^+$ and $X^-$, while the remaining
2 generators, together with $\ub_1$ constitute the $g$-generators.
Furthermore, the generators $\uxi^\indvone$ and $\ub_\indvone$ deliver
the $Y^\pm$-, $Z^\pm$-generators.

\subsubsection{Isometry algebras for homogeneous very special
quaternionic spaces}

After dimensional reduction from 4 to 3 dimensions, the very special
K{\"a}hler spaces of the previous section are enlarged to very special
quaternionic manifolds. The corresponding isometry algebras are likewise
extended and now have the following form (the index $M$ runs over $q+2$
values):
\begin{eqnarray}
{\cal V}&=& {\cal V}'_0 +{\cal V}'_{1} + {\cal V}'_2 \ ,\nonumber\\
{\cal V}'_0 &=&\so(1,1)\oplus \so(q+3,3)\oplus {\cal
S}_q(P,\dot P), \nonumber\\
{\cal V}'_1 &=& (\underline{\xi}^\indvone,\underline{b}_\indvone)
\oplus(\underline{\alpha}{}_\indvone,\underline{\beta}{}^\indvone) =
(1,\mathrm{spinor},\mathrm{vector}) ,
\nonumber\\
{\cal V}'_2 &=& \underline{\epsilon}{}_+\oplus
(\underline{\alpha}{}_1,\underline{\beta}{}^M, \underline{\beta}{}^0)
\oplus \underline{b}{}_1 = (2,\mathrm{vector},0) ,
\end{eqnarray}
where we indicated the representation of ${\cal V}'_1$ and ${\cal V}'_2$
according to the three subalgebras of ${\cal V}'_0$. As in the previous
cases, we decomposed the isometry algebra in terms of the gradings with
respect to the Cartan generator $\uepsilon^\prime$. Again, for the
symmetric spaces, the isometry algebra will be extended with additional
generators, with gradings $-1$ and $-2$ with respect to
$\uepsilon^\prime$, such that the algebra is semisimple. The solvable
algebra in the quaternionic case is
\begin{equation}
\Solv_Q = \{ \uepsilon', \Solv(\so(q+3,3)), \uxi^\indvone, \ub_\indvone,
\ub_1, \underline{\alpha}{}_\indvone,\underline{\beta}{}^\indvone,
\underline{\epsilon}{}_+, \underline{\alpha}{}_1,\underline{\beta}{}^M,
\underline{\beta}{}^0\} \,.
\end{equation}

\subsection{Six dimensional origin of special geometries}
 \label{ss:6dorigin}
$\mathrm{G}/\mathrm{H}$ sigma-models in 3 dimensions can be interpreted
as toroidal dimensional reduction of a higher-dimensional theory
\cite{Cremmer:1999du}. This reinterpretation, known as oxidation, relies
on a suitable embedding of $\Sl(D-2,\mathbb{R})\subset \mathrm{G}$, that
represents the restoration of the higher dimensional gravity sector
\cite{Keurentjes:2002xc}. Then the remaining generators of $\mathrm{G}$
are organized in $\Sl(D-2,\mathbb{R})$ irreps and give rise to form
fields in higher dimensions.
\par
The maximal dimension $D$ to which the original $3$-dimensional coset
sigma-model can be oxidized is called "oxidation endpoint", and is
limited by the split rank $r_3$ of the quaternionic isometry algebra
$\mathrm{G}$. This can already be seen from table \ref{allLpq} for the
spaces with $r_3\leq 2$. The general rule is
\begin{equation}
D=r_3+2, \qquad\mbox{or}\qquad  D=r_3 + 3, \label{Dendpoint}
\end{equation}
where the latter only occurs for theories that have an empty scalar
manifold in $D$ dimensions. This is the case with $D=4$ for $SG_4$, for
$D=5$ for $SG_5$ and we will see below that it is also true for $D=6$
coupled only to vector multiplets.
 \par
Since the toroidal reduction/oxidation preserves the number of
supersymmetries, the resulting theories in $D=6$ are still theories with
$\mathcal{N}_Q=8$. These are also $\mathcal{N}=2$ theories, which have
chiral generators and are therefore often denoted as $(1,0)$
supergravities. As was shown in \cite{deWit:1991nm,Andrianopoli:2004xu},
all rank 3 and rank 4 homogeneous quaternionic spaces (not only the
symmetric ones) have the same oxidation endpoint, namely $D=6$ and they
all can be described from reduction on a $T^3$ torus. The only
ingredients that distinguish one quaternionic space from another are the
numbers of tensor and vector multiplets coupled to the gravity multiplet
in six dimensions.

This unified higher-dimensional description of very special real, K{\"a}hler
and quaternionic-K{\"a}hler spaces allows us to understand the origin of
their isometry groups. In particular, it allows us to single out the part
of their symmetries not related to the very procedure of dimensional
reduction, i.e. the isometries already present in $D=6$ that simply
permute matter multiplets of the same type.
 \par
These $D=6$ supergravities have  an obligatory gravitational multiplet,
consisting of metric, anti-selfdual 2-form and two gravitinos
$(g_{MN},\psi_{i|M},B^-_{MN})$, where  $M=0,\ldots , 5$, $i=1,2$. Then
also a number of matter multiplets can be coupled, namely tensor
multiplets containing each a self--dual 2-form, 2 spinor fields and a
scalar $(B^+_{MN},\chi^i,\phi)$, vector multiplets containing a
six-dimensional vector and two gauginos $(A_M,\lambda^i)$, and
hypermultiplets containing only scalar fields and spinor fields. The full
theory is given in \cite{Riccioni:2001bg}. Hypermultiplets are not
relevant in our construction, since hyperscalars form a normal
quaternionic space already in six dimensions that does not become
enlarged when stepping down to $D=5,4,3$, so they cannot give rise to
chains of manifolds connected with ${\bf r}$- and ${\bf c}$-maps.
 \par
The total bosonic field content of the remaining gravity-matter system of
$D=6$, $(1,0)$ supergravity is:
\begin{equation}\label{bfieldcont}
    (g_{MN},B^I_{MN},A_M^\Lambda,\phi^\alpha),\qquad I=1,\ldots ,n_T+1,
    \quad \Lambda = 1,\ldots , n_V, \qquad \alpha = 1,\ldots ,n_T,
\end{equation}
where we allow to have $n_V$ vector multiplets and $n_T$ tensor
multiplets.

Scalars of the theory come only from tensor multiplets and parametrize
the coset manifold \cite{Romans:1986er,Andrianopoli:1996ve}:
\begin{equation}\label{scalD6}
    \mathcal{M}^{D=6}=\frac{\SO(n_T,1)}{\SO(n_T)}.
\end{equation}

The isometry group of the six-dimensional scalar manifold,
$\mathrm{SO}(n_T,1)$, acts not only on scalars, but on the two-forms
$B^I$ in a way that preserves their coupling to six-dimensional gravity.
In order to promote this scalar isometry to be a symmetry of the whole
theory, we should consider also the topological term
\cite{Nishino:1984gk,Bergshoeff:1986mz,VanProeyen:1985ib} that describes
the coupling of tensor multiplets to vector multiplets
\begin{equation}\label{lagr}
   \mathcal{L}_{\rm CS}=C_{I\Lambda\Sigma}B^IF^\Lambda F^\Sigma,
\end{equation}
where $F^\Lambda$ are the field strength of the vectors. This generic
form was already conjectured in \cite{Romans:1986er,deWit:1991nm} and was
found in \cite{Riccioni:1999xq}. This term breaks explicitly the
$\SO(n_T,1)$ symmetry, unless the vectors transform under a suitable
$n_V$--dimensional representation $R_V$ with the property that the
symmetric product of two of these representations contains the vectorial
representation of $\SO(n_T,1)$, that is the one under whose action the
tensors transform. In the particular case in which $R_V$ is a spinorial
representation of $\SO(n_T,1)$, the corresponding supergravity theories
in $D=6$ give rise, after dimensional reduction, to the chains of
manifolds with special geometry $L(q,P,\dot{P})$.
 \par
The parameters $(q,P,\dot{P})$ are related to the number of tensor and
vector multiplets of $D=6$ supergravity as follows:
\begin{equation}\label{rel}
n_T=q+1, \qquad n_V=(P+\dot{P})\mathcal{D}_{q+1},
\end{equation}
where $\mathcal{D}_{q+1}=\mathcal{D}_{n_T}$ is also the dimension of an
irreducible real representation\footnote{This uses the equivalence of the
even elements of the Clifford algebra ${\cal C}^+(q+1,1)$ with ${\cal
C}(q+1,0)$. Only the even elements are needed for the spinor
representation.} of $\Spin(q+1,1)$. This allows their use for an
$\SO(q+1,1)$ invariant coupling in (\ref{lagr}).
 \par
Following this interpretation, homogeneous $L(-1,0)$ and $L(-1,P)$ in
$D=6$ correspond, respectively, to pure supergravity and to supergravity
coupled to $P$ vector multiplets only. These are the exceptions that
follow the second case of (\ref{Dendpoint}).
 \par
This allows to understand the origin of the $\mathcal{S}_q(P,\dot{P})$
part of the isometry groups of the scalar manifolds in lower dimensions.
The representation $n_V$ of six-dimensional vector fields can be big
enough to admit more symmetries than $\mathrm{SO}(q+1,1)$, that are in
the centralizer of the reducible (if $P+\dot{P}>1$) real spinor
representation of $\mathrm{SO}(q+1,1)$. The new symmetries constitute the
$\mathcal{S}_q(P,\dot{P})$ group, which becomes part of the isometry
groups after dimensional reduction, and we see that it is "decoupled"
from scalars in $D=6$. This tells us also that the scalars $\phi^\alpha$
and more generally, all fields coming from tensor multiplets and that do
not transform under $\mathcal{S}_q(P,\dot{P})$ group, are related to the
$X$ space and singlets $(h_I,g_I,p_I,q_I)$ in Alekseevsky's
classification, whereas vector fields $A_M^\Lambda$ under dimensional
reduction give rise to the spaces $Y$ and $Z$.
\par
To make this identification more precise, we should anticipate the role
of the paint group that is an important compact part of the isometry
groups given by $\mathrm{G}_{\rm paint} = \mathrm{SO}(q)\times
\mathcal{S}_q(P,\dot{P})$, as we will show later, see (\ref{Gpaintgen}).
All fields of the $D=6$ supergravities considered here fall into
representations of the paint group. For example, the $(q+1)$ scalars
$\phi^\alpha$ do not transform under the $\mathcal{S}_q(P,\dot{P})$
subgroup. Under the action of the $\SO(q)\subset \SO(1,q+1)$ part of the
paint group they split in two subsets: a singlet $\alpha$ and a vector
that we call $X^-$ in order to make contact with Alekseevsky's
classification. We can give a solvable description to the scalar manifold
they parametrize, identifying the scalar field $\alpha$ with a noncompact
Cartan generator $\ualpha$ of $\mathrm{SO}(q+1,1)$ and fields from $X^-$
with a space of remaining generators, all having the same grading with
respect to $\ualpha$. The Alekseevsky diagram that represents this
solvable algebra $\Solv(\mathrm{SO}(q+1,1))$, following the scheme of
table \ref{tbl:genV} is
\begin{equation}
\begin{array}{l||c|c|c|c|c|c|c||} \cline{2-8}
            &   \phantom{p_0} & \phantom{p_1} & \phantom{p_2} & \phantom{p_3} &
            \phantom{\tilde{X}^-} & \phantom{\tilde{Y}^-} & \phantom{\tilde{Z}^-}
   \\ \cline{2-8}
             &    &   &  &   &   &  & \\ \cline{2-8}
  \mbox{ \raisebox{1.5ex}[0cm][0cm]{$ \Solv_{\SO(n_T,1)} = $ }}
             &   &  &   &   & X^- &   & \\ \cline{2-8}
             &   & & & \underline{\alpha} &   &   &  \\ \cline{2-8}
\end{array}
%\begin{array}{|c|c|c|c|c|c|c|}
%\hline
% & & & & X^- & & \\
% \hline
% & & &\underline{\alpha} & & & \\
% \hline
%\end{array}.
\end{equation}
The $(q+2)$ two-form fields $B_{MN}^I$ do not transform under the
$\mathcal{S}_q(P,\dot{P})$ subgroup. The index $I$ is in the fundamental
representation of $\SO(q+1,1)$, which under the $\SO(q)$ subgroup splits
into $1+1+q$. $B^1$ and $B^2$ constitute two singlets under the paint
group while $B^i$ transform in the vector ($q$--dimensional)
representation of $\SO(q)$.

The full paint group $\SO(q)\times \mathcal{S}_q(P,\dot{P})$ acts
linearly on the vectors $A_M^\Lambda$ in $D=6$. In particular, as we
mentioned before, the vector fields constitute a $n_V$--dimensional, real
spinor representation of $\SO(q+1)$ that is split into two irreducible
representations under the subgroup $\SO(q)$. Under the
$\mathcal{S}_q(P,\dot{P})$ subgroup vectors transform as a real
representation of dimension
$(P+\dot P)\mathcal{D}_{q+1}/2$.
\par
After dimensional reduction of $D=6$ supergravity on a circle the scalars
$A_5^\Lambda$ coming from vector fields and the one coming from the
metric, $g_{55}$, enlarge the scalar manifold of $D=6$ to a real special
version of $L(q,P,\dot{P})$. The metric scalar is represented, in the
Alekseevsky formalism, by a singlet $\underline{\lambda}$, while those
coming from the vectors of the two different spin representations span
the subspaces $Y^-$ and $Z^-$.
\begin{table}[!htb]
  \caption{\it The bosonic fields of $D=6$ supergravity theories, their
  representation under the paint
group and their contribution to scalars of
  $D=5$ and $D=4$  supergravities.
  \label{tbl:D6reps}}
\begin{center}
  $\begin{array}{|c|c|c|c|c|}\hline
 & \SO(q) & \mathcal{S}_q(P,\dot P)& D=5\mbox{ scalars} & \mbox{additional scalars in }D=4 \\
\hline
g_{MN} & - & - & g_{55}\rightarrow 1& g_{44},g_{45}\rightarrow 2\\
B^{1,2}_{MN} & - & - & -&B^{1,2}_{45}\rightarrow 2 \\
B^i_{MN} & q & - & -&B^i_{45}\rightarrow q\\
A^\Lambda_M & %\frac{1}{e_q}
\mathcal{D}_{q+1}& %e_q
(P+\dot P) &
A^\Lambda_5\rightarrow (P+\dot P)\mathcal{D}_{q+1}
& A^\Lambda _4\rightarrow (P+\dot P)\mathcal{D}_{q+1} \\
\phi^\alpha  & q+1 & - & q+1& 0\\
 \hline
\end{array}$
\end{center}
\end{table}
 \par
Reducing the theory further to $D=4$ and dualizing all the tensors to
vectors (Einstein--Maxwell supergravity \cite{Gunaydin:1983mi}), we
obtain a special K{\"a}hler geometry defined by new copies of $X$, $Y$ and
$Z$ spaces and four new singlets. From the transformation properties of
the $D=6$ fields under the paint group it is clear that the new copy of
the $X$ space is spanned by the scalars coming from the $B^i_{MN}$
tensors and that the new copies of $Y$ and $Z$ spaces are necessarily
identified with the fifth components of the vectors. From the new
singlets, instead, two are coming from the metric ($g_{44}$ and $g_{45}$)
and two from the $B^{1,2}_{MN}$ tensors.
\par
We summarize the above statements in table \ref{tbl:D6reps}.

%%%%%%%%%%%%%%%%%%%%%%%%%%%%%%%%%%%%%%%%%%%%%%%%%%%%%%%%%%%%%%%%%%
\section{The Tits-Satake projection}
\label{Titsatesection}
%%%%%%%%%%%%%%%%%%%%%%%%%%%%%%%%%%%%%%%%%%%%%%%%%%%%%%%%%%%%%%%%%%

In this section we explain the Tits-Satake projection of a metric
solvable Lie algebra and how it is related to the notions of
\textit{paint} group $\mathrm{G}_{\mathrm{paint}}$ and \textit{subpaint}
group $\mathrm{G}_{\mathrm{subpaint}} \subset
\mathrm{G}_{\mathrm{paint}}$. We will extract these notions from the case
of the Tits-Satake projections of solvable Lie algebras associated with
symmetric spaces $\Solv(\mathrm{G/H})$ and generalize them to the case of
metric solvable Lie algebras associated with nonsymmetric spaces such as
homogeneous special geometries.

\subsection{TS projection for non maximally split symmetric spaces}
\label{genediscussa}

Following the discussion of \cite{Fre':2005sr} let us recall that if the
scalar manifold of supergravity is a \textit{non maximally noncompact
manifold} $\mathrm{G/H}$ the Lie algebra of the numerator group is some
appropriate real form $\mathbb{G}_R$ of a complex Lie algebra
$\mathbb{G}$.
The Lie algebra $\mathbb{H}$ of the denominator $\mathrm{H}$ is the
maximal compact subalgebra $\mathbb{H} \subset \mathbb{G}_R$. Denoting,
as usual, by $\mathbb{K}$ the orthogonal complement of $\mathbb{H}$ in
$\mathbb{G}_R$:
\begin{equation}
  \mathbb{G}_R = \mathbb{H} \, \oplus \,\mathbb{K}
\label{Grdecompo}
\end{equation}
and defining as noncompact rank or rank of the coset $\mathrm{G/H}$ the
dimension of the noncompact Cartan subalgebra:\footnote{The chosen CSA is
a maximally noncompact one. We thank A. Keurentjes and Ph. Spindel for
this remark.}
\begin{equation}
  r_{\rm nc}\, = \, \mbox{rank} \left( \mathrm{G/H}\right)  \, \equiv \, \mbox{dim} \,
  \mathcal{H}^{\rm nc} \quad ; \quad \mathcal{H}^{\rm nc} \, \equiv \,
  \mbox{CSA}_{\mathbb{G}} \, \bigcap \, \mathbb{K}
\label{rncdefi}
\end{equation}
we obtain that $r_{\rm nc} \leq \mbox{rank}(\mathbb{G})$, where the
equality is the statement that the manifold is \emph{maximally
noncompact} (or `\emph{maximally split}').
\par
When the equality is strict, the manifold $\mathrm{G_R/H}$ is still
metrically equivalent to a solvable group manifold but the form of the
solvable Lie algebra $\Solv(\mathrm{G_R/H})$, whose structure constants
define the Nomizu connection, is now more complicated than in the
maximally noncompact case. The Tits-Satake theory of noncompact cosets
and split subalgebras is a classical topic in Differential Geometry and
appears in textbooks. Yet, within such a mathematical framework there is
a peculiar universal structure of the solvable algebra
$\Solv(\mathrm{G_R/H})$ that, up to our knowledge, had not been observed
before \cite{Fre':2005sr} namely that of paint and subpaint groups and
which extends beyond symmetric spaces as we demonstrate in the present
paper.
\par
Explicitly we have the following scheme. One can split the Cartan
subalgebra into its compact and noncompact subalgebras:
\begin{equation}
\begin{array}{rcccc}
  \mathrm{CSA}_{\mathbb{G}_{R}} &=& {\rm i} \mathcal{H}^{\rm comp} & \oplus
  &
  \mathcal{H}^{\rm nc} \\
  \null & \null &\Updownarrow & \null & \Updownarrow  \\
  \mathrm{CSA}_{\mathbb{G}} &=& \mathcal{H}^{\rm comp} & \oplus
  & \mathcal{H}^{\rm nc} \
  \end{array},
\label{hcompnoncom}
\end{equation}
and these parts are orthogonal using the Cartan-Killing metric.
Therefore, every vector in the dual of the full Cartan subalgebra, in
particular every root $\alpha$, can be decomposed into its transverse and
parallel part to $\mathcal{H}^{\rm nc}$:
\begin{equation}
  \alpha = \alpha_{\bot} \, \oplus \,\alpha_{||} .
\label{splittus}
\end{equation}
The Tits-Satake projection consists of two steps. First one sets all
$\alpha_{\bot} = 0 $, projecting the original root system
$\Delta_\mathbb{G}$ onto a new system of vectors $\overline\Delta$ living
in a Euclidean space of dimension equal to the noncompact rank $r_{\rm
nc}$. The set $\overline{\Delta}$ is called a restricted root system. It
is not an ordinary root system in the sense that roots can occur with
multiplicities different from one and $2\alpha_{||}$ can be a root if
$\alpha_{||}$ is one. In the second step, one deletes the multiplicities
of the restricted roots. Thus we have
\begin{equation}
  \Pi_{\mathrm{TS}} \quad : \quad \Delta_\mathbb{G} \, \mapsto  \, \Delta_{\rm TS},;\qquad
  \Delta_\mathbb{G}\ \stackrel{\alpha_{\bot} = 0}{\longmapsto}\ \overline{\Delta}\
 % \stackrel{\mbox{deleting multiplicities}}{\longmapsto}
 \begin{array}{c}
   \mbox{\small deleting} \\[-4mm]
   \longmapsto \\[-4mm]
   \mbox{\small multiplicities}
 \end{array}\
  \Delta_{\rm TS}.
 \label{projecto}
\end{equation}
\textit{If $\overline{\Delta}$ contains no restricted root that is the
double of another one}, then $\Delta_{\rm TS}$ is a root system of simple
type. We will show later that this root subsystem defines a Lie algebra
$\mathbb{G}_{\mathrm{TS}}$, \emph{the Tits-Satake subalgebra} of
$\mathbb{G}_R$:
\begin{equation}
  \Delta_{\rm TS} = \mbox{root system of } \mathbb{G}_{\mathrm{TS}}, \qquad
\mathbb{G}_{\mathrm{TS}} \, \subset \,\mathbb{G}_R. \label{TitsSatake}
\end{equation}
The Tits-Satake subalgebra $\mathbb{G}_{\mathrm{TS}}$ is, as a
consequence of its own definition, the maximally noncompact real section
of its own complexification. For this reason, considering its maximal
compact subalgebra $\mathbb{H}_{\mathrm{TS}} \, \subset \,
\mathbb{G}_{\mathrm{TS}}$ we have a new smaller coset
$\mathrm{G_{\mathrm{TS}} / H_{\mathrm{\mathrm{TS}}}}$ which is maximally
split and whose associated solvable algebra
$\Solv(\mathrm{G_{\mathrm{TS}} / H_{\mathrm{TS}}})$ has the standard
structure utilized in \cite{Fre:2005bs} to  prove complete integrability
of supergravity compactified to 3 dimensions. By itself this result
demonstrates the relevance of the Tits-Satake projection of a
supergravity theory, which, as we already emphasized, certainly extends
much beyond cosmic billiards.
 \par
\textit{In the case when the doubled restricted roots are present in}
$\overline{\Delta}$, the projection cannot be expressed in terms of a
simple Lie algebra, but the concept remains the same. The root system is
the so-called $bc_r$ system, with $r=r_{\rm nc}$ the noncompact rank of
the real form $\mathbb{G}$. It is the root system of a group
$\mathrm{G_{\mathrm{TS}}}$, which is now non-semisimple. The manifold is
similarly defined as $\mathrm{G_{\mathrm{TS}}/H_{TS}}$, where
$\mathrm{H_{TS}}$ is the maximal compact subgroup of
$\mathrm{G_{\mathrm{TS}}}$.
\par
The next question is: what is the relation between the two solvable Lie
algebras $\Solv(\mathrm{G_{R} / H})$ and $\Solv(\mathrm{G_{\mathrm{TS}} /
H_{\mathrm{TS}}})$? The answer can be formulated through the following
statements A-E.
 \par
\paragraph{A]}
In a projection more than one higher dimensional vector can map to the
same lower dimensional one. This means that in general there will be
several roots of $\Delta_\mathbb{G}$ that have the same image in
$\Delta_{\rm TS}$. The compact roots vanish under this projection.
Therefore, apart from these compact roots, there are two types of roots:
those that have a distinct image in the projected root system and those
that arrange in multiplets with the same projection. We can split the
root spaces in subsets according to whether there is such a degeneracy or
not. Calling $\Delta^+_\mathbb{G}$ and $\Delta_{\rm TS}^+$ the sets of
positive roots of the two root systems, we have the following scheme:
\begin{eqnarray}
 && \begin{array}{lclclcc}
 \Delta^+_\mathbb{G} & = & \Delta^\eta& \bigcup &\Delta^\delta& \bigcup & \Delta_{\mathrm{comp}} \\
    \downarrow \Pi_{\mathrm{TS}}& & \downarrow \Pi_{\mathrm{TS}} &   & \downarrow \Pi_{\mathrm{TS}} &   &   \\
\Delta_{\mathrm{TS}}^+ &=& \Delta_{\mathrm{TS}}^\ell & \bigcup & \Delta_{\mathrm{TS}}^s &&  \\
  \end{array} \nonumber\\[3mm]
&& \forall \alpha^\ell  \in \Delta_{\mathrm{TS}}^\ell \ : \ \dim
\Pi^{-1}_{\mathrm{TS}} \left[ \alpha^\ell \right]  = 1, \qquad %\mbox{ and
 \forall \alpha^s  \in  \Delta_{\mathrm{TS}}^s \ : \ \dim
\Pi^{-1}_{\mathrm{TS}} \left[ \alpha^s \right]  =   m[\alpha^s]>1.
 \label{SubsetsRootspaces}
\end{eqnarray}
The $\delta  $ part thus contains all the roots that have multiplicities
under the Tits-Satake projection while the roots in the $\eta $ part have
no multiplicities. These roots of type $\eta  $ are orthogonal to $\Delta
_{\mathrm{comp}}$. Indeed, this follows from the fact that for any two
root vectors $\alpha $ and $\beta $ where there is no root of the form
$\beta +m\alpha $ with $m$ a non-zero integer, the inner product of
$\beta $ and $\alpha $ vanishes. It also follows from this definition
that in maximally split symmetric spaces, in which case $\Delta
_{\mathrm{comp}}=\emptyset$, all root vectors are in $\Delta ^\eta $ or
$\Delta ^\ell $ (as the Tits-Satake projection is then trivialized).

These subsets moreover satisfy the following properties under addition of
root vectors:
\begin{equation}
\begin{array}{|l|l|}
\hline \mathbb{G}&  \mathbb{G}_{\mathrm{TS}} \\ \hline \Delta ^\eta
+\Delta^\eta \subset\Delta ^\eta &
\Delta_{\mathrm{TS}}^\ell+\Delta_{\mathrm{TS}}^\ell
  \subset\Delta_{\mathrm{TS}}^\ell \\
 \Delta ^\eta +\Delta^\delta  \subset\Delta ^\delta  &
 \Delta_{\mathrm{TS}}^\ell+\Delta_{\mathrm{TS}}^s
  \subset\Delta_{\mathrm{TS}}^s \\
  \Delta ^\delta  +\Delta^\delta  \subset\Delta ^\eta\bigcup\Delta ^\delta
   & \Delta_{\mathrm{TS}}^s+\Delta_{\mathrm{TS}}^s
  \subset\Delta_{\mathrm{TS}}^\ell \bigcup\Delta_{\mathrm{TS}}^s\\
   \Delta_{\mathrm{comp}}+ \Delta ^\eta =\emptyset &   \\
    \Delta_{\mathrm{comp}}+ \Delta ^\delta \subset \Delta ^\delta  &\\
    \hline
\end{array}
\end{equation}
Because of this structure we can enumerate the generators of the solvable
algebra $\Solv(\mathrm{G_{R} / H})$ in the following way:
\begin{eqnarray}
\Solv(\mathrm{G_{R} / H})&=&\left\{H_i,
\Phi_{\alpha^\ell},\Omega_{\alpha^s | I}\right\} \nonumber\\
H_i & \Rightarrow & \mbox{Cartan generators} \nonumber\\
\Phi_{\alpha^\ell} & \Rightarrow & \eta -\mbox{roots} \nonumber\\
\Omega_{\alpha^s | I} &\Rightarrow & \delta -\mbox{roots} \quad ; \quad
(I=1,\dots , m[\alpha^s] ). \label{enumerato}
\end{eqnarray}
The index $I$ enumerating the $m$--roots of $\Delta_{\mathbb{G}_R}$ that
have  the same projection in $\Delta_{\mathrm{TS}}$ is named the
\textit{paint index}.
\paragraph{B]}
There exists a \textit{compact subalgebra} $\mathbb{G}_{\mathrm{paint}} \, \subset
  \, \mathbb{G}_R$ which acts as  an algebra of outer automorphisms ({\it i.e.} outer derivatives) of the  solvable
  algebra $\Solv_{\mathbb{G}_R} \equiv \Solv(\mathrm{G_{R} /
  H}) \subset \mathbb{G}_R$, namely:
\begin{equation}
  \left[ \mathbb{G}_{\mathrm{paint}} \, , \, \Solv_{\mathbb{G}_R} \right]  \subset
  \Solv_{\mathbb{G}_R}.
\label{linearrepre}
\end{equation}
\paragraph{C]}
The Cartan generators $H_i$ and the generators $\Phi_{\alpha^\ell} $ are
singlets under the action of  $\mathbb{G}_{\rm paint}$, {\it i.e.} each
of them  commutes with the whole of $\mathbb{G}_{\rm paint}$:
\begin{equation}
  \left[ H_i \, , \,\mathbb{G}_{\mathrm{paint}}\right] \, = \, \left[ \Phi_{\alpha^\ell} \, ,
  \,\mathbb{G}_{\rm paint}\right]\, = \, 0
\label{hiphicommuti}
\end{equation}
On the other hand, each of the multiplets of generators $\Omega_{\alpha^s
| I}$ constitutes an orbit under the adjoint action of the paint group
${G}_{\mathrm{paint}}$, {\it i.e.} a linear representation
$\mathbf{D}{[\alpha^s]}$ which, for different roots $\alpha^s$ can be
different:
\begin{equation}
  \forall \, X \, \in \, \mathbb{G}_{\mathrm{paint}} \quad : \quad  \left[ X \, , \,
  \Omega_{\alpha^s |
I}\right] \, = \, \left( D^{[\alpha^s]}[X]\right) _I^{\phantom{I}J} \,
\Omega_{\alpha^s | J} \label{replicas}
\end{equation}
\paragraph{D]}
The \textit{paint algebra} $\mathbb{G}_{\mathrm{paint}}$ contains a
subalgebra
\begin{equation}
  \mathbb{G}^0_{\rm subpaint} \, \subset \, \mathbb{G}_{\mathrm{paint}}
\label{paint0}
\end{equation}
such that with respect to $\mathbb{G}^0_{\mathrm{subpaint}}$,  each
$m[\alpha^s]$--dimensional representation $\mathbf{D}{[\alpha^s]}$
branches as follows:
\begin{equation}
  \mathbf{D}{[\alpha^s]} \,
  \stackrel{\mathbb{G}^0_{\rm subpaint}}{\Longrightarrow} \,
  \underbrace{{\mathbf{1}}}_{\mbox{singlet}}
  \, \oplus \, \underbrace{{\mathbf{J}}}_{(m[\alpha^s]-1)-\mbox{dimensional}
  }
\label{splittatonibus}
\end{equation}
Accordingly we can split the range of the multiplicity index $I$ as
follows:
\begin{equation}
  I = \left\{ 0,x \right\},\qquad  x=1,\dots,m[\alpha^s]-1.
\label{Irango}
\end{equation}
The index $0$ corresponds to the singlet, while $x$ ranges over the
representation $\mathbf{J}$.
\paragraph{E]}
The tensor product $\mathbf{J} \otimes \mathbf{J} $ contains both the identity
representation $\mathbf{1}$ and the representation $\mathbf{J}$ itself.
Furthermore, there exists, in the representation $\bigwedge ^3
\mathbf{J}$ a $\mathbb{G}^0_{\mathrm{subpaint}}$-invariant tensor
$a^{xyz}$ such that the two solvable Lie algebras $\Solv_{\mathbb{G}_R}$
and $\Solv_{\mathbb{G}_{\mathrm{TS}}}$ can be written as follows
\begin{equation}
\begin{array}{|l|l|}
\hline
  \Solv_{\mathbb{G}_R}  & \Solv_{\mathbb{G}_{\mathrm{TS}}} \\
  \hline
 \left[ H_i \, , \, H_j \right] = 0 & \left[ H_i \, , \, H_j \right] =0\\
\left[ H_i \, , \, \Phi_{\alpha^\ell}
  \right] = \alpha^\ell_i \, \Phi_{\alpha^\ell}
  &\left [ H_i \, , \, E^{\alpha^\ell} \right ] = \alpha^\ell_i \\
\left[ H_i \, , \, \Omega_{\alpha^s|I}
  \right] = \alpha^s_i \, \Omega_{\alpha^s|I} &
\left [ H_i \, , \, E^{\alpha^s} \right ] = \alpha^s_i \, E^{\alpha^s}
\\
\left [ \Phi_{\alpha^\ell} \, , \, \Phi_{\beta^\ell} \right ]
=N_{\alpha^\ell\beta^\ell} \, \Phi_{\alpha^\ell + \beta^\ell}
 &
 \left [ E^{\alpha^\ell} \, , \, E^{\beta^\ell} \right ] = N_{\alpha^\ell\beta^\ell} \,
E^{\alpha^\ell + \beta^\ell}    \\
\left [ \Phi_{\alpha^\ell} \, , \, \Omega_{\beta^s|I} \right ]
=N_{\alpha^\ell\beta^s} \, \Omega_{\alpha^\ell + \beta^s|I} &
 \left [ E^{\alpha^\ell} \, , \, E^{\beta^s} \right ] = N_{\alpha^\ell\beta^s} \,
E^{\alpha^\ell + \beta^s}   \\
\mbox{If } \alpha^s + \beta^s \in \Delta_{\mathrm{TS}}^\ell: & \\
\phantom{.}\quad\left [ \Omega_{\alpha^s|I} \, , \, \Omega_{\beta^s|J}
\right ] =\delta^{IJ} \,
 N_{\alpha^s\beta^s} \Phi_{\alpha^s +\beta^s}
  &
 \left [ E^{\alpha^s} \, , \, E^{\beta^s} \right ] = N_{\alpha^s\beta^s} E^{\alpha^s +\beta^s}
  \\
\mbox{If }
 \alpha^s + \beta^s \in \Delta_{\mathrm{TS}}^s: & \\
\phantom{.}\quad \cases{ \left [ \Omega_{\alpha^s|0} \, , \,
\Omega_{\beta^s|0} \right ] =
 N_{\alpha^s\beta^s} \Omega_{\alpha^s +\beta^s|0} \cr
 %\null \cr
 \left [ \Omega_{\alpha^s|0} \, , \, \Omega_{\beta^s|x} \right ]
=
 N_{\alpha^s\beta^s} \Omega_{\alpha^s +\beta^s|x} \cr
 %\null \cr
 \left [ \Omega_{\alpha^s|x} \, , \, \Omega_{\beta^s|y} \right ]
=
 N_{\alpha^s\beta^s} \left( \delta^{xy} \Omega_{\alpha^s +\beta^s|0}
 + \,  a^{xyz} \, \Omega_{\alpha^s +\beta^s|z} \right ) \cr} &
 \left [ E^{\alpha^s} \, , \, E^{\beta^s} \right ] = N_{\alpha^s\beta^s} E^{\alpha^s +\beta^s}
  \\
   \hline
\end{array}
 \label{paragonando}
\end{equation}
where $N_{\alpha \beta }=0$ if $\alpha + \beta \notin
\Delta_{\mathrm{TS}}$.

\subsection{Paint and subpaint groups in an example} \label{illustroexe}

We now want to illustrate the general structure described in the previous
subsection through the analysis of a specific example of a non maximally
split symmetric space that, under the name of $\mathrm{L(8,1)}$, pertains
to the homogeneous $\mathcal{N}=2$ special geometries. This will be both
educational in order to clarify the notion of Tits-Satake projection and
instrumental to extract a general systematics for the paint and subpaint
groups, which we will later recognize in the entire classification of
homogeneous special geometries.
\par
As we already discussed, the quaternionic member of the so named
$\mathrm{L(8,1)}$ family of special geometries is the following coset:
\begin{equation}\label{L81quat}
    \frac{\mathrm{G}_R}{\mathrm{H}}=\frac{\mathrm{E_{8(-24)}}}{\mathrm{E_{7(-133)}\times \SU(2)}}
\end{equation}
The quaternionic nature of this non maximally split symmetric space is
signaled by the presence of the $\mathrm{SU(2)}$ factor in the
denominator group and it is confirmed by the decomposition of the adjoint
representation of the numerator group:
\begin{equation}
  \mathbf{248} \, \stackrel{\mathrm{E_{7(-133)}\times
  \SU(2)}}{\Longrightarrow} \, (\mathbf{133},\mathbf{1}) \oplus
  (\mathbf{1},\mathbf{3}) \oplus (\mathbf{56},\mathbf{2})
\label{248Qsplit}
\end{equation}
Indeed the $4\times 28 = 112$ coset generators being in the
$(\mathbf{56},\mathbf{2})$ of ${\mathrm{E_{7(-133)}\times \SU(2)}}$ are
$\mathrm{SU(2)}$ doublets and transform symplectically under $\USp(56)$
transformations due to the symplectic embedding of the $\mathbf{56}$
representation of the compact $\mathrm{{E_7}}$ group.
\par
The quaternionic structure, however, is not relevant to our present
discussion that focuses on the mechanisms of the Tits-Satake projection.
By means of this latter we obtain the following result:
\begin{equation}
  \Pi_{\mathrm{TS}} \quad : \quad \frac{\mathrm{E_{8(-24)}}}{\mathrm{E_{7(-133)}\times
  SU(2)}} \, \longrightarrow \, \frac{\mathrm{F_{4(4)}}}{\mathrm{USp(6)}\times \mathrm{SU(2)}}
\label{TSexemplo}
\end{equation}
and we just note that the projected manifold is still quaternionic for
similar reasons to those of (\ref{248Qsplit}). Indeed it is the
quaternionic member of the $\mathrm{L(1,1)}$ family in the classification
of homogeneous special geometries. So the maximal noncompact Lie algebra
$\mathrm{F_{4(4)}}$ is the Tits-Satake subalgebra of
$\mathrm{E_{8(-24)}}$. Let us see how this happens, following step by
step the scheme described in the previous section.
 \par
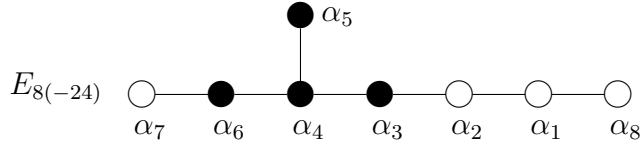
\begin{figure}[!hbt]
\caption{\it The Tits-Satake diagram of $E_{8(-24)}$, rank = 8, split
rank =  4, $\mathrm{G}_{\mathrm{TS}} = \mathrm{F}_{4(4)}.$
\label{e8dink}}\centering
\begin{picture}(100,100)
      \put (-70,35){$E_{8(-24)}$} \put (-20,35){\circle {10}} \put
(-23,20){$\alpha_7$} \put (-15,35){\line (1,0){20}} \put (10,35){\circle*
{10}} \put (7,20){$\alpha_6$} \put (15,35){\line (1,0){20}} \put
(40,35){\circle* {10}} \put (37,20){$\alpha_4$} \put (40,65){\circle*
{10}} \put (48,62.8){$\alpha_5$} \put (40,40){\line (0,1){20}} \put
(45,35){\line (1,0){20}} \put (70,35){\circle* {10}} \put
(67,20){$\alpha_{3}$} \put (75,35){\line (1,0){20}} \put (100,35){\circle
{10}} \put (97,20){$\alpha_{2}$} \put (105,35){\line (1,0){20}} \put
(130,35){\circle {10}} \put (127,20){$\alpha_1$} \put (135,35){\line
(1,0){20}} \put (160,35){\circle {10}} \put (157,20){$\alpha_8$}
\end{picture}
%\vskip 1cm
\end{figure}

The rank of the complex $\mathrm{E}_8$ algebra is $8$ and, and in its
real section $\mathrm{E}_{8(-24)}$ we can distinguish $4$ compact and $4$
noncompact Cartan generators. In a Euclidean orthonormal basis the
complete $\mathrm{E_8}$ root system is composed of the following $240$
roots:
\begin{equation}
  \Delta_{E_8} \, \equiv \, \left\{\begin{array}{cll}
    \pm \epsilon _i \pm \epsilon _j
    & \left( i \ne j\right)  & \mathbf{112} \\
     \underbrace{\pm \ft 12 \epsilon _1 \,  \pm \ft 12 \epsilon _2 \,
     \pm \ft 12 \epsilon _3\, \pm \ft 12\epsilon _4 \, \pm \ft 12\epsilon _5 \, \pm \ft 12\epsilon _6 \,
     \pm \ft 12\epsilon _7 \, \pm \ft 12 \epsilon _8\,}_{\mbox{even number of minus signs}}&\null & \mathbf{128} \\
     \hline
    \null & \null & \mathbf{240} \
  \end{array} \right\},
\label{e8system}
\end{equation}
and a convenient choice of the simple roots is provided by the following
ones:
\begin{eqnarray}
 \alpha_1 &=& \{0,1,-1,0,0,0,0,0\}, \nonumber \\
 \alpha_2 &=& \{0,0,1,-1,0,0,0,0\}, \nonumber \\
 \alpha_3 &=& \{0,0,0,1,-1,0,0,0\}, \nonumber \\
 \alpha_4 &=& \{0,0,0,0,1,-1,0,0\}, \nonumber \\
 \alpha_5 &=& \{0,0,0,0,0,1,-1,0\}, \nonumber \\
 \alpha_6 &=& \{0,0,0,0,0,1,1,0\},  \nonumber\\
 \alpha_7 &=&
  \{-\frac{1}{2},-\frac{1}{2},-\frac{1}{2},-\frac{1}{2},-\frac{1}{2},-\frac{1}{2},-\frac{1}{2},-\frac{1}{2}\}, \nonumber \\
  \alpha_8 &=& \{1,-1,0,0,0,0,0,0\}.
\end{eqnarray}
The corresponding Dynkin diagram is displayed in fig. \ref{e8dink}.
where the roots $\alpha_3 , \, \alpha_4 , \, \alpha_5 , \, \alpha_6$ have
been marked in black. This indicates that these simple roots are compact,
and Cartan generators as e.g. $\alpha _3^i\mathcal{H}_i$ belong to
$\mathcal{H}^{\mathrm{comp}}$.  In this way these diagrams define both
the real form $E_{8(-24)}$ and the corresponding Tits-Satake projection
of the root system.  The noncompact CSA $\mathcal{H}^{\rm nc}$ is the
orthogonal complement of $\mathcal{H}^{\mathrm{comp}}$. Let us also note
that the black roots form the Dynkin diagram of a $D_4$ algebra,
\textit{i.e} in its compact form the Lie algebra of $\mathrm{SO(8)}$.
This is the origin of the paint group
\begin{equation}
  \mathrm{G}_{\mathrm{paint}} = \mathrm{SO(8)},
\label{Gpaintus}
\end{equation}
pertaining to this example. We shall identify it in a moment, but let us
first perform the Tits-Satake projection on the root system. This case is
particularly simple since the span of the simple compact roots
$\mathbf{\alpha_3,\alpha_4,\alpha_5,\alpha_6}$ is just given by the
Euclidean space along the orthonormal axes
$\epsilon_4,\epsilon_5\,\epsilon_6,\epsilon_7$. The Euclidean space along
the orthonormal axes $\epsilon_1,\epsilon_2\,\epsilon_3,\epsilon_8$ span
the noncompact CSA. Note that this is not the same as the span of
$\mathbf{\alpha_1,\alpha_2,\alpha_7,\alpha_8}$. Denoting the components
of root vectors in the basis $\epsilon _i$ by $\alpha ^i$, the splitting
(\ref{splittus}) is  very simple. We just have:
\begin{equation}
  \alpha_\bot = \left\{ \alpha^4 \, , \, \alpha^5 \, , \,\alpha^6 \, , \,\alpha^7
  \right\} \quad ; \quad  \alpha_\| = \left\{ \alpha^1 \, , \, \alpha^2 \, , \,\alpha^3 \, ,
  \,\alpha^8 \right \},
\label{exesplittus}
\end{equation}
and the projection (\ref{projecto}) immediately yields the following
restricted root system:
\begin{equation}
\Delta _{\mathrm{TS}}=
  \left\{\begin{array}{clr}
    \pm \epsilon _i \pm \epsilon _j
    & \left( i \ne j \quad ; \quad i,j=1,2,3,8 \right)  & \mathbf{24} \\
    \pm \epsilon _i
    & \left( i=1,2,3,8 \right)  & \mathbf{8} \\
     \pm \ft 12 \epsilon _1 \,  \pm \ft 12 \epsilon _2 \,  \pm \ft 12 \epsilon _3 \,
     \pm \ft 12 \epsilon _8 &\null & \mathbf{16} \\
     \hline
    \null & \null & \mathbf{48} \
  \end{array} \right\},
\label{F4system}
\end{equation}
which can be recognized to be the root system of the  simple complex
algebra $\mathrm{F_{4}}$.
\par
With reference to the notations introduced in the previous section let us
now identify the subsets $\Delta^{\eta}$ and  $\Delta^\delta$ in the
positive root
subsystem of $\Delta^+_{E_8} %\subset \Delta_{E_8}
$ and their
corresponding images in the projection, namely
$\Delta_{\mathrm{TS}}^{\ell}$ and $\Delta_{\mathrm{TS}}^s$.
\par
Altogether, performing the projection the following situation is
observed:
\begin{itemize}
\item There are 24 roots that have null
projection on the noncompact space, namely
\begin{equation}
 \alpha_\|= 0 \, \Leftrightarrow \, \alpha=   \pm \epsilon _i \pm \epsilon
 _j \quad ; \quad i,j=4,5,6,7.
\label{compatterutte}
\end{equation}
These roots, together with the four compact Cartan generators, form the
root system of a $D_4$ algebra, whose dimension is exactly 28. In the
chosen real form such a subalgebra of $\mathrm{E_{8(-24)}}$ is  the
compact algebra $\mathrm{SO(8)}$ and its exponential acts as the paint
group, as already mentioned in (\ref{Gpaintus}).  All the remaining roots
have a non--vanishing projection on the compact space. In particular:
\item There are 12 positive roots of $\mathrm{E}_8$ that are exactly projected
on the 12 positive long roots of $\mathrm{F}_4$, namely the first line of
(\ref{F4system}), which we therefore identify with
$\Delta_{\mathrm{TS}}^\ell$. For these roots we have $\alpha_\bot=0$ and
they constitute the $\Delta^\eta$ system mentioned above:
\begin{equation}
{\Delta}^+_{E_8} \, \supset \,  \Delta_{\mathrm{TS}}^\eta  = \left\{
\epsilon_i \pm \epsilon_j
  \right\}= \Delta_{\mathrm{TS}}^\ell \quad ; \quad i <j  \quad ; \quad \, i,j=1,2,3,8
\label{Deltaeta}
\end{equation}
\item There are 8 different positive roots of $\mathrm{E}_8$ that have the
same projection on each of the $12=4 \oplus 8 $ positive short roots of
$\mathrm{F}_4$, i.e. the second and third line of (\ref{F4system}).
Namely all the remaining $12 \times 8=96$ roots of $\mathrm{E}_8$ are all
projected on short roots of $\mathrm{F}_4$. The set of $\mathrm{F}_4$
positive short roots can be split as follows:
\begin{equation}
\begin{array}{||lclc|r||}
\hline
   \Delta^s_{\mathrm{TS}} & = & \Delta^s_{\mathrm{vec}} \, \bigcup \,
\Delta^s_{\mathrm{spin}} \, \bigcup \, \Delta^s_{\overline{\mathrm{spin}}} & \null & \null \\
\Delta^s_{\mathrm{vec}}  & = &  \left\{ \epsilon_i  \right\}  & i=1,2,3,8 & \mathbf{4} \\
 \Delta^s_{\mathrm{spin}} & = & \underbrace{\pm \ft 12 \epsilon _1 \,  \pm \ft 12 \epsilon _2 \,
 \pm \ft 12 \epsilon _3 \,
     +\,  \ft 12 \epsilon _8}_{\mbox{even number of minus signs}} &\null & \mathbf{4}  \\
      \Delta^s_{\overline{\mathrm{spin}}} & = & \underbrace{\pm \ft 12 \epsilon _1 \,  \pm \ft 12 \epsilon _2 \,
      \pm \ft 12 \epsilon _3 \,
     +\,  \ft 12 \epsilon _8}_{\mbox{odd number of minus signs}} &\null & \mathbf{4}  \\
     \hline
     \null & \null & \null & \null & \mathbf{12} \\
     \hline
     \hline
\end{array}
\label{deltashortF4}
\end{equation}
Correspondingly the subset $\Delta^\delta \subset \Delta_{E_8} $ defined
by its projection property $\Pi_{\mathrm{TS}} \left( \Delta^\delta
\right) = \Delta_{\mathrm{TS}}^s$ is also split in three subsets as
follows:
\begin{equation}
\begin{array}{||lcl|c|r||}
\hline
   \Delta^\delta_+ & = & \Delta^\delta _{\mathrm{vec}} \, \bigcup \,
\Delta^\delta_{\mathrm{spin}}   &\null &  \null \\
\Delta^\delta_{\mathrm{vec}}  & = &  \left\{
\underbrace{\epsilon_i}_{\alpha_\|} \oplus \underbrace{\left( \pm
\epsilon _j\right) }_{\alpha_\bot}\right\} \quad , \quad \left(
\begin{array}{c}
  i=1,2,3,8 \\
  j=4,5,6,7
\end{array}\right)  & \mathrm{4}  \times \mathbf{ 8} &  32 \\
 \Delta^\delta_{\mathrm{spin}} & = & \left\{\underbrace{\left( \pm \ft 12 \epsilon _1 \,  \pm \ft 12 \epsilon _2 \,
 \pm \ft 12 \epsilon _3 \,
     +\,  \ft 12 \epsilon _8\right)}_{\alpha_\|\quad \mbox{even $\#$ of $-$ signs}} \, \oplus \,
     \underbrace{\left(\pm \ft 12 \epsilon _4  \, \pm \ft 12 \epsilon _5 \,
     \pm \ft 12 \epsilon _6 \, \pm \ft 12 \epsilon _7\right)}_{\alpha_\bot\quad \mbox{even $\#$ of $-$ signs}}\right\}
       & \mathrm{4}  \times  \mathbf{8}  &  32  \\
       %%%%%%%%%%%%%%%%%%%%%%%%%%%%%%%%%%%%%%%%%%%%%%%%%%%%
       \Delta^\delta_{\overline{\mathrm{spin}}} & = & \left\{\underbrace{\left( \pm \ft 12 \epsilon _1 \,
       \pm \ft 12 \epsilon _2 \,
 \pm \ft 12 \epsilon _3 \,
     +\,  \ft 12 \epsilon _8\right)}_{\alpha_\|\quad \mbox{odd $\#$ of $-$ signs}} \, \oplus \,
     \underbrace{\left(\pm \ft 12 \epsilon _4  \, \pm \ft 12 \epsilon _5 \,
     \pm \ft 12 \epsilon _6 \, \pm \ft 12 \epsilon _7\right)}_{\alpha_\bot \quad \mbox{odd $\#$ of $-$}}\right\}
       & \mathrm{4}  \times  \mathbf{8}  &  32  \\
       %%%%%%%%%%%%%%%%%%%%%%%%%%%%%%%%%
     \hline
  \null & \null &  \null & \null & \null \mathbf{96} \\
     \hline
     \hline
\end{array}
\label{deltashortE8}
\end{equation}
\end{itemize}
We can now verify the general statements made in the previous sections
about the paint group representations to which the various roots are
assigned. First of all we see that, as we claimed,  the long roots of
$\mathrm{F_4}$, namely those $12$ given in (\ref{Deltaeta}) are singlets
under the paint group $\mathrm{G}_{\mathrm{paint}} = \mathrm{SO(8)}$. All
other roots fall into multiplets with the same Tits-Satake projection and
each of these latter has always the same multiplicity, in our case $m=8$
(compare with (\ref{replicas})). So the short roots of
$\mathrm{F_{4(4)}}$ fall into $8$--dimensional representations of
$\mathrm{G}_{\rm paint}=\mathrm{SO(8)}$. But which ones? $\mathrm{SO(8)}$
has three kinds of octets $\mathbf{8_v}$, $\mathbf{8_s}$ and
$\mathbf{8_{\bar s}}$ and, as we stated, not every root $\alpha_s$ of the
Tits-Satake algebra $\mathbb{G}_{\mathrm{TS}}$ falls in the same
representation $\mathbf{D}$ of the paint group although in this case all
$\mathbf{D}[\alpha^s]$ have the same dimension. Looking back at our
result we easily find the answer. The $4$ positive roots in the subset
$\Delta^\delta_{\mathrm{vec}}$ have as compact part $\alpha_\bot$ the
weights of the vector representation of $\mathrm{SO(8)}$. Hence the roots
of $ \Delta^\delta_{\mathrm{vec}}$ are assigned to the $\mathbf{8_v}$ of
the paint group. The $4$ positive roots in
$\Delta^\delta_{\mathrm{spin}}$ have instead as compact part the weights
of the spinor representation of $\mathrm{SO(8)}$ and so they are assigned
to the $\mathbf{8_s}$ irreducible representation. Finally, with a similar
argument, we see that the $4$ roots of
$\Delta^\delta_{\overline{\mathrm{spin}}}$ are in the conjugate spinor
representation $\mathbf{8_{\bar s}}$. The last part of the general
discussion of section \ref{genediscussa} is now easy to verify in the
context of our example, namely that relevant to the subpaint group
$\mathrm{G}^0_{\mathrm{subpaint}}$ (we will omit sometimes the `subpaint'
indication for convenience). According to
(\ref{paint0})-(\ref{splittatonibus}) we have to find a subgroup
$\mathrm{G}^0 \subset \mathrm{SO(8)}$ such that under reduction with
respect to it, the three octet representations branch simultaneously as :
\begin{eqnarray}
\mathbf{8_v} & \stackrel{\mathrm{G}^0}{\longrightarrow} & \mathbf{{1}} \oplus \mathbf{7}, \nonumber\\
\mathbf{8_s} & \stackrel{\mathrm{G}^0}{\longrightarrow} & \mathbf{{1}} \oplus \mathbf{7}, \nonumber\\
\mathbf{8_{\bar s}} & \stackrel{\mathrm{G}^0}{\longrightarrow} &
\mathbf{{1}} \oplus \mathbf{7}. \label{three8in1plus7}
\end{eqnarray}
Such group $\mathrm{G}^0$ exists and it is uniquely identified as the
$14$ dimensional $\mathrm{G_{2(-14)}}$. Hence the subpaint group is
$\mathrm{G_{2(-14)}}$. Considering now (\ref{paragonando}) we see that
the commutation relations of the solvable Lie algebra $\Solv\left(
\mathrm{E_{8(-24)}}/\mathrm{E_{7(-133)}\times SU(2)} \right) $ precisely
fall into the general form displayed in the first column of that table
with the index $x=1,\dots,7$ spanning the fundamental $7$-dimensional
representation of $\mathrm{G_{2(-14)}}$ and the invariant antisymmetric
tensor $a^{xyz}$ being given by the $\mathrm{G_{2(-14)}}$-invariant
octonionic structure constants. Indeed the representation $\mathbf{J}$
mentioned in section \ref{genediscussa} is the fundamental $\mathbf{7}$
and we have the decomposition:
\begin{equation}
  \mathbf{7} \, \times \, \mathbf{7}  \, = \, \underbrace{\mathbf{14} \, \oplus \, \mathbf{7}}_{\mathrm{antisymmetric}} \,
  \oplus \underbrace{\mathbf{27} \, \oplus \, \mathbf{1}}_{\mathrm{symmetric}}.
\label{7x7=49}
\end{equation}
This shows that, as claimed in point [E] of the general discussion, the
tensor product $\mathbf{J}\times \mathbf{J}$ contains both the singlet
and $\mathbf{J}$.
\par
 In the example studied in paper \cite{Fre':2005sr}, namely
\begin{equation}
  \Pi_{\mathrm{TS}} \, : \, \frac{\mathrm{E_{7(-5)}}}{\mathrm{SO(12) \times
  SU(2)}} \, \longrightarrow \, \frac{\mathrm{F_{4(4)}}}{\mathrm{USp(6)}\times
\mathrm{SU(2)}} \label{oldexample}
\end{equation}
the image of the Tits-Satake projection yields the same maximally split
coset as in the case presently illustrated, although the original
manifold is a different one. The only difference that distinguishes the
two cases resides in the paint group. There we have
$\mathrm{G}_{\mathrm{paint}} = \mathrm{SO(3) \times SO(3) \times SO(3)}$
and the subpaint group was identified as $\mathrm{G}^0_{\rm subpaint}
=\mathrm{SO(3)_{diag}}$. Correspondingly the index $x=1,2,3$ spans the
triplet representation of $\mathrm{SO(3)}$ which is the $\mathbf{J}$
appropriate to that case and the invariant tensor $a^{xyz}$ is given by
the Levi-Civita symbol $\varepsilon^{xyz}$.
\par
Let us now consider the group theoretical meaning of the splitting of
$\mathrm{F_{4(4)}}$ roots into the three subsets
$\Delta^s_{\mathrm{vec}}$, $ \Delta^s_{\mathrm{spin}}$,
$\Delta^s_{\mathrm{TS},\overline{\mathrm{spin}}}$, which are assigned to
different representations of the paint group $\mathrm{SO(8)}$. This is
easily understood if we recall that there exists  a subalgebra
$\mathrm{SO(4,4)} \subset \mathrm{F_{4(4)}}$ with respect to which we
have the following branching rule of the adjoint representation of
$\mathrm{F_{4(4)}}$:
\begin{equation}\label{buh}
    \mathbf{52}\stackrel{\mathrm{SO(4,4)}}\rightarrow\mathbf{28}^{\rm nc}
    \oplus\mathbf{8}_v^{\rm nc}\oplus\mathbf{8}_s^{\rm nc}\oplus\mathbf{8}_{\bar{s}}^{\rm nc}
\end{equation}
The superscript $nc$ is introduced just in order to recall that these are
representations of the noncompact real form $\mathrm{SO(4,4)}$ of the
$D_4$ Lie algebra. By $\mathbf{28}$, $\mathbf{8}_v$, $\mathbf{8}_s$ and
$\mathbf{8}_{\bar{s}}$ we have already denoted and we continue to denote
the homologous representations in the compact real form $\mathrm{SO(8)}$
of the same Lie algebra. The algebra $\mathrm{SO(4,4)}$ is regularly
embedded and therefore its Cartan generators are the same as those of
$\mathrm{F_{4(4)}}$. The $12$ positive long roots of $\mathrm{F_{4(4)}}$
are  the only positive roots of $\mathrm{SO(4,4)}$, while the three sets
$ \Delta^s_{\mathrm{vec}}$, $\Delta^s_{\mathrm{spin}}$,
$\Delta^s_{\overline{\mathrm{spin}}}$ just correspond to the positive
weights of the three representations $\mathbf{8}_v^{\rm nc}$,
$\mathbf{8}_s^{\rm nc}$ and $\mathbf{8}_{\bar{s}}^{\rm nc}$,
respectively. This is in agreement with the branching rule (\ref{buh}).
So the conclusion is that the different paint group representation
assignments of the various root subspaces correspond to the decomposition
of the Tits-Satake algebra $\mathrm{F_{4(4)}}$ with respect to what we
can call the \textit{sub Tits-Satake algebra}\footnote{This concept
corresponds to the algebra $G_s$ in \cite{Keurentjes:2002rc}.}
$\mathbb{G}_{\mathrm{subTS}} = \mathrm{SO(4,4)}$. We can just wonder how
the concept of sub Tits-Satake algebra can be defined. This is very
simple and obvious from our example. $\mathrm{G}_{\mathrm{subTS}}$ is the
normalizer of the paint group $\mathrm{G}_{\mathrm{paint}}$ within the
original group $\mathrm{G}_\mathbb{R}$. Indeed there is a maximal
subgroup:
\begin{equation}
  \mathrm{SO(4,4)} \times \mathrm{SO(8)}  \, \subset \,
  \mathrm{E_{8(-24)}},
\label{subTSgruppo}
\end{equation}
with respect to which the adjoint of $\mathrm{E_{8(-24)}}$ branches as
follows:
\begin{equation}\label{decE8So44}
    \mathbf{248} \stackrel{\mathrm{SO(4,4)} \times \mathrm{SO(8)}}{\longrightarrow}
    (1,\mathbf{28})\oplus (\mathbf{28}^{\rm nc},\mathbf{1}) \oplus (\mathbf{8_v}^{\rm nc},\mathbf{8_v})
\oplus (\mathbf{8_s}^{\rm nc},\mathbf{8_s})\oplus (\mathbf{8_{\bar
s}}^{\rm nc},\mathbf{8_{\bar s}})
\end{equation}
and the last three terms in this decomposition display the pairing
between representations of the paint group and representations of the sub
Tits-Satake group. Alternatively we can view the \textit{subpaint group}
$\mathrm{G}^0_{\mathrm{subpaint}} = \mathrm{G_{2(-14)}}$ as the
\textit{normalizer} of the Tits-Satake subgroup $\mathrm{G}_{\mathrm{TS}}
= \mathrm{F_{4(4)}}$ within the original group $\mathrm{G}_\mathbb{R} =
\mathrm{E_{8(-24)}}$. Indeed we have a subgroup
\begin{equation}
  \mathrm{F_{4(4)}} \times \mathrm{G_{2(-14)}}  \, \subset \,
  \mathrm{E_{8(-24)}},
\label{subpaintgruppo}
\end{equation}
such that the adjoint of $\mathrm{E_{8(-24)}}$ branches as follows:
\begin{equation}\label{decE8F4}
    \mathbf{248}\stackrel{\mathrm{F_{4(4)}} \times \mathrm{G_{2(-14)}}}{\longrightarrow}
    (\mathbf{52},\mathbf{1})\oplus (\mathbf{1},\mathbf{14}) \oplus (\mathbf{26},\mathbf{7})
\end{equation}
The two decompositions (\ref{decE8So44}) and (\ref{decE8F4}) lead to the
same decomposition with respect to the intersection group:
\begin{eqnarray}
  G_{\rm intsec} &   \equiv &
  \left(\mathrm{G}_{\mathrm{TS}} \times \mathrm{G}^0_{\mathrm{subpaint}} \right)\bigcap
  \left(\mathrm{G}_{\mathrm{subTS}} \times \mathrm{G}_{\mathrm{paint}} \right)
  =\mathrm{G}_{\mathrm{subTS}} \times \mathrm{G}^0_{\mathrm{subpaint}}\nonumber\\
   & = &  \left( \mathrm{F_{4(4)}} \times
  \mathrm{G_{2(-14)}}\right)\bigcap \left(  \mathrm{SO(4,4)} \times
  \mathrm{SO(8)}\right) %\nonumber\\
  =\mathrm{SO(4,4)} \times \mathrm{G_{2(-14)}}.
\label{porcellino}
\end{eqnarray}
We find
\begin{eqnarray}\label{decona}
 \mathbf{248} & \rightarrow &
(\mathbf{1},\mathbf{14})\oplus
(\mathbf{1},\mathbf{7})\oplus(\mathbf{1},\mathbf{7})\oplus
(\mathbf{8}_v^{\rm nc},\mathbf{7})
\oplus (\mathbf{8}^{\rm nc}_s,\mathbf{7})\oplus(\mathbf{8}^{\rm nc}_{\bar{s}},\mathbf{7})\nonumber\\
& \null & \oplus(\mathbf{28}^{\rm nc},\mathbf{1}) \oplus (\mathbf{8}^{\rm
nc}_v,\mathbf{1}) \oplus (\mathbf{8}^{\rm nc}_s,\mathbf{1})\oplus
(\mathbf{8}^{\rm nc}_{\bar{s}},\mathbf{1}).
\end{eqnarray}
The adjoint of the Tits-Satake subalgebra $\mathrm{G}_{\mathrm{TS}} =
\mathrm{F_{4(4)}}$  is reconstructed by collecting together all the
singlets with respect to the subpaint group $\mathrm{G}^0_{\rm
subpaint}$. Alternatively the adjoint of the paint algebra
$\mathrm{G}_{\rm paint} =\mathrm{SO(8)}$ is reconstructed by collecting
together all the singlets with respect to the \textit{sub Tits-Satake
algebra} $\mathrm{G}_{\mathrm{subTS}} = \mathrm{SO(4,4)}$.
 \par
Finally, we can recognize the sub Tits-Satake algebra as the algebra
generated by the CSA and roots $\Delta ^\ell $ (and their negatives) in
the decomposition \ref{SubsetsRootspaces}.

\subsection{TS projection for homogeneous special geometries} \label{Tsforsolvable}

After our detailed discussion of the Tits-Satake projection in the case
of symmetric spaces we can extract a general scheme that will apply also
to more general solvable Lie algebras as those appearing in the context
of $\mathcal{N}=2$ homogeneous special geometries. Let us discuss how the
Tits-Satake projection can be reformulated relying on the paint and
subpaint group structures. In section \ref{genediscussa} our starting
point was the geometrical projection of the root system
$\Delta_\mathbb{G}$ onto the noncompact Cartan subalgebra by setting, for
each root $\alpha \in \Delta_\mathbb{G}$ its compact part $\alpha_\bot$
to zero. This is the operation that is no longer available in the general
case of a solvable algebra as that mentioned in (\ref{genestrucca}). We
now only have the solvable algebra, which corresponds to the noncompact
part $\alpha_\|$. Indeed at the level of the solvable Lie algebra there
is no notion of the compact Cartan generators. However, the structures
that still persist and allow us to define the \textit{Tits-Satake
projection} are those of paint and subpaint groups. Indeed for all the
solvable Lie algebras $\Solv_{\mathcal{M}}$ considered in the
classification of homogeneous special geometries the following statements
A-E will be true:
\paragraph{A1]}
There exists a \textit{compact algebra} $\mathbb{G}_{\rm paint} $ which
acts as  an algebra of outer automorphisms ({\it i.e.} outer derivatives)
of the  solvable algebra $\Solv_{\mathcal{M}}$. The algebra
$\mathbb{G}_{\rm paint}$ is rigorously defined as follows. Given the
solvable Lie algebra $\Solv_{\mathcal{M}}$ the corresponding Riemannian
manifold ${\mathcal{M}}= \exp \left[ \Solv_{\mathcal{M}} \right]$ has an
algebra of isometries $\mathbb{G}^{\rm iso}_{\mathcal{M}}$, which is
normally larger than $\Solv_{\mathcal{M}}$, and for all special
homogeneous manifolds $\mathcal{M}$ such algebras were studied and
completely classified in \cite{deWit:1993wf,deWit:1995tf}, see section
\ref{isomhomspecgeom}. Obviously $\Solv_{\mathcal{M}} \, \subset \,
\mathbb{G}^{\rm iso}_{\mathcal{M}}$. Let us define the subalgebra of
automorphisms of the solvable Lie algebra in the standard  way:
\begin{equation}
 \mathbb{G}^{\rm iso}_{\mathcal{M}}  \, \supset \,
 \mathrm{Aut} \, \left[ \Solv_\mathcal{M}\right] \,  \quad = \quad \left\{  X \, \in
 \, \mathbb{G}^{\rm iso}_{\mathcal{M}} \, \mid \, \forall \, \Psi \,
 \in \, \Solv_\mathcal{M}\ :\ \left[ X\, ,\, \Psi \right] \, \in \,
 \Solv_\mathcal{M} \, \right\}.
\label{automorfismi}
\end{equation}
By its own definition the algebra $\mathrm{Aut} \, \left[
\Solv_\mathcal{M}\right] $ contains $\Solv_\mathcal{M}$ as an ideal.
Hence  we can define the algebra of external automorphisms as the
quotient:
\begin{equation}
  \mathrm{Aut}_{\mathrm{Ext}} \, \left[ \Solv_\mathcal{M}\right] \,
  \equiv \, \frac{\mathrm{Aut} \, \left[
  \Solv_\mathcal{M}\right]}{\Solv_\mathcal{M}},
\label{outerauto}
\end{equation}
and we identify $\mathbb{G}_{\mathrm{paint}}$ as the maximal compact
subalgebra of $\mathrm{Aut}_{\mathrm{Ext}} \, \left[
\Solv_\mathcal{M}\right]$. Actually we immediately see that
\begin{equation}
  \mathbb{G}_{\mathrm{paint}} \, = \, \mathrm{Aut}_{\mathrm{Ext}} \, \left[
\Solv_\mathcal{M}\right]. \label{pittureFuori}
\end{equation}
Indeed, as a consequence of its own definition the algebra
$\mathrm{Aut}_{\mathrm{Ext}} \, \left[ \Solv_\mathcal{M}\right]$ is
composed of isometries which belong to the stabilizer subalgebra
$\mathbb{H}  \, \subset  \, \mathbb{G}^{\rm iso}_{\mathcal{M}}$ of any
point of the manifold, since $\Solv_\mathcal{M}$ acts transitively. In
virtue of the Riemannian  structure of $\mathcal{M}$ we have $\mathbb{H}
\subset \so(n) $ where $n = \mbox{dim} \left(\Solv_{\mathcal{M}} \right)$
and hence also $\mathrm{Aut}_{\mathrm{Ext}} \, \left[
\Solv_\mathcal{M}\right] \, \subset \, \so(n)$ is a compact Lie algebra.
\paragraph{A2]}
We can now reformulate the notion of maximally noncompact or maximally
split algebras in such a way that it applies to the case of all
considered solvable algebras,  independently whether they come from
symmetric spaces or not. \textit{The algebra $\Solv_\mathcal{M}$ is
maximally split if the paint algebra is trivial, namely:}
\begin{equation}
  \Solv_\mathcal{M} = \mbox{maximally split} \, \Leftrightarrow \, \mathrm{Aut}_{\mathrm{Ext}} \, \left[
\Solv_\mathcal{M}\right] = \emptyset. \label{maxsplit}
\end{equation}
For maximally split algebras there is no Tits-Satake projection, namely
the Tits-Satake subalgebra is the full algebra.
\paragraph{B]}
Let us now consider non maximally split algebras such that
$\mathrm{Aut}_{\mathrm{Ext}} \, \left[ \Solv_\mathcal{M}\right] \ne
\emptyset$. Let $r$ be the rank of $\Solv_{\mathcal{M}},$ namely the number
of its Cartan generators $H_i$ and $n$ the
number of its nilpotent generators $\mathcal{W}_\alpha$, namely the
number of generalized roots $\vec{\alpha}$. The whole set of Cartan generators $H_i$
plus a subset of $p$ nilpotent generators $\mathcal{W}_{\alpha^\ell} $ associated with roots
$\vec{\alpha}^\ell$ that we name \textit{long} close a
solvable subalgebra $\Solv_{\mathrm{subTS}} \subset \Solv_{\mathcal{M}}$
that is made of singlets  under the action of the paint Lie algebra
$\mathbb{G}_{\rm paint}$, {\it i.e.}
\begin{eqnarray}
\Solv_{\mathrm{subTS}} &  =& \mbox{span} \, \left \{ H_i, \mathcal{W}_{\alpha^\ell}  \right\}, \nonumber\\
\left[ \Solv_{\mathrm{subTS}} \, , \, \Solv_{\mathrm{subTS}} \right] &
\subset
 & \Solv_{\mathrm{subTS}},\nonumber\\
\forall \, X  \in \, \mathbb{G}_{\rm paint} \, , \, \forall
 \Psi \, \in \, \Solv_{\mathrm{subTS}} & : &
[X,\Psi ]=0.\label{subTsalgebrageneral}
\end{eqnarray}
We name $\Solv_{\mathrm{subTS}}$ the \textit{sub Tits-Satake algebra}. By
definition $\Solv_{\mathrm{subTS}}$ has the same rank as the original
solvable algebra $\Solv_{\mathcal{M}}$. We show in later sections that
there is a very short list of possible cases for
$\Solv_{\mathrm{subTS}}$. In all possible cases, it is the solvable Lie
algebra of a symmetric maximally split coset
$\mathbb{G}_{\mathrm{subTS}}/\mathbb{H}_{\mathrm{subTS}}$. In this way,
eventually, we have the notion of a semisimple Lie algebra
$\mathbb{G}_{\mathrm{subTS}}$.
\begin{table}[!htb]
  \caption{\it $\mathbb{G}_{\mathrm{subTS}}$. The solvable algebra of the
  (maximally split) coset $\mathbb{G}_{\mathrm{subTS}}/\mathbb{H}_{\mathrm{subTS}}$   is the
  sub Tits-Satake algebra. The lines distinguish spaces of different rank, similar
  to the scheme in table \ref{allLpq}. The inverse $\mathbf{c}$--map
  leads from the last column to the middle one, and the inverse
  $\mathbf{r}$--map to the first column, each time reducing the rank
  with~1.
  \label{tbl:subTS}}
\begin{center}
$  \begin{array}{|ccc|} \hline
\mbox{real} & \mbox{K{\"a}hler} & \mbox{quaternionic}   \\
\hline
  &  & \begin{array}{c}
    \SO(1,1) \\
    \SU(1,1)
  \end{array}  \\
\hline
  & \SU(1,1)  & \begin{array}{c}
    \SO(2,2) \\
    \mathrm{G}_{2(2)}
  \end{array}     \\
\hline
\SO(1,1)  & [\SU(1,1)]^2 & \SO(3,4) \\
\hline
 [\SO(1,1)]^2 & [\SU(1,1)]^3 & \SO(4,4) \\
\hline
\end{array}$
\end{center}
\end{table}
These are given in table \ref{tbl:subTS}, and correspond to the notion of
sub Tits-Satake algebra as it was used for symmetric spaces. However, for
homogeneous spaces we start only with the solvable algebras, and as such
$\Solv_{\mathrm{subTS}}$ is the algebra that is intrinsically defined as
the sub Tits-Satake algebra. This subtlety becomes more relevant for the
Tits-Satake algebra itself, where the solvable algebra is not in all
cases the solvable algebra of a symmetric space, and thus a corresponding
group $G_{\mathrm{TS}}$ is not well defined.
 \par
A first very basic grouping of the algebras $\Solv_{\mathcal{M}}$ is as
such done on the basis of their sub Tits-Satake algebra. It suffices to
compare tables \ref{allLpq} and \ref{tbl:subTS} to recognize the groups,
that are, up to one exception, just groups according to the rank of the
solvable algebras.

\paragraph{C1]}
Considering the orthogonal decomposition of the original solvable Lie
algebra with respect to its
 \textit{sub Tits-Satake algebra}:
\begin{equation}
  \Solv_{\mathcal{M}} \, = \, \Solv_{\mathrm{subTS}} \, \oplus \,
  \mathbb{K}_{\rm short}.
\label{Kshort}
\end{equation}
We find that the orthogonal subspace $\mathbb{K}_{\rm short}$ necessarily
decomposes into a sum of $q$ subspaces:
\begin{equation}
   \mathbb{K}_{\rm short} \, = \, \bigoplus_{\wp=1}^{q} \,
   \mathbb{D}\left [ \mathcal{P}^+_\wp,\mathbf{Q}_\wp \right],
\label{shortoni}
\end{equation}
where each $\mathbb{D}\left [ \mathcal{P}^+_\wp,\mathbf{Q}_\wp \right ]$
is the tensor product:
\begin{equation}
  \mathbb{D}\left [ \mathcal{P}^+_\wp,\mathbf{Q}_\wp \right ] \, =
  \, \mathcal{P}^+_\wp \, \otimes \, \mathbf{Q}_\wp
\label{tensoreprodutto}
\end{equation}
of an irreducible module $\mathbf{Q}_\wp$ (i.e. representation) of the
compact paint algebra $\mathbb{G}_{\mathrm{paint}}$ with an irreducible
module $\mathcal{P}^+_\wp$ of the solvable sub Tits-Satake algebra
$\Solv_{\mathrm{subTS}}$. As we already noticed, $\Solv_{\mathrm{subTS}}$
is the maximal Borel subalgebra of the maximally split, semisimple, real
Lie algebra $\mathbb{G}_{\mathrm{subTS}}$. Hence an irreducible module
$\mathcal{P}^+_\wp$ of $\Solv_{\mathrm{subTS}}$ necessarily decomposes in
the following way:
\begin{equation}
  \mathcal{P}^+_\wp \, = \, \bigoplus_{s=1}^{n_\wp} \,
  \mathbb{W}[\vec{\alpha}^{(\wp,s)}],\qquad n_\wp=\dim\mathcal{P}^+_\wp,
\label{Wspazi}
\end{equation}
where each $\mathbb{W}[\vec{\alpha}^{(\wp,s)}]$ is an eigenspace of the
CSA of $\mathbb{G}_{\mathrm{subTS}}$,  which coincides with that of
$\Solv_{\mathrm{subTS}}$ and eventually with the $\mathrm{CSA}$ of the
original $\Solv_\mathcal{M}$. Explicitly this means:
\begin{equation}
  \forall \, H_i \, \in \, \mathrm{CSA}\left( \Solv_\mathcal{M}\right) \, , \,
  \forall \Psi \, \in \, \mathbb{W}[\vec{\alpha}^{(\wp,s)}] \, \otimes \,
   \mathbf{Q}_\wp \quad : \quad \left[
  H_i \, , \, \Psi\right]  \, = \, {\alpha}^{(\wp,s)}_i \, \Psi.
\label{eigenspazione}
\end{equation}
Furthermore the $r$-vectors of eigenvalues, which are roots of
$\Solv_{\mathcal{M}}$ (see (\ref{genestrucca})), are identified by
(\ref{Wspazi}) as the non negative weights of some irreducible module
$\mathcal{P}_\wp$ of the simple Lie algebra
$\mathbb{G}_{\mathrm{subTS}}$:
\begin{equation}
  \mathcal{P}_\wp  =  \mathcal{P}^+_\wp \, \oplus \,
  \mathcal{P}^-_\wp,\qquad
  \mathcal{P}^-_\wp  =  \bigoplus_{s=1}^{n_\wp} \,
  \mathbb{W}[-\vec{\alpha}^{(\wp,s)}].
\label{halrepresentation}
\end{equation}
Indeed for the solvable Lie algebras $\Solv (\mathrm{G/H})$ of maximally
split cosets the irreducible modules are easily constructed as
\textit{half-modules} of the full algebra $\mathbb{G}$, namely by taking
the eigenspaces associated with  non negative weights.
\paragraph{C2]}
The decomposition of $\mathbb{K}_{\rm short}$ mentioned in
(\ref{shortoni}) has actually a general form depending on the rank. We
will discuss this here for the quaternionic-K{\"a}hler manifolds, as the
other ones can be obtained by restriction of the generators using the
inverse $\mathbf{c}$-- and $\mathbf{r}$--maps as discussed in section
\ref{ss:invrcmap}.
\begin{description}
  \item[r = 4)] In this case there are just three modules of
  $\mathbb{G}_{\mathrm{subTS}} = \mathrm{SO(4,4)}$ involved in the sum of (\ref{shortoni})
  namely $\mathcal{P}_{\mathbf{8_v}}$,
  $\mathcal{P}_{\mathbf{8_s}}$, $\mathcal{P}_{\mathbf{8_{\bar s}}}$,
where $\mathbf{8_{v,s,{\bar s}}}$ denotes the vector, spinor and
conjugate spinor representation, respectively. All these three modules
are $8$ dimensional, which means that for all of them there are $4$
positive weights and $4$ negative ones. Denoting these half spaces by
$\mathbf{4^+_{v,s,{\bar s}}}$, we can write:
\begin{equation}
  \mathbb{K}_{\rm short} = \left ( \mathbf{4^+_v}, \mathbf{Q_v} \right )
  \oplus \left ( \mathbf{4^+_s}, \mathbf{Q_s} \right )\oplus \left ( \mathbf{4^+_{\bar s}}, \mathbf{Q_{\bar s}} \right
  ),
\label{shortonR4}
\end{equation}
where $\mathbf{Q_{v,s,{\bar s}}}$ are three different irreducible modules
of $\mathbb{G}_{\mathrm{paint}}$ that we will discuss in later sections.
The generic case is that where all three representations
$\mathbf{Q_{v,s,{\bar s}}}$ are non vanishing and this corresponds to
$L(q,P)$ or $L(4m,P,\dot P)$ with $q,P,m\ge 1$. Special cases where two
of the three representations $\mathbb{G}_{\mathrm{paint}}$ vanish
correspond to the classes $L(0,P)$, while for $L(0,P,\dot{P})$ only one
of these representations vanishes. The limiting case is that where all
three representations are deleted and the full algebra is just $\Solv
\left (\mathrm{\frac{\SO(4,4)}{\SO(4) \times \SO(4)}}\right)$, which is
$L(0,0)$. Note that (\ref{shortonR4}) is the generalization of the
decomposition (\ref{decE8So44}) applying to the case $L(8,1)$. There we
have $\mathbb{G}_{\rm paint} =\mathrm{SO(8)}$ and the aforementioned
irreducible modules are:
\begin{equation}
  \mathbf{Q_{v}} =\mathbf{ 8_v} \quad ; \quad \mathbf{Q_{s}} =
  \mathbf{8_s} \quad ; \quad \mathbf{Q_{{\bar s}}} = \mathbf{8_{\bar s}}
\label{loreipsum}
\end{equation}
  \item[r = 3)] In this case there is only one module of
  $\mathbb{G}_{\mathrm{subTS}} = \mathrm{SO(3,4)}$ involved in the sum of (\ref{shortoni})
  namely $\mathcal{P}_{\mathbf{8_s}}$ where  $\mathbf{8_{s}}$ denotes the $8$ dimensional spinor
   representation of $\mathrm{SO(3,4})$. With a notation completely analogous
   to that employed above let $\mathrm{4^+_s}$ denote the space spanned by the eigenspaces
   pertaining to positive spinor weights. Then we can write:
\begin{equation}
  \mathbb{K}_{\rm short} = \left ( \mathbf{4^+_s}, \mathbf{Q_s} \right ),
\label{shortonR3}
\end{equation}
where the representation $\mathbf{Q_s}$ of the paint group will be
discussed in later sections. When $\mathbf{Q_s}$ is non vanishing we
describe the $L(-1,P)$ spaces which at the quaternionic level are never
given by symmetric spaces. When $\mathbf{Q_s}$ vanishes, we degenerate in
the case $L(-1,0)$, which is already maximally split.
  \item[r = 2)] In this case, there is one exceptional case, namely $\mathrm{SG}_5$,
  where $G_R=G_{\mathrm{subTS}}=G_{2(2)}$. In all other cases, there are
  two modules of $\mathrm{SO(2,2)}$ involved in the sum  of
  (\ref{shortoni}) and these are the spinor module
  $\mathcal{P}_{\mathbf{4_s}}$ and the vector module $\mathcal{P}_{\mathbf{4_v}}$.
  Both modules
  are $4$-dimensional and in our adopted notations we can write:
\begin{equation}
  \mathbb{K}_{\rm short} = \left ( \mathbf{2^+_s}, \mathbf{Q_s} \right ) \oplus \left ( \mathbf{2^+_v}, \mathbf{Q_v} \right )\,.
\label{shortonR2}
\end{equation}
Later on in this section, we will discuss the representations
$\mathbf{Q_s}$, $\mathbf{Q_v}$ of the paint group and show how the coset
manifolds in the series $L(-2,P)$ can be reconstructed. When $P=0$, only
the representation $\mathbf{Q_v}$ is non-vanishing.
\item[r = 1)] In
this case we have to distinguish between $G_{\mathrm{subTS}} = \SO(1,1)$
or $G_{\mathrm{subTS}} = \SU(1,1)$. When $G_{\mathrm{subTS}} = \SU(1,1)$
we have:
\begin{equation}
  \mathbb{K}_{\rm short} = \left ( \mathbf{1^+_s}, \mathbf{Q_s} \right),
 \label{shortonR11}
\end{equation}
where $\mathbf{1^+_s}$ denotes the positive weight subspace of the spinor
representation of $\so(1,2)$, i.e. the fundamental of $\su(1,1)$, which
is two-dimensional. The representation $\mathbf{Q_s}$ will be discussed
later. When $G_{\mathrm{subTS}} = \SO(1,1)$ on the other hand, we have:
\begin{equation}
  \mathbb{K}_{\rm short} = \left ( \mathbf{1^+_s}, \mathbf{Q_s} \right) \oplus \left ( \mathbf{1^+_v}, \mathbf{Q_v}
  \right)\,.
 \label{shortonR12}
\end{equation}
In this case, $\mathbf{1^+_s}$ denotes a subspace of weight $1/2$ with
respect to $\mathbb{G}_{\mathrm{subTS}} = \so(1,1)$, while the subspace
$\mathbf{1^+_v}$ has weight 1. When $\mathbf{Q_s}$ is non-vanishing, we
describe the spaces $L(-3,P)$, $P\geq 1$. When $\mathbf{Q_s}$ vanishes,
we are describing the space $L(-3,0)$. The representations $\mathbf{Q_s}$
and $\mathbf{Q_v}$ of the paint group that appear here will be discussed
later.
\end{description}
We can now note a regularity in the decomposition of $\mathbb{K}_{\rm
short}$. For all values of the rank we always have the space
$(\mathcal{S}^+,\mathbf{Q_s})$ that associates a representation of the
paint group to the half spinor representation of the sub Tits-Satake
algebra. In the case of rank $r=4$ in addition to this we also have the
representations $\mathbf{Q_v}$ and $\mathbf{Q_{\bar s}}$, which we
associate to what we can name the $\mathcal{V}^+$ and ${\bar
\mathcal{S}}^+$ half modules. These latter  exist in rank $4$ and can
vanish in lesser rank. The reader may compare this with the remarks at
the end of section \ref{ss:classhomquatK}, where the generators for the
manifolds of rank lower than 4 were recognized as subset of those of rank
4. With this proviso we have established a notation covering all the
cases which enables us to proceed to the next point and give a general
definition of the Tits-Satake projection.
\paragraph{D]}
The \textit{paint algebra} $\mathbb{G}_{\rm paint}$ contains a subalgebra
\begin{equation}
  \mathbb{G}^0_{\mathrm{subpaint}} \, \subset \,
  \mathbb{G}_{\mathrm{paint}},
\end{equation}
such that with respect to $\mathbb{G}^0_{\mathrm{subpaint}}$,  each of
the three irreducible representations $\mathbf{Q_{v,s,{\bar s}}}$
branches as:
\begin{equation}
  \mathbf{Q_{v,s,{\bar s}}} \,  \stackrel{\mathbb{G}^0_{\mathrm{subpaint}}}{\Longrightarrow} \,
  \underbrace{{\mathbf{1}}}_{\mbox{singlet}}
  \, \oplus \, {\mathbf{J_{{v,s,{\bar s}}}}},
\label{splittatonibussolv}
\end{equation}
where the representation ${\mathbf{J_{{v,s,{\bar s}}}}}$ is in general
reducible.
\paragraph{E]}
The restriction to the singlets of $\mathbb{G}^0_{\mathrm{subpaint}}$
defines a Lie subalgebra of $\Solv_\mathrm{M}$, namely, if we set:
\begin{equation}
  \Solv_{\mathrm{TS}} \, \equiv \, \Solv_{\mathrm{subTS}} \, \oplus \, \left( \mathcal{V}^+, \mathbf{1}\right)
  \, \oplus \, \left( \mathcal{S}^+, \mathbf{1}\right) \, \oplus \, \left( \overline{\mathcal{S}}^+,
  \mathbf{1}\right),
\label{finalmente}
\end{equation}
we get:
\begin{equation}
 \left[  \Solv_{\mathrm{TS}} \, , \, \Solv_{\mathrm{TS}} \right] \,
 \subset \, \Solv_{\mathrm{TS}}.
\label{subalgebraclosure}
\end{equation}
%\end{description}
\bigskip
\par
Relying on all the above properties and structures described in points
A], B], C], D] and  E], which turn out to hold true for every
$\Solv_{\mathcal{M}}$ considered in the sequel, irrespectively whether it
is associated with a symmetric space or not, we can define the
Tits-Satake projection at the level of solvable algebras by stating:
\begin{eqnarray}
&&\Pi_{\mathrm{TS}} \quad : \quad \Solv_{\mathcal{M}} \, \longrightarrow
\, \Solv_{\mathrm{TS}}
\, \subset \, \Solv_{\mathcal{M}} \nonumber\\
&&\Psi \, \in \, \Solv_{\mathrm{TS}} \quad \mbox{if and only if} \quad :
\quad \forall X \, \in \, \mathbb{G}^{0}_{\mathrm{subpaint}} \ :\ [X ,
\Psi ]= 0. \label{genraladefi}
\end{eqnarray}
In other words, we define the Tits-Satake solvable subalgebra
$\Solv_{\mathrm{TS}}$ as spanned by all the \textit{singlets} under the
\textit{subpaint group} $\mathrm{G}_{\rm subpaint}$. By its very
definition the Tits-Satake subalgebra contains the \textit{sub
Tits-Satake algebra} $\Solv_{\mathrm{subTS}} \, \subset \,
\Solv_{\mathrm{TS}}$ which is made of singlets with respect to the full
paint group $\mathrm{G}_{\rm paint}$  The subtle points in the above
definition of the Tits-Satake projection is given by point D] and E].
Namely it is a matter of fact, which is not obvious a priori, that the
addition of the three modules (occasionally vanishing) $\mathcal{V}^+,
\mathcal{S}^+,  \overline{\mathcal{S}}^+$ to the sub Tits-Satake algebra
$\Solv_{\mathrm{subTS}}$ always defines a new Lie algebra. Being true
this implies that a subalgebra $\Solv_{\mathrm{TS}}$ with the structure
(\ref{finalmente}) exists in $\Solv_{Q}$ and
$\mathbb{G}_{\mathrm{subpaint}}$ is its stability subalgebra. Vice versa,
the existence of a subpaint algebra such that the decomposition
(\ref{splittatonibussolv}) is true, implies that the subspace
(\ref{finalmente}) closes a subalgebra since the kernel of a subalgebra
of automorphisms is necessarily a closed subalgebra.

\subsection{Results for the TS projection of homogeneous special manifolds}
The discussion of section \ref{Tsforsolvable} outlined the scheme of
Tits-Satake projections. We will now demonstrate how indeed the
generators of table~\ref{tbl:genV} and their weights given in
(\ref{weightsV}) \cite{D'Auria:2004cu} fit in this picture. In section
\ref{pitturagruppo}, these results will be developed starting from the
paint group as required by the systematic procedure outlined above.
 \par
It is convenient to change the basis of the CSA in the following way:
\begin{equation}
    H_1  =  h_0 + h_1,\qquad H_2  =  h_0 - h_1,\qquad
    H_3  =  h_2 + h_3,\qquad H_4  =  h_2 - h_3.
\label{newCSA}
\end{equation}
This leads to
\begin{equation}
 \label{weightsVnewbasis}
\begin{array}{|llll|}
\hline
 H_1\,:\,(0,0,0,0)  &g_0\,:\,(1, 1, 0, 0) & q_0\,:\,(0, 1, -1, 0) & p_0\,:\,(1, 0, 1, 0) \\
 H_2\,:\,(0,0,0,0) & g_1\,:\,(1, -1, 0, 0) & q_1\,:\,(0, 1, 1, 0) & p_1\,:\,(1, 0, -1, 0) \\
 H_3\,:\,(0,0,0,0) & g_2\,:\,(0, 0, 1, 1) & q_2\,:\,(1, 0, 0, -1) & p_2\,:\,(0, 1, 0, 1) \\
 H_4\,:\,(0,0,0,0) & g_3\,:\,(0, 0, 1, -1) & q_3\,:\,(1, 0, 0, 1) & p_3\,:\,(0, 1, 0, -1) \\
 X^+\,:\,(0, 0, 1, 0) & X^-\,:\,(0, 0, 0, 1) & \tilde X^+\,:\,(1, 0, 0, 0) & \tilde X^-\,:\,(0, 1, 0, 0) \\
Y^+\,:\,(\frac{1}{2}, - \frac{1}{2}, \frac{1}{2},- \frac{1}{2} ) &
Y^-\,:\,(\frac{1}{2},- \frac{1}{2},- \frac{1}{2},\frac{1}{2}) & \tilde
Y^+\,:\,(\frac{1}{2},\frac{1}{2},\frac{1}{2},
  \frac{1}{2}) & \tilde Y^-\,:\,(\frac{1}{2},\frac{1}{2}, - \frac{1}{2}, - \frac{1}{2}) \\
Z^+\,:\,(\frac{1}{2},-\frac{1}{2}
     ,\frac{1}{2},\frac{1}{2}) & Z^-\,:\,(\frac{1}{2},-\frac{1}{2}
     ,- \frac{1}{2},
  -\frac{1}{2}) & \tilde Z^+\,:\,(\frac{1}{2},\frac{1}{2},\frac{1}{2},
  -\frac{1}{2} ) & \tilde Z^-\,:\,(\frac{1}{2},\frac{1}{2},
  -\frac{1}{2},
  \frac{1}{2}) \\
\hline
\end{array}
\end{equation}
The subset $(H_i,g_i,q_i,p_i)$ can be recognized as the CSA and the 12
positive roots of the $D_4$ simple root system. They generate the
solvable sub Tits-Satake algebra:
\begin{equation}
  \mbox{span} \left\{H_i,g_i,q_i,p_i\right\} = \Solv_{\mathrm{subTS}}
  \, \equiv \, \Solv \left[ \frac{\mathrm{SO(4,4)}}{\mathrm{SO(4)} \times
  \mathrm{SO(4)}}\right].
\label{subso44}
\end{equation}
With reference to the general decomposition of $\mathbb{K}_{\rm short}$
mentioned in (\ref{shortonR4}) the following identification is henceforth
evident:
\begin{eqnarray}
 \mbox{span} \left\{X^+,X^-,\widetilde{X}^+,\widetilde{X}^-\right\}
 & = & \left(\mathbf{ 4^+_v},\mathbf{Q_v}\right), \nonumber\\
\mbox{span} \left\{Y^+,Y^-,\widetilde{Y}^+,\widetilde{Y}^-\right\}
 & = & \left(\mathbf{ 4^+_s},\mathbf{Q_s}\right), \nonumber\\
 \mbox{span} \left\{Z^+,Z^-,\widetilde{Z}^+,\widetilde{Z}^-\right\}
 & = & \left(\mathbf{ 4^+_{\bar s}},\mathbf{Q_{\bar s}}\right).
\end{eqnarray}
Indeed, the weights assigned to the $4$ operators of type $X$ are the $4$
positive weights of the $8$--dimensional vector representation of
$\mathrm{SO(4,4)}$. The weights assigned to the $4$ operators of type $Y$
are the $4$ positive weights of the $8$--dimensional spinor
representation $\mathbf{s}$ of $\mathrm{SO(4,4)}$, as there are an
\textit{even} number of minus signs in the eigenvalues $\pm \ft 12$. The
odd number of minus signs for the operators of type $Z$ identifies them
with the positive weights of the representation $\overline{\mathbf{s}}$
of $\mathrm{SO(4,4)}$.
\par
Before turning to that let us examine the cases of lower rank. To this
effect observe that the case $q=-3$ is not a \textit{special}
quaternionic manifold. Its symmetry structure is not of the form of table
\ref{tbl:genV}. For the other cases, we have explained at the end of
section \ref{ss:classhomquatK} how they can be obtained from truncating
the general structure of rank 4 spaces. There is an anomaly for the case
$\mathrm{SG}_5$ where the weights do not follow the scheme of
(\ref{weightsV}), but were given in (\ref{weightsSG5}). These can be
recognized as the CSA and 6 positive roots of G$_2$. In this case we have
a maximally split algebra and the Tits-Satake projection is trivial.
\par
In general the Tits-Satake projection on the algebras is a subalgebra
consisting of only one root with the same root vector. Therefore it
removes the redundancy of the $X$, $Y$ and $Z$ columns indicated in the
last row of table \ref{tbl:genV}. In the generic case, the Tits-Satake
projected algebra is thus just the algebra with one entry in any entry of
table~\ref{tbl:genV} that is present. This reduction is performed group
theoretically in the following way. Once the representations of the paint
group to which each of the $X,Y$ and $Z$ spaces have been assigned have
been identified one has to single out the subpaint group such that each
of these representations splits into a singlet plus more and the
collection of the three singlets plus the sub Tits-Satake algebra makes
the Tits-Satake one.

%%%%%%%%%%%%%%%%%%%%%%%%%%%%%%%%%%%%%%%%%%%%%%%%%%%%%%%%%%
% Systematics of Paint Groups %%%%%%%%%%%%%%%%%%%%%%%%%%%%
%%%%%%%%%%%%%%%%%%%%%%%%%%%%%%%%%%%%%%%%%%%%%%%%%%%%%%%%%%%
\section{The systematics of paint groups}
\label{pitturagruppo}

As we explained in section \ref{Tsforsolvable}, the Tits-Satake
projection originally defined for symmetric spaces in terms of a
geometrical projection of the root space, can be generalized to all
solvable algebras of special geometries in terms of the paint and
subpaint group structures. The systematic procedure outlined there,
started as step A] with the identification of the paint group. This is
what we do now, unveiling a very elegant pattern of such paint groups.
\par
As we claimed in the introduction, the specially fascinating property of
the paint group is that it is invariant under the $\mathbf{r}$-map or
$\mathbf{c}$-map, namely under dimensional reduction. For this reason it
is not relevant whether we identify it at the level of the quaternionic,
special K{\"a}hler or real member of a given family. It is a property of the
entire family. We begin by reviewing the case of the symmetric spaces.

\subsection{The paint group for noncompact symmetric spaces}
 \label{pitturasimmetrica}

In section \ref{Tsforsolvable}, we defined the paint group as the group
of external automorphisms of the solvable algebra associated with a
certain homogeneous space (\ref{pittureFuori}). For noncompact symmetric
spaces there exists another, more common, definition of the paint group.
Referring to the presentation in the beginning of section
\ref{genediscussa},
the paint group is defined as a subgroup of $\mathbb{H}$, whose Cartan
generators are those in $\mathcal{H}^{\rm comp}$ and the roots are those
in $\Delta _{\mathrm{comp}}$ (and their negatives), i.e. those that have
no component $\alpha_{||}$ in the decomposition (\ref{splittus}).

\begin{figure}[!hbt]
 \caption{\it Satake diagram of $E_{6(2)}$. The paint group can be
seen to be $\SO(2)^2$.\label{diagE62}}
\begin{center}
\iffigs
 \includegraphics[height=25mm]{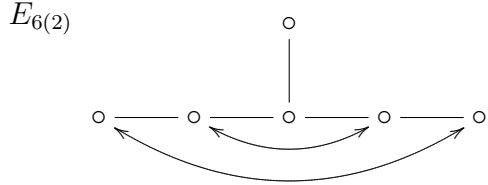}
\else
 $E_{6(2)}$ \xymatrix{ & & {\circ} \ar@{-}[d] \\ {\circ}
\ar@{-}[r] \ar@/_2pc/@{<->}[rrrr] & {\circ} \ar@{-}[r]
\ar@/_1pc/@{<->}[rr] & {\circ} \ar@{-}[r] & {\circ} \ar@{-}[r] & {\circ}
}
\end{center}
 \fi
 \iffigs
 \hskip 1.5cm \unitlength=1.1mm
 \end{center}
  \fi
\end{figure}

As we mentioned already in the example in section \ref{illustroexe}, a
real form ${\mathbb{G}}_\mathbb{R}$ of the Lie algebra $\mathbb{G}$ is
represented by the so-called Satake diagrams (see for instance
\cite{Helgason}), which are Dynkin diagrams with the following extra
decorations:
\begin{itemize}
\item Compact simple roots (those in $\Delta _{\mathrm{comp}}$) are denoted by filled circles.
\item Simple roots that result in the same restricted root setting $\alpha _\bot=0$ are
connected with a two-sided arrow. These are simple roots that necessarily
belong in $\Delta ^\delta $.
\end{itemize}
Given the Satake diagram the paint group can then be read  from it in the
following way. The black dots form a Dynkin diagram of the semi-simple
type. The paint group then contains a factor corresponding to this
painted subdiagram. This corresponds to the roots in $\Delta
_{\mathrm{comp}}$ and the elements of $\mathcal{H}^{\rm comp}$ for which
these roots have non-vanishing components. Furthermore, for every arrow,
there is one additional $\SO(2)$-factor that commutes with the rest of
the paint group. These correspond to the additional generators in
$\mathcal{H}^{\rm comp}$. An example of this is given in figures
\ref{e8dink} and \ref{diagE62}.
For the symmetric quaternionic spaces of rank 4, the paint groups are
summarized in table~\ref{paintgrsymm}.
\begin{table}[htb]
\caption{\it Symmetric very special real spaces and their corresponding
K{\"a}hler and quaternionic spaces. The last two columns indicate the paint
and subpaint groups respectively. The spaces corresponding to $L(1,1)$
(above the line) are maximally noncompact and do not have any paint
group. \label{paintgrsymm}}
\begin{center}
\begin{tabular}{|c|ccc|cc|}
\hline $C(h)$ & $\mbox{real}$ & $\mbox{K{\"a}hler}$ &
$\mbox{quaternionic}$ & $\mathrm{G}_{\rm paint}$ & $\mathrm{G}^0_{\mathrm{subpaint}}$\\
\hline $L(1,1)$ & $\frac{\Sl(3,\mathbb{R})}{\mathrm{SO}(3)}$ &
$\frac{\mathrm{Sp}(6)}{\mathrm{U}(3)}$ &
$\frac{\mathrm{F}_{4(4)}}{\mathrm{USp}(6)\times \mathrm{SU}(2)}$ & -- & -- \\
\hline $L(2,1)$ & $\frac{\Sl(3,\mathbb{C})}{\mathrm{SU}(3)}$ &
$\frac{\mathrm{SU}(3,3)}{\mathrm{SU}(3)\times
\mathrm{SU}(3)\times\mathrm{U}(1)}$ &
$\frac{\mathrm{E}_{6(2)}}{\mathrm{SU}(2)\times \mathrm{SU}(6)}$ &
$\mathrm{SO}(2)^2$ & $\mathbf{1}$ \\
$L(4,1)$ & $\frac{\mathrm{SU}^*(6)}{\mathrm{Sp}(3)}$ &
$\frac{\mathrm{SO}^*(12)}{\mathrm{SU}(6)\times\mathrm{U}(1)}$ &
$\frac{\mathrm{E}_{7(-5)}}{\mathrm{SO}(12)\times\mathrm{SU}(2)}$ &
$\mathrm{SO}(3)^3$ & $\mathrm{SO(3)}_{\rm diag}$ \\ $L(8,1)$ &
$\frac{\mathrm{E}_{6(-26)}}{\mathrm{F}_{4(-52)}}$ &
$\frac{\mathrm{E}_{7(-25)}}{\mathrm{E}_{6(-78)}\times\mathrm{U}(1)}$ &
$\frac{\mathrm{E}_{8(-24)}}{\mathrm{E}_{7(-133)}\times\mathrm{SU}(2)}$ &
$\mathrm{SO}(8)$ & $\mathrm{G_{2(-14)}}$\\ \hline
\end{tabular}
\end{center}
\end{table}
The cases $L(4,1)$ and $L(8,1)$ have already been extensively discussed,
at the level of their quaternionic member in \cite{Fre':2005sr} and in
section \ref{illustroexe}. Here we can briefly explain the group theory
of the case $L(2,1)$. It suffices to note that the $\mathrm{E_{6(2)}}$
Lie algebra contains $\mathrm{F_{4(4)}}$ as a maximal subalgebra and that
the adjoint has the following branching rule:
\begin{equation}
  \mathbf{78} \, \stackrel{\mathrm{F_{4(4)}}}{ \longrightarrow} \, \mathbf{52}
  \oplus \mathbf{26}.
\label{78splitto}
\end{equation}
This shows that the subpaint group is empty since the normalizer of the
Tits-Satake subalgebra  $\mathrm{F_{4(4)}}$ is null. On the other hand,
recalling the decomposition of the fundamental representation of
$\mathrm{F_{4(4)}}$ with respect to the subalgebra $\mathrm{SO(4,4)}$
\begin{equation}\label{guh}
     \mathbf{26}\stackrel{\mathrm{SO(4,4)}}{ \longrightarrow}
     \mathbf{1}\oplus\mathbf{1}\oplus\mathbf{8_v}^{\rm nc}\oplus\mathbf{8_s}^{\rm nc}
     \oplus\mathbf{8_{\bar{s}}}^{\rm nc},
\end{equation}
together with the branching rule of the adjoint given in (\ref{buh}), we
conclude that under the subgroup $\mathrm{SO(4,4) \times \SO(2)^2}$ we
have:
\begin{equation}
  \mathbf{78} \, \stackrel{\mathrm{SO(4,4)} \times
\mathrm{SO(2)^2}}{ \longrightarrow} \, \left( \mathbf{28}^{\rm
nc},\mathbf{1},\mathbf{1} \right)\, \oplus \, \left( \mathbf{8_v}^{\rm
nc},\mathbf{2},\mathbf{1}\right) \oplus \, \left( \mathbf{8_s}^{\rm
nc},\mathbf{1},\mathbf{2}\right) \oplus\left( \mathbf{8_{\bar s}^{\rm
nc}},\mathbf{1},\mathbf{2}\right)
  \oplus \left( \mathbf{1},\mathbf{1},\mathbf{1}\right)\oplus \left(
  \mathbf{1},\mathbf{1},\mathbf{1}\right),
\label{78splitto2}
\end{equation}
which shows that the paint group is indeed $\mathrm{SO(2)^2}$ as claimed.
From (\ref{78splitto2}) we also read off the  representations
$\mathbf{Q_{v,s,{\hat s}}}$ defined by (\ref{shortonR4}) that pertain to
this case:
\begin{equation}
  \mathbf{Q_{v}} \, = \, \mathbf{(2,1)} \quad ; \quad \mathbf{Q_{s}} \, = \,\mathbf{ (1,2)} \quad ;
  \quad\mathbf{Q_{{\hat s}}} \, = \, \mathbf{(1,2)}.
\label{casoe6Qrep}
\end{equation}

%%%%%%%%%%%%%%%%%%%%%%%%%%%%%%%%%%%%%%%%%%%%%%%%%%%%%%%%%%%%
% Paint groups in homogeneous special manifolds %%%%%%%%%%%%
%%%%%%%%%%%%%%%%%%%%%%%%%%%%%%%%%%%%%%%%%%%%%%%%%%%%%%%%%%%%

\subsection{The paint group in homogeneous special geometries}
\label{homopainted} In the previous section we mentioned how the paint
group for non-compact symmetric spaces can be inferred from the
corresponding Satake diagrams. In this section we shall determine the
paint groups for general homogeneous special geometries.

{}From table \ref{allLpq}, one can see that all homogeneous quaternionic
spaces of rank less than 3 are symmetric. For these it is thus possible
to use Satake diagrams to determine the paint groups.
 \par
\begin{figure}[!hbp]
\caption{\it The Satake diagram of $\mathrm{SO}(q+3,3)$  for $q$ odd. The
paint group is represented by the subdiagram made of filled circles and
is seen to be $\mathrm{SO}(q)$.\label{SoSatake}} \centering
\begin{picture}(90,50)
     \put(-50,35){\circle {10}} \put (-45,35){\line
(1,0){20}}\put(-20,35){\circle {10}} \put (-15,35){\line
(1,0){20}}\put(10,35){\circle {10}} \put (15,35){\line (1,0){20}} \put
(40,35){\circle* {10}}\put (45,35){\line (1,0){20}}\put (70,35){\circle*
{2}} \put (75,35){\circle* {2}} \put (80,35){\circle* {2}} \put
(85,35){\line (1,0){20}} \put (110,35) {\circle* {10}} \put
(115,38){\line (1,0){20}} \put (130,35){\line (-1,1){10}} \put
(130,35){\line (-1,-1){10}}\put (115,33){\line (1,0){20}} \put
(140,35){\circle* {10}}
\end{picture}
%\vskip 1cm
\end{figure}
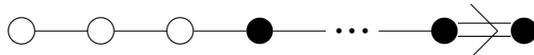

The spaces of rank 3 and 4 all have a five-dimensional origin. We can
thus use the structure of their isometry groups, as exhibited in section
\ref{isomhomspecgeom} to find the corresponding paint groups. Focusing at
the isometry algebra of the quaternionic spaces, one can immediately
recognize that the part $\mathcal{S}_q(P,\dot{P})$ (if non-trivial) will
always be part of the paint group. Indeed, it acts as a group of external
automorphisms on $\mathcal{V}_1$ and $\mathcal{V}_2$. Since moreover it
commutes with the rest of $\mathcal{V}_0$ it also acts as a group of
external
automorphisms on the solvable part of $\mathcal{V}_0$. \\
Moreover, also the $\so(q+3,3)$ part of $\mathcal{V}_0$ acts as a group
of automorphisms on $\mathcal{V}_1$ and $\mathcal{V}_2$. This implies
that the part of $\so(q+3,3)$ that acts as a group of external
automorphisms on its own solvable algebra will also be part of the paint
group. But this is nothing but the paint group of $\mathrm{SO}(q+3,3)$
and can be inferred from the corresponding Satake diagram fig.
\ref{SoSatake}.
It can be easily
verified that this contributes an $\mathrm{SO}(q)$-factor to the paint
group.
 \par
The argument for the K{\"a}hler and real spaces is completely analogous. One
should just replace $\so(q+3,3)$ by $\so(q+2,2)$, $\so(q+1,1)$
respectively. In each case one can conclude that the paint group for a
general homogeneous special geometry is given by
\begin{equation}
\label{Gpaintgen} \mathrm{G}_{\rm paint} = \mathrm{SO}(q) \times
\mathcal{S}_q(P,\dot{P}).
\end{equation}
%%%%%%
Accidentally, this structure is insensitive to the sign of $q$. Indeed
although the structure of the solvable algebra is very different, for say
$q=3$ and for $q=-3$, yet the paint group is the same in both cases and
the same irreducible representations are present. Hence also the subpaint
group and the relevant decompositions will be the same for $\pm q$.
\par
The action of the paint group on the solvable algebra of the
corresponding manifolds can also be induced from section
\ref{isomhomspecgeom}. Let us focus on the real spaces for a moment.
\begin{itemize}
\item The Cartan generators of the solvable algebra are singlets
under the paint group.
\item The $q$ nilpotent generators of
$\Solv(\so(q+1,1))$ (corresponding to the generators of type $X$ in
Alekseevsky's notation) transform as a vector under the $\SO(q)$ part of
the paint group, while they are inert under $\mathcal{S}_q(P,\dot{P})$.
\item The generators $\uxi^\indvone$ transform in a (in general reducible)
spinor representation of $\SO(q+1)$. The space $\uxi^\indvone$ however
splits into two subspaces $Y$ and $Z$, with different gradings with
respect to the non-compact Cartan generator of $\SO(q+1,1)$ (denoted by
$\ualpha$). Since the $\SO(q)$ part of the paint group commutes with this
Cartan generator, the $\uxi^\indvone$ split into two (in general
reducible) spinor representations under the $\SO(q)$ factor of the paint
group. These correspond to the spaces $Y$ and $Z$ in Alekseevsky's
notation. Since $\mathcal{S}_q(P,\dot{P})$ commutes with $\ualpha$ as
well, the spaces $Y$ and $Z$ will also separately transform in (vector)
representations of the $\mathcal{S}_q(P,\dot{P})$ factor of the paint
group.
\end{itemize}
Since for the K{\"a}hler and quaternionic-K{\"a}hler spaces, the generators of
the solvable algebra occur in the same representations as in the case of
the real spaces, the story for them is essentially the same as described
above. The only difference, is that the number of singlets is increased,
giving rise to non-trivial $\mathrm{G}_{\mathrm{subTS}}$ algebras.
 \par
We can now easily match the above findings with the general discussion of
section \ref{Tsforsolvable}. There we worked at the level of the
quaternionic member of each family, since this allowed to include all
cases, also those that are not in the image of the $\mathbf{c}$-map or of
the $\mathbf{r}$-map. Yet the invariance of the paint group with respect
to these maps precisely means that the $\mathbf{Q_{v,s,{\bar s}}}$
representations remain the same in real, in special K{\"a}hler and in
quaternionic-K{\"a}hler algebras. What changes is just the
$\mathrm{G}_{\mathrm{subTS}}$ algebra which climbing up from quaternionic
to real geometry (dimensional oxidation) is progressively reduced of
rank. The result was anticipated in item B] of section
\ref{Tsforsolvable}, and is given in table \ref{tbl:subTS}.
 \par
The information contained in the above discussion is what was needed in
order to determine the desired representations $\mathbf{Q_{v,s,{\bar
s}}}$ of the paint group, respectively associated with the vector, spinor
and conjugate spinor weights of the sub Tits-Satake algebra. Our
findings, which are an immediate consequence of the real Clifford algebra
representations discussed in appendix \ref{realcliffalg}, are summarized
in table \ref{QVSSB}.
\begin{table}[!htb]
 \caption{\it The assignments of paint group representations in homogeneous special geometries.
 \label{QVSSB}}
\begin{center}
\begin{tabular}{|c|c|c|c|c|}
\hline
\textbf{Family}
& \textbf{Paint group} & $\mathbf{Q_V}$ &$\mathbf{Q_S}$ & $\mathbf{Q_{\bar{S}}}$ \\
\hline\hline
$\mathrm{L}(-3,P)$
&$\mathrm{SO(3)} \times\mathrm{USp(2P)}$ & (3,1) & (4,P) &--\\\hline
$\mathrm{SG}_4$
&$\mathrm{SO(2)}$ & -- & 2 & -- \\\hline\hline

$\mathrm{L}(-2,P)$ &
$\mathrm{SO(2)}\times\mathrm{U(P)}$ & (2,1) & (2,P) & -- \\\hline

$\mathrm{SG}_5$ &
$\mathbf{1}$ & -- & -- & -- \\\hline\hline

$\mathrm{L}(-1,P)$ &
$\mathrm{SO(P)}$ & -- & P &
-- \\\hline\hline

$\mathrm{L}(0,P,\dot{P})$ &
%$\mathrm{L}(0,P,\dot{P})_Q$ &
$\mathrm{SO(P)}\times\mathrm{SO(\dot{P})}$ & --& (P,1) & (1,$\dot{P}$)
\\\hline

\multicolumn{5}{|c|}{\emph{normal}}
\\ \hline

$\mathrm{L}(q,P)$
 &$\mathrm{SO}(q)\times\mathrm{SO}(P)$
& $(q,1)$ & $(\ft12\mathcal{D}_{q+1},P)$ & $(\ft12\mathcal{D}_{q+1},P)$ \\
$q=1,7\,\, \mbox{mod}\, 8$ %&
& & & &\\ \hline

$\mathrm{L}(q,P,\dot{P})$ &
$\mathrm{SO}(q)\times\mathrm{SO}(P)\times\mathrm{SO}(\dot{P})$ &
$(q,1,1)$ & $(\ft12\mathcal{D}_{q+1},P,1) +$ &
 $(\ft12\mathcal{D}_{q+1},P,1) +$ \\
$q=8\,\, \mbox{mod}\, 8$
& & & $(\ft12\mathcal{D}_{q+1},1,\dot{P}) $ &
$(\ft12\mathcal{D}_{q+1},1,\dot{P}) $\\\hline

\multicolumn{5}{|c|}{\emph{complex}}
\\ \hline
$\mathrm{L}(q,P)$
& $\mathrm{SO}(q)\times\mathrm{U}(P)$ & $(q,1)$ & $(\ft12\mathcal{D}_{q+1},P)$ & $(\ft12\mathcal{D}_{q+1},P)$ \\
$q=2,6\,\, \mbox{mod}\, 8$ % &
 & & & &\\ \hline

 \multicolumn{5}{|c|}{\emph{quaternionic}}
 \\ \hline

$L(q,P)$
 & $\mathrm{SO}(q)\times\mathrm{USp}(2P)$ & $(q,1)$
& $(\ft12\mathcal{D}_{q+1},P)$ & $(\ft12\mathcal{D}_{q+1},P)$
\\
$q=3,5\,\, \mbox{mod} \, 8$
 & & &  & \\ \hline

$L(q,P,\dot{P})$
& $\mathrm{SO}(q)\times\mathrm{USp}(2P)\times\mathrm{USp}(2\dot{P})$ &
$(q,1,1)$ & $(\ft12\mathcal{D}_{q+1},P,1) + $ &
$(\ft12\mathcal{D}_{q+1},P,1) + $
\\
$q=4\,\, \mbox{mod} \, 8$
& & & $(\ft12\mathcal{D}_{q+1},1,\dot{P})$ &
$(\ft12\mathcal{D}_{q+1},1,\dot{P})$\\ \hline
\end{tabular}
\end{center}
\end{table}
In writing the spinor representations, one may comment about the way that
the spinor representations are denoted. In real components, the
representations of $\SO(q)$ are of dimension $\ft12\mathcal{D}_{q+1}$.
The complex or quaternionic structure acts on the same components. A
notation $(\ft12\mathcal{D}_{q+1},P)$ as representation of $\SO(q)\times
\U(P)$ for the complex case  means that it is a complex $P$ dimensional
representation for $\U(P)$ but the complex structure is taken into
account for the counting of real components in the first factor.
Alternatively, we could have written it as $(\ft14\mathcal{D}_{q+1},2P)$
when we take real components for the representation of $\U(P)$ and
complex spinor representations. Similarly, in the quaternionic case we
can write the representations of $\SO(q)\times \USp(2P)$ either as
$(\ft12\mathcal{D}_{q+1},P)$, as $(\ft14\mathcal{D}_{q+1},2P)$ (dividing
the complex structures over the two sides) or as
$(\ft18\mathcal{D}_{q+1},4P)$.

Note that for $q=-2$ the $\mathbf{Q_V}$ representation does not originate
from the $X$-generators as for $q\geq 1$, but from the equality of the
roots $q_0$ and $q_1$, of $p_0$ and $p_1$ as explained at the end of
section \ref{ss:classhomquatK}. On the other hand, for $q=-3$ we do not
have the scheme of table \ref{tbl:genV}, but the result follows from the
known scheme for symmetric spaces.
%%%%%%%%%%%%%%%%%%%%%%%%%%%%%%%%%%%%%%%%%%%%%%%%%%%%%%%
% Subpaint group
%%%%%%%%%%%%%%%%%%%%%%%%%%%%%%%%%%%%%%%%%%%%%%%%%%%%%%%
\subsection{The subpaint group}
 \label{sottopitture}
The subpaint group whose Lie algebra we denoted
$\mathrm{G}_{\mathrm{subpaint}}$  was defined in section
\ref{Tsforsolvable} through its property (\ref{splittatonibussolv})
relative to the decomposition of the representations
$\mathbf{Q_{v,s,{\bar s}}}$ or alternatively as the subgroup of the paint
group that commutes with the Tits-Satake subalgebra. As we emphasized
there, the very existence of a solvable Lie subalgebra with the structure
(\ref{finalmente}) implies the existence of the subpaint group with the
property (\ref{splittatonibussolv}) and vice versa. Hence the search for
subpaint subalgebras is the group theoretical formulation of the
Tits-Satake projection.
 \par
To single out the subpaint group $\mathrm{G}_\mathrm{subpaint} \subset
\mathrm{G}_{\mathrm{paint}}$ for a homogeneous space $L(q,P,\dot{P})$,
whose paint  group is $\mathrm{G}_{\mathrm{paint}} = \mathrm{SO}(q)\times
\mathcal{S}_q(P,\dot{P})$ the following strategy can be adopted:
\begin{enumerate}
\item Since the representation $\mathbf{Q_v}$ corresponding to the $X$--space
generators  is always of the form
\begin{equation}
  \mathbf{Q_v} \, = \, \left(\mathbf{q},\mathbf{1}\right)
\label{oppalala}
\end{equation}
where $\mathbf{q}$ denotes the vector representation of $\mathrm{SO}(q)$
and $\mathbf{1}$ the singlet representation of
$\mathcal{S}_q(P,\dot{P})$, first one decomposes this representation with
respect to the subgroup
\begin{equation}
  \mathrm{SO}(q-1)\subset
\mathrm{SO}(q) \label{soqmeno1}
\end{equation}
as $\mathbf{q} \rightarrow \mathbf{1} + \mathbf{(q-1)}$. The singlet,
named $X_{\bullet}$ is the only element of the $X$--space which survives
the TS projection.
\item Next, one looks for a subgroup $\mathrm{G}_{\mathrm{subpaint}} \subset \mathrm{SO}(q-1)\times \mathcal{S}_q(P, \dot{P})$
such that the decomposition of the spinor representations $\mathbf{Q_s}$
and $\mathbf{Q_{\bar s}}$, respectively associated with the spaces  $Y$
and $Z$, contain each at least one singlet. Then one chooses one of the
singlets $Y_\bullet$ and defines $Z_\bullet= [X_\bullet, Y_\bullet]$ to
complete the Tits-Satake projection.
\end{enumerate}
 \par
Thus in order to figure out the subpaint group one should consider the
explicit spinor representations of the Clifford algebra
$\mathcal{C}_{q+1}$. Here we explore one by one the cases $-3 \leq q \leq
9$ with arbitrary $P$, and we give also some examples for $q>9$. The
results depend on the structure of the $\mathcal{S}_q(P,\dot{P})$ part of
the paint group, and as such the cases with different $q$ are divided
into three groups: \textit{normal}, \textit{almost complex} and
\textit{quaternionic} as explained in appendix \ref{realcliffalg}.
\par
There are also two quaternionic spaces that are outside of the
$L(q,P,\dot{P})$ families, namely  pure $\mathcal{N}= 2$ supergravities
in $D=4$ and $D=5$ dimensions. They are both symmetric spaces, out of
which only the first one, $SG_4$, is non--split and has a non--trivial
paint group: $\mathrm{SO}(2)$. The subpaint group in this case is empty,
because in order to get singlets we have to break the paint group
completely.
%%%%%%%%%%%%%%%%%%%%%%%%%%%%%%%
\subsubsection{The normal case}
\label{sub_normal}
\begin{description}
\item [$q=\pm 1$.]
The paint group is $\mathrm{G}_{\mathrm{paint}} = \mathrm{SO}(P)$. In the
solvable algebra are only $P$--dimensional vector representations
present. These decompose as $\mathbf{P} \,\rightarrow \, \mathbf{1} +
\left( \mathbf{P-1}\right) $ under $\mathrm{SO}(P-1)$, which is therefore
identified as the subpaint group.
\item {\bf $q=0$.}\hskip 0.2cm  The paint group is $\mathrm{G}_{\mathrm{paint}} =
\mathrm{SO}(P)\times\mathrm{SO}(\dot{P})$, and there are again only
vector representations. Analogously as in the previous case, we thus find
that the subpaint group is
$\mathrm{G}_{\mathrm{subpaint}}=\mathrm{SO}(P-1)\times
\mathrm{SO}(\dot{P} - 1)$.
 \item {\bf $q=7$.} The paint group is $\mathrm{G}_{\mathrm{paint}} =
 \mathrm{SO}(7)\times\mathrm{SO}(P)$. The two representations that
 are involved are  $(\mathbf{7},\mathbf{1}) = \mathbf{Q_v}$ and
 $(\mathbf{8},\mathbf{P})=\mathbf{Q_s} = \mathbf{Q_{\bar s}} $, the real 8--dimensional spinor representation of
 $\SO(7)$. The subpaint group that allows to find singlets in both is $\mathrm{G}_{\mathrm{subpaint}} =
 \mathrm{SU}(3)\times\mathrm{SO}(P-1)$, where $\mathrm{SU}(3)\subset\mathrm{G}_2\subset\mathrm{SO(7)}$.
 Indeed, the representations split as follows
\begin{eqnarray} && \nonumber
(\mathbf{7},\mathbf{1}) \rightarrow (\mathbf{1},\mathbf{1}) + (\mathbf{3},\mathbf{1})+ (\overline{\mathbf{3}},\mathbf{1}), \\
&& (\mathbf{8},\mathbf{P}) \rightarrow (\mathbf{1},\mathbf{1}) +
(\mathbf{1},\mathbf{1}) +
 (\mathbf{1},\mathbf{P-1}) + (\mathbf{1},\mathbf{P-1}) + (\mathbf{6},\mathbf{1})
 +(\mathbf{6},\mathbf{P-1}).
 \end{eqnarray}
\item {\bf $q=8$.} The paint group is $\mathrm{G}_{\mathrm{paint}} =
 \mathrm{SO}(8)\times\mathrm{SO}(P)\times\mathrm{SO}(\dot{P})$.
 In this case there are two inequivalent real spinor representations of $\mathrm{SO}(8)$
 involved: $\mathbf{8_s}$ and $\mathbf{8_{\bar{s}}}$. The representations of the full paint group are
 $\mathbf{Q_v} = (\mathbf{8_v},\mathbf{1},\mathbf{1})$, $\mathbf{Q_s} = (\mathbf{8_s},\mathbf{P},\mathbf{1})
 \oplus (\mathbf{8_s},\mathbf{1},\dot{\mathbf{P}})$ and
 $\mathbf{Q_{\bar s}} = (\mathbf{8_{\bar{s}}},\mathbf{P},\mathbf{1})\oplus
 \mathbf{(8_{\bar{s}}},\mathbf{1},\dot{\mathbf{P}})$. Following the strategy described above, we select first
  the subgroup $\mathrm{SO}_0(7) \subset \mathrm{SO}(8)$ , which splits the
8--dimensional vector representation (space $X$) into a singlet plus a
7--dimensional vector. Both spinor representations remain irreducible
under the action of this subgroup. Hence, we have to look for a smaller
subgroup inside $\mathrm{SO}_0(7)$. This is $\mathrm{G}_2$, which can be
defined  as the intersection $\mathrm{SO}_0(7) \bigcap
\mathrm{SO}_+(7)\bigcap \mathrm{SO}_-(7)$, where $\SO(7)_\pm$ are the
stability subgroups of the spinor and conjugate spinor representations,
respectively. In this case,  in order to obtain
 singlets  in the spinor representation it suffices to split just one
 of the two vector representations, either $\mathbf{P}$, or
 $\dot{\mathbf{P}}$. Indeed, e.g. under $\mathrm{G}_2\times\mathrm{SO}(P-1)\times \SO(\dot P)$
 the representations split as follows
\begin{eqnarray}
(\mathbf{8_v,1,1}) & \rightarrow & \mathbf{(1,1,1)} + \mathbf{(7,1,1)}, \nonumber \\
 \mathbf{(8_s,P,1)} + \mathbf{(8_s,1,\dot{P})} & \rightarrow &
  \mathbf{(1,1,1)} + \mathbf{(7,1,1)} + \mathbf{(1,P-1,1)}\nonumber\\
  &&  + \mathbf{(7,P-1,1)} + \mathbf{(1,1,\dot{P})}
  +\mathbf{(7,1,\dot{P})}, \nonumber\\
  \mathbf{(8_{\bar s},P,1)} + \mathbf{(8_{\bar s},1,\dot{P})} & \rightarrow &
  \mathbf{(1,1,1)} + \mathbf{(7,1,1)} + \mathbf{(1,P-1,1)} \nonumber\\
  &&  + \mathbf{(7,P-1,1)} + \mathbf{(1,1,\dot{P})}
  +\mathbf{(7,1,\dot{P})} .
 \end{eqnarray}
  In this way we obtain singlets in the decomposition of all
three  involved representations. The subpaint group is thus either
$\mathrm{G}_{\mathrm{subpaint}} =
 \mathrm{G}_2\times\mathrm{SO}(P-1)\times \SO(\dot P)$ or $\mathrm{G}_{\mathrm{subpaint}} =
 \mathrm{G}_2\times\SO(P)\times \mathrm{SO}(\dot{P}-1)$.
 In case $\dot P=0$, the subpaint group is $\mathrm{G}_{\mathrm{subpaint}} =
 \mathrm{G}_2\times\mathrm{SO}(P-1)$.

 \item  {\bf $q=9$.} The paint group is $\mathrm{G}_{\mathrm{paint}} =
 \mathrm{SO}(9)\times\mathrm{SO}(P)$. Real spinor representations of
 $\mathrm{SO}(9)$ are $16$--dimensional.
 The involved representations of the paint group are $\mathbf{Q_v} = \mathbf{(9,1)}$ and  $\mathbf{Q_{s}} =
 \mathbf{Q_{\bar s}}=\mathbf{(16,P)}$. The subpaint group is
 $\mathrm{G}_{\mathrm{subpaint}} = \mathrm{SO}(7)_+\times\mathrm{SO}(P-1)$, where
 $\mathrm{SO}(7)_+\subset\mathrm{SO}(8) \subset \mathrm{SO}(9)$. The subpaint
 group induces the following splitting
\begin{eqnarray}
\mathbf{(9,1)} & \rightarrow & \mathbf{(1,1)} + \mathbf{(8_s,1)}  \\
 \mathbf{(16,P)} & \rightarrow & \mathbf{(1,1)} + \mathbf{(1,P-1)} + \mathbf{(7_v,1)} + \mathbf{(7_v,P-1)} +
 \mathbf{(8_s,1)} +\mathbf{(8_s,P-1)}\nonumber
 \end{eqnarray}
\end{description}
In general, we thus conclude that we started from paint groups of the
form $\SO(q)\times \SO(P)\times \SO(\dot P)$. The first factor is broken
to the common stability subgroup of the vector and spinor
representations. Further we break $\SO(P)$ to $\SO(P-1)$. In case $P,\dot
P\geq 1$ we have to break only one of the factors in $\SO(P)\times
\SO(\dot P)$ in this way, except for $q=0$, which is special due to the
fact that in that case the two factors belong to different restricted
roots of the solvable algebra.
%%%%%%%%%%%%%%%%%%%%%%%%%%%%%%%%%%%%%%%%%%%
\subsubsection{The almost complex case}
The search for the paint algebra is analogous to the one for the real
case. We only need a little bit more of care about the treatment of the
complex structure. The paint group is $\SO(q)\times U(P)$. We will
consider this complex structure as part of the unitary group. In order to
find a singlet in the representation $\mathbf{Q_s}$ or
$\mathbf{Q_{\overline{s}}}$, we have to consider the stability subgroup
of a vector of $\U(P)$, which is $\U(P-1)$. Furthermore we have to find
as in the real case the common stability group of a vector and spinor
representation of $\SO(q)$. This we will do explicitly for $q=-2,2$ and
6. The subpaint group is then the product of the latter with $\U(P-1)$.
\begin{description}
  \item[$q=\pm 2$.] $\SO(2)$ is already broken by the vector
  representation. Hence the analysis is finished at this point and we obtain $\mathrm{G}_{\mathrm{subpaint}} = \mathrm{U}(P-1)$. However,
  note that the vector as well as the 2-dimensional spinor representations therefore split in 2 singlets.
  \item[$q=6$.] The vector representation breaks $\SO(6)$ to $\SO(5)$. The spinor
  representation of $\SO(6)$, which is as a real 8-dimensional representation,
  is the same as the one of $\SO(5)$ . To analyse the latter, it is convenient
  to use the isomorphism $\so(5) \sim \usp(4)\sim \su(2,\mathbb{H})$. Then the spinor
  representation becomes a vector representation. A typical vector can be put along one quaternion,
  such that it is left
  invariant under the transformations of the other quaternions. Therefore
  the stability group is $\su(1,\mathbb{H})\sim\su(2)$. Note that we do
  not have to consider here the generator associated with the complex
  structure on the Clifford algebra as this has been taken into account
  on the side of the $\U(P)$ factor.
  When we consider the decompositions
  of the vector and spinor representations under $\SO(6)\rightarrow
  \SO(5)\rightarrow \SU(2)$ we can consider e.g. the $\mathbf{5}$ as the
  antisymmetric traceless representation of $\usp(4)$, and obtain as such
\begin{equation}
  \mathbf{6}\rightarrow
  \mathbf{1}+\mathbf{5}\rightarrow\mathbf{1}+\mathbf{1}+\mathbf{2}+\mathbf{2}, \qquad
  \mathbf{8}\rightarrow
  \mathbf{8}\rightarrow\mathbf{1}+\mathbf{1}+\mathbf{1}+\mathbf{1}+\mathbf{4}.
 \label{decompSO6SU2}
\end{equation}
\end{description}
Hence ultimately, in this case $\mathrm{G}_{\mathrm{subpaint}} =
\mathrm{SU}(2)\times \mathrm{U}(P-1)$, and there are just more singlets
in each of the spaces $X$, $Y$ and $Z$. In order to define the
Tits-Satake projection we have to select among the $X$ generators one,
commute it with one of $Y$'s and single out the corresponding $Z$.
%%%%%%%%%%%%%%%%%%%%%%%%%%%%%%%%%%%%%%%%%
\subsubsection{The quaternionic case}
\begin{description}
  \item[$q=\pm 3$.] In this case the paint group is
$\mathrm{G}_{\mathrm{paint}} = \mathrm{SO}(3)\times{\mathrm{USp}}(2P)$.
The representations present are $\mathbf{Q_v}=\mathbf{(3,1)}$ and
$\mathbf{Q_s}=\mathbf{Q_{\bar s}}=\mathbf{(2,2P)}$. The last can
(sticking to real notations) also be denoted as $\mathbf{(4,P)}$.
Splitting the paint group first as
$\mathrm{SO}(3)\times{\mathrm{USp}}(2)\times{\mathrm{USp}}(2P-2)$, and
then taking the diagonal of the $\mathrm{SO}(3)\times{\mathrm{USp}}(2)$
factor, the representation $\mathbf(2,2P)$ splits as follows under this
$\mathrm{SO}(3)_{\mathrm{diag}}\times{\mathrm{USp}}(2P-2)$:
\begin{equation}
\mathbf{(2,2P)} \quad  \rightarrow \quad \mathbf{(1,1)} + \mathbf{(3,1)}
+ \mathbf{(2,2P-2)}.
\end{equation}
In order to obtain a singlet in the vector representation, we then take
an $\mathrm{SO}(2)$ subgroup of $\mathrm{SO}(3)_{\mathrm{diag}}$. The
subpaint group is thus
\begin{equation}
\mathrm{G}_{\mathrm{subpaint}} = \mathrm{SO}(2)_{\mathrm{diag}} \times
{\mathrm{USp}}(2P-2).
\end{equation}
  \item[$q=4$.]
The story here is similar to the previous case. The paint group is
$\mathrm{G}_{\mathrm{paint}} =
\mathrm{SO}(4)\times\mathrm{USp}(2P)\times\mathrm{USp}(2\dot{P})$. One
can choose either to break the $P$ or the $\dot P$ sector, we will do the
former, i.e. breaking $\mathrm{USp}(2P)$ to
$\mathrm{USp}(2)\times\mathrm{USp}(2P-2)$. It is useful to consider
$\mathrm{SO}(4)$ as $\mathrm{SO}(3)_L \times \mathrm{SO}(3)_R$. The
vector representation breaks into one singlet and one triplet under the
diagonal subgroup of the two $\mathrm{SO}(3)_{L/R}$. The subpaint group
is the further diagonal with $\mathrm{USp}(2)$, and is thus given by
\begin{equation}
\mathrm{G}_{\mathrm{subpaint}} =
\mathrm{SO}(3)_{\mathrm{diag}}\times\mathrm{USp}(2P-2)\times{\mathrm{USp}}(2\dot{P}).
\end{equation}
\item[$q=5$.] The paint group is $\mathrm{G}_{\mathrm{paint}} =
\mathrm{SO}(5)\times\mathrm{USp}(2P)$. The vector representation $
\mathbf{Q_v}=\mathbf{(5,1)}$ breaks $\SO(5)$ to $\SO(4)$. We split it as
usual into two subgroups $\mathrm{SO}(3)_{L/R}$. The 8-dimensional spinor
representation then splits as $\mathbf{4}+\overline{\mathbf{4}}$, where
each one transforms only under one of the factors $\SU(2)_{L/R}$
mentioned above, such that only one of these factors have to be broken to
get the subpaint group. Then we consider again the subgroup
$\mathrm{USp}(2) \times \mathrm{USp}(2P-2)\subset \mathrm{USp}(2P)$ and
define the subpaint group as the product:
\begin{equation}
  \mathrm{G}_{\mathrm{subpaint}}
  =\mathrm{SO}(3)_{\mathrm{diag}}\times\mathrm{SO}(3)_R\times\mathrm{USp}(2P-2),
\label{Gpittoq5}
\end{equation}
where the diagonal is taken between $\mathrm{SO}(3)_L$ and
$\mathrm{USp}(2)$. The representations have the following splittings
\begin{eqnarray}
 \mathbf{(5,1)} &\rightarrow& \mathbf{(1,1,1)} + \mathbf{(2,2,1)},  \nonumber \\
\mathbf{(4,2P)} &\rightarrow&  \mathbf{(1,1,1)} +
\mathbf{(3,1,1)} + \mathbf{(2,1,2P-2)} + \mathbf{(2,2,1)} \nonumber\\
&& + \mathbf{(2,1,2P-2)}.
\end{eqnarray}
\end{description}
We summarize the results in table~\ref{subpittati}.
\begin{table}[!htb]
  \caption{\it The paint and subpaint algebras of  special manifolds for the first $10$ values of the parameter
  $q$.}\label{subpittati}
\begin{center}
\begin{tabular}{|c|l|r|}\hline
$q$ & $\mathrm{G_{paint}}$ & $\mathrm{G}^0_{\rm subpaint}$\\
 \hline
 $0$ & $\SO(P)\times \SO(\dot{P})$ & $\SO(P-1)\times \SO(\dot{P}-1)$ \\
 $1$ & $\SO(P)$ & $\SO(P-1)$ \\
 $2$ & $\SO(2)\times \U(P)$ & $\U(P-1)$ \\
 $3$ & $\SO(3)\times\USp(2P)$ & $\SO(2)\times \USp(2P-2))$ \\
 $4$ & $\SO(4)\times\USp(2P)\times\USp(2\dot{P})$ & $\SO(3)\times \USp(2P-2)\times \USp(2\dot{P})$\\
 $5$ & $\SO(5)\times \USp(2P)$ & $\SO(4) \times \USp(2P-2))$ \\
 $6$ & $\SO(6)\times \U(P)$ & $\SU(2)\times \U(P-1)$ \\
 $7$ & $\SO(7)\times \SO(P)$ & $\SU(3)\times \SO(P-1)$ \\
 $8$ & $\SO(8)\times \SO(P)\times \SO(\dot{P})$ & $G_2\times \SO(P-1)\times\SO(\dot{P})$ \\
 $9$ & $\SO(9)\times \SO(P)$ & $\SO_+(7)\times\SO(P-1)$ \\
  \hline
\end{tabular}
\end{center}
\end{table}

\section{Universality classes of homogeneous special geometries}
 \label{universaclassa}
Relying on the lore presented in previous sections we can finally come to
the main point of our paper, namely the grouping of homogeneous special
geometries into a few \textit{universality classes} according to their
Tits-Satake projections. The relevance of this organization of
supergravity models was already stressed in the introduction, in view of
various aspects of supergravity dynamics, foremost among them being the
\textit{cosmic billiard dynamics}. Indeed the Tits-Satake projected model
captures the relevant features of this dynamics, \textit{i.e.} the
positions of the walls and the ensuing evolution of the cosmological
scale factors. Another, probably even more compelling reason to consider
these universality classes should have emerged from the mathematical
discussion presented in previous sections and concerns the relation of
the effective supergravity models corresponding to each special manifold
with its microscopic string theory origin. Pivotal in this respect is the
concept of \textit{sub Tits-Satake algebra} which we have introduced in
section \ref{Tsforsolvable} and defined in (\ref{subTsalgebrageneral}).
The attentive reader has certainly noted that the sub Tits-Satake algebra
is not only universal for the various rank cases but the manifold it
spans, for instance
\begin{equation}
 \mathcal{M}_{\mathrm{subTS}} \, = \, \frac{\mathrm{SO(4,4)}}{\mathrm{SO(4)} \times \mathrm{SO(4)}}
\label{subTS44}
\end{equation}
is the standard moduli space of a toroidal compactification in
perturbative string theory, $\SO(n,n)$ being the invariance group of the
Narain lattice, for a $\mathrm{T^n}$ torus. It is therefore suggestive
that the decomposition of the full scalar manifold according to the sub
Tits-Satake algebra and the paint groups corresponds to an algebraic
counterpart of the organization of degrees of freedom into string
perturbative ones and non-perturbative ones due to orbifold
singularities, branes, anti-branes and the like. The various possible
choices of paint groups, within the same universality class probably
distinguish different brane systems whose main dynamical features are
similar, notably the cosmic billiard evolution. An example of such a
microscopic interpretation of the TS projection can be found in
appendix~\ref{d3d7}.
\par
Here we collect our results for the TS projections of all Lie algebras
representing isometries of special homogeneous manifolds of
$\mathcal{N}=2$ supergravities. Applying the scheme developed in section
\ref{Titsatesection} that, in short, consists in modding out the original
isometry group by the paint group, we get a list of TS projected algebras
that we gather in two separate tables: one for the manifolds that are
very special, and the other for the manifolds that are not very special,
which we call `exotic cases'.
\subsection{Description of the Tits-Satake projections}
Let us start by analysing the very special manifolds that include the
$L(q,P,\dot{P})$ families with $q\geq -1$ and pure supergravity in five
dimensions: $\mathrm{SG}_5$. In table \ref{TSvsp} we give the
corresponding real, special K{\"a}hler and quaternionic versions of the TS
projected isometry algebras. What is important to note here is that the
original infinite set of isometry algebras, extensively discussed in
section \ref{isomhomspecgeom}, projects onto a finite set of algebras.
This means that infinite families share all the same TS projections,
which reflects the fact that each family has the same system of
restricted roots and the only difference between different members of a
family comes from multiplicities of these restricted root spaces that are
removed in the TS projection.
\begin{table}[!htb]
  \caption{\it The TS projections for isometry algebras of the very special manifolds.
  \label{TSvsp}}
\begin{center}$
\begin{array}{|l|l|l|l|}
\hline\hline
 \mbox{name}  & \mbox{Real} & \mbox{K{\"a}hler} & \mbox{quaternionic-K{\"a}hler}\\
   &    G_{\rm TS} & G_{\rm TS} & G_{\rm TS} \\
\hline\hline
  \mathrm{SG}_5  & -  & \su(1,1) &  \mathfrak{g}_{2(2)}
  \\ \hline
 L(-1,0)  & \so(1,1)  & \su(1,1)^2 & \so(3,4)
 \\ \hline
 L(-1,P)  & \so(2,1) & \Solv_{SK}(-1,1) & \Solv_Q(-1,1) \\
 \hline
 L(0,0)  & \so(1,1)\oplus \so(1,1) & \su(1,1)^3 & \so(4,4)   \\
\hline
 L(0,P)  & \so(1,1)\oplus \so(2,1) & \su(1,1)\oplus \so(3,2) & \so(5,4)  \\
\hline
L(0,P,\dot P)   & \Solv_R(0,1,1) & \Solv_{SK}(0,1,1) & \Solv_Q(0,1,1)  \\
\hline
 \hspace{-2mm} \begin{array}{l}
   L(q,P) \\
   L(4m,P,\dot P)
 \end{array}    & \mathfrak{s\ell }(3,\mathbb{R}) & \mathfrak{sp}(6) & \mathfrak{f_{4(4)}}  \\
 \hline\hline
\end{array}$
\end{center}
\end{table}

Analysing table \ref{TSvsp} one finds that the TS projected algebras
given here correspond to the isometry algebras of seven very special
manifolds: $L(1,1)$, $L(0,1,1)$, $L(0,1)$, $L(0,0)$, $L(-1,0)$,
$L(-1,1)$, $\mathrm{SG}_5$ that represent the already announced
universality classes of $\mathcal{N}=2$ supergravity models. These are
shown in table \ref{universala}.
\begin{table}[!htb]
\caption{\it The seven universality classes of very special homogeneous
geometries.\label{universala}}
\begin{center}
\begin{tabular}{||c|l||}\hline
Universality Class& Members of the class \\
%    &      \\
\hline $L(1,1)$& $\begin{array}{lcl}
  L(q,P) &\mbox{for}& q,P\geq 1
   \\
  L(4m,P,\dot{P}) &\mbox{for}& m\,P\dot P\neq 0 %}
   \\
\end{array}$\\[2mm]
\hline $L(0,1,1)$ & \ $L(0,P,\dot{P})$\hspace{6mm} for\ \ \ $ P\,\dot{P}\neq 0$ \\
\hline $L(0,1)$  & $\begin{array}{lcl}
  L(0,P,0) &\hspace{5mm} \mbox{for} & P\ne 0 \\
  L(q,0) &\hspace{5mm} \mbox{for} & q >0
\end{array}$\\
\hline $L(0,0)$& \ $L(0,0)$\\ \hline
 $L(-1,0)$&\ $L(-1,0)$\\
\hline
$L(-1,1)$& \ $L(-1,P)$\hspace{8mm} for\ \ \ $P\ge1$ \\ %[2mm]
\hline $SG_5$&\ $SG_5$\\
 \hline
\end{tabular}
\end{center}
\end{table}
\par
The remaining spaces that are not very special are of lower rank $r\leq
2$, namely $L(-3,0)$, $L(-3,P)$, $SG_4$, $L(-2,0)$, $L(-2,P)$. Since they
are all symmetric spaces, one can determine the corresponding paint
groups and TS projections, using Satake diagrams.

The reason why we call these cases exotic is because the resulting TS
projected algebras do not correspond to $\mathcal{N}=2$ supergravity
models any more, so performing the TS projection of, say, a special
K{\"a}hler space one arrives at a space that is no longer special K{\"a}hler. We
analyse the resulting TS projected spaces in detail below.
\par
{\bf rank = 1}
\begin{itemize}
\item $L(-3,0)$ corresponds to the symmetric
coset $\frac{\mathrm{USp}(2,2)}{\mathrm{USp}(2)\times \mathrm{USp}(2)}$.
The paint group in this case can be found from the Satake diagram
$\mathrm{G}_{\rm paint} = \mathrm{SO}(3)$. So, the space is not split.
The Tits--Satake projection of $\mathrm{USp}(2,2)$ leads to
$\mathrm{SU}(1,1)$ and the projected manifold
\begin{equation}
\mathcal{M}_{\rm TS} = \frac{\mathrm{SU}(1,1)}{\mathrm{U}(1)}
\end{equation}
is not quaternionic! Indeed, the space $L(-3,0)$ encodes four scalars
belonging to just one hypermultiplet, so it cannot be further restricted
to a quaternionic submanifold.
\item By $SG_4$ one denotes the quaternionic space generated by
the canonical quaternionic subalgebra
$\frac{\mathrm{SU}(2,1)}{\mathrm{SU}(2)\times\mathrm{U}(1)}$. This
manifold of dimension four is not split and the paint group is
$\mathrm{G}_{\rm paint}=\mathrm{SO}(2)$. The Tits-Satake projection of
the $\mathrm{SU}(2,1)$ restricted root system gives a so--called $bc_1$
root system\footnote{The names $bc_1$ and $bc_2$ that are used to denote
some of the the TS projected algebras denote, respectively, rank one and
rank two non-simple Lie algebras, see e.g. \cite{Helgason} for more
detailed explanations.}, given by the solvable algebra
\begin{equation}\label{bc1}
\Solv_{\rm TS} = \mbox{Span}\{h,\lambda,2\lambda\},
\end{equation}
where $(h,\lambda)$ are generators of $\Solv(\mathrm{SU}(1,1))$ and
$2\lambda$ is a generator corresponding to a doubled root. The resulting
space $\mathcal{M}_{\mathrm{TS}}$ is 3--dimensional.
\item The $L(-3,P)$ family consists of the symmetric quaternionic spaces
$\frac{\mathrm{USp}(2,2P+2)}{\mathrm{USp}(2)\times\mathrm{USp}(2P+2)}$.
Their paint group is given by $\mathrm{G}_{\rm paint} =
\mathrm{SO}(3)\times\mathrm{USp}(2P)$. The solvable algebras of the
$L(-3,P)$ manifolds are described in the following way:
\begin{equation}
\mathcal{M}_{\rm TS}: \quad \bigcirc \,\,\,\lambda, \quad m_\lambda =
4\,P, \quad m_{2\lambda} = 3
\end{equation}
where $\bigcirc$ represents the Dynkin diagram of $A_1$, $\lambda$ is the
positive root of $A_1$, which in this solvable algebra occurs $m_\lambda$
times, and $2\lambda$ is the doubled root, which has multiplicity
$m_{2\lambda}$. The Tits--Satake projection is not a symmetric space and
gives again  a $bc_1$ root system, \ref{bc1}.
\end{itemize}
 \par
{\bf rank = 2}
\begin{itemize}
\item $L(-2,0)$ is in its quaternionic version given by the symmetric
coset $\frac{\mathrm{SU}(2,2)}{\mathrm{SU}(2)\times\mathrm{SU}(2)}$. The
paint group in this case can be found from the Satake diagram and it is
$\mathrm{G}_{\rm paint} = \mathrm{SO}(2)$. So, the space is not split.
The Tits--Satake projection in this case leads to the six-dimensional
manifold:
\begin{equation}
\mathcal{M}_{\mathrm{TS}} =
\frac{\mathrm{SO}(2,3)}{\mathrm{SO}(2)\times\mathrm{SO}(3)}\,.
\end{equation}
\item $L(-2,P)$. The quaternionic version is a
symmetric space
$\frac{\mathrm{SU}(2,P+2)}{\mathrm{SU}(2)\times\mathrm{U}(P+2)}$ with
paint group $\mathrm{G}_{\rm paint} =
\mathrm{SO}(2)^2\times\mathrm{SU}(P)$.
The solvable algebra is described as follows
\begin{equation}
\mathcal{M}_{\rm TS}:  \begin{picture}(100,20)
      \put (40,15){\circle {10}} \put (37,0){$\lambda_2$}\put (45,18){\line (1,0){20}}
\put (55,15){\line (1,1){10}} \put (55,15){\line (1,-1){10}}\put
(45,13){\line (1,0){20}} \put (70,15){\circle {10}} \put
(67,0){$\lambda_{1}$}
\end{picture} \quad
m_{\lambda_1} = 2, \quad m_{2\lambda_1} = 0, \quad m_{\lambda_2} = 2P,
\quad m_{2\lambda_2} = 1\,.
\end{equation}
We have denoted the restricted root system by its Dynkin diagram, also
mentioning the multiplicities by which the (simple) restricted roots (and
their possible doubles) occur. In this case the projected space, obtained
by disregarding all multiplicities has a $bc_2$ root system.
\end{itemize}
These results are summarized in the table \ref{exotic}, where we organize
exotic spaces by just denoting the type of restricted root system that
characterize their Tits-Satake projections.
\begin{table}[!htb]
\caption{\it Exotic universality classes of homogeneous geometries and
their Tits Satake root systems. \label{exotic}}
\begin{center}
\begin{tabular}{||c|c|c||l||}\hline
Universality & SK & Q & Members of the class \\
Class    & \null & \null & \null      \\
\hline
$bc_2$ & $bc_1$&$bc_2$&  $L(-2,P), \, P>0$\\[2mm]
\hline
$b_2$ & $\frac{\mathrm{SU}(1,1)}{\mathrm{U}(1)}$ & $\frac{\mathrm{SO}(3,2)}{\mathrm{SO}(3)\times\mathrm{SO}(2)}$ & $L(-2,0)$ \\
\hline
$bc_1$ & -- & $bc_1$ & $\mathrm{SG}_4, L(-3,P), \,P>0$\\
\hline $a_1$ & -- & $\frac{\mathrm{SU}(1,1)}{\mathrm{U}(1)}$ & $L(-3,0)$\\
[2mm] \hline
\end{tabular}
\end{center}
\end{table}
%%%%%%%%%%%%%%%%%%%%%%%%%%%%%%%%%%%%%
\subsection{The universality classes}
\label{classasetta} Apart from the exotic cases discussed above, we
conclude that the other special homogeneous manifolds corresponding to
infinite families of supergravity theories with $8$ supercharges in
dimensions $D=5$, $D=4$ or $D=3$, fall into altogether $7$ universality
classes, displayed in table \ref{TSvsp}. Members within the same class
are distinguished by different choices of the paint group and of its
representations $\mathbf{Q_{v,s,{\bar s}}}$. The maximal split
representative of the class, namely the Tits-Satake projected manifold is
in five out of seven cases a symmetric coset manifold. The only cases
where the Tits-Satake manifold is not symmetric are the families that
project onto $L(-1,1)$ and $L(0,1,1)$. The most populated universality
class, which encompasses most of the homogeneous special geometries is
the class $L(1,1)$, whose split representative has been studied, from the
point of view of cosmic dynamics in \cite{Fre':2005sr}. According to the
results of \cite{Fre:2005bs}, for all the five classes where the
Tits-Satake projected manifold is symmetric, the evolution of the cosmic
billiard corresponds to an integrable system for which the solution of
the evolutionary equations can be given in closed analytic form depending
on a complete set of initial conditions.
 \par
{}From the point of view of cosmic billiards the open problem is that of
understanding the  relation between the complete integral of the
evolutionary equations in the non projected case with respect to those of
the projected one. As explained in \cite{Fre':2005sr} all solutions of
the TS projected model, which is completely integrable, are solutions of
the non projected one and can be further arbitrarily rotated by means of
the paint group to new more general solutions. Yet, counting of the
integration constants shows that the non projected model contains still
more solutions that are not of this type. It is a challenge for future
investigations to understand the structure of the missing solutions and
display their relation with the bulk of solutions produced by the
Tits-Satake projection.
%%%%%%%%%%%%%%%%%%%%%%%%%%%%
\section{Summary and perspectives}
 \label{ss:summary}
In this paper we have shown that the classical mathematical construction
of the Tits-Satake projection can be extended from the case of simple
groups that act as isometries on symmetric spaces to all solvable Lie
groups corresponding to homogeneous scalar manifolds of supergravity
models with 8 supercharges. In constructing this extension the key notion
is that of paint group. It turns out that this construction allows to
organize solvable Lie algebras and hence the corresponding supergravity
models in a small list of universality classes, where all the members of
the same class share essential basic features of their dynamics, notably
their basic cosmic billiard dynamics. We have also emphasized that the
value of such grouping in universality classes goes beyond the scope of
cosmic billiards and it is relevant to the full scope of the considered
supergravity models. We argue that the structuring of the solvable Lie
algebra according to what we named the sub Tits-Satake algebra and the
paint group is most probably a powerful tool to make contact with the
microscopic string interpretation of the model in terms of orbifold
compactification plus brane degrees of freedom, the paint group being as
the name suggests related to the permutations of coloured branes.
\par
We have emphasized that, within each universality class of special
manifolds, what distinguishes the different members, is just a compact
part of the isometry group, that we named the \emph{paint group}. In this
paper we found a general formula, for all homogeneous special geometries,
encoding the systematics of the paint groups which are preserved by the
$\mathbf{r}$- and $\mathbf{c}$- maps, which respectively connect the
real, the special K{\"a}hler and the quaternionic-K{\"a}hler members of the same
family $L(q,P,\dot{P})$ of homogeneous manifolds.
\par
Furthermore, we investigated the structure of the \emph{subpaint group}
($\mathrm{G_{subpaint}}\subset \mathrm{G_{paint}}$), defined as the
stabilizer of the Tits--Satake solvable subalgebra. This allows to
reformulate the very notion of Tits--Satake projection  in terms of Lie
algebra representations and little groups.
 \par
We now summarize to the reader how to find these results in the paper.
Homogeneous special manifolds exist in 3 varieties as real, special
K{\"a}hler and quaternionic-K{\"a}hler manifolds (and they have also an origin
from 6-dimensional supergravity, as we explained). The $\mathbf{r}$-map
defines for any such real manifold a special K{\"a}hler manifold and the
$\mathbf{c}$-map defines for any special K{\"a}hler manifold a
quaternionic-K{\"a}hler manifold. The $\mathbf{r}$- and $\mathbf{c}$-map also
each increase the rank of the manifolds by one. We consider series of
special manifolds connected by these maps. The list of homogeneous spaces
has been given in table \ref{allLpq}. Apart from two exceptional cases,
related to pure supergravity in 4 and 5 dimensions, and denoted by
$\mathrm{SG}_4$ and $\mathrm{SG}_5$, they are characterized by numbers
$q\geq -3$, $P$ and $\dot P$ with the restrictions as in that table.
 \par
We explained, following the ideas of \cite{Alekseevsky1975}, how these
homogeneous manifolds are in 1-to-1 relation with solvable algebras of
rank at most 4, that describe the translations in the manifolds. The
generators of these spaces are systematically represented by table
\ref{tbl:genV}, and the components of the related root vectors are given
in (\ref{weightsV}), where the entries are turned by 90$^\circ$ w.r.t.
table \ref{tbl:genV}. Observe that the case $q=-3$ is not a
\textit{special} quaternionic manifold. Its symmetry structure is not of
the form of table \ref{tbl:genV}. In a suitable basis, the solvable
algebra of the special K{\"a}hler manifold, denoted as $\Solv_{SK}$, is
obtained by just deleting some of the generators of the solvable algebra
of the quaternionic version, denoted by $\Solv_{Q}$. The solvable algebra
of the real version, $\Solv_R$, is obtained by further deleting
generators of $\Solv_{SK}$. In this way the (inverse) $\mathbf{c}$-map is
the definition of a subalgebra $\Solv_{SK}\subset \Solv_{Q}$ and the
(inverse) $\mathbf{r}$-map is the definition of a subalgebra
$\Solv_R\subset \Solv_{SK}$. This is illustrated in table \ref{tbl:genV}.
These new definitions allow to extend these maps to manifolds that are
not quaternionic-K{\"a}hler, as is the case for the Tits-Satake projected
ones. With these definitions we obtain the commutativity of these maps
with the Tits-Satake projection, as illustrated in (\ref{diagrammo}).
 \par
The Tits-Satake projection on an algebra is a subalgebra that includes
only one nilpotent generator for each root vector. Therefore it removes
the redundancy of the $X$, $Y$ and $Z$ columns indicated in the last row
of table \ref{tbl:genV}. In the generic case, the Tits-Satake projected
algebra is thus just the algebra with one entry in any entry of table
\ref{tbl:genV} that is present. Tables \ref{tbl:SATS1} and
\ref{tbl:SATS2} give the resulting Tits-Satake and sub Tits-Satake
manifolds. They also mention in the second column which columns of table
\ref{tbl:genV} are present in each case, and this is sufficient to obtain
the Tits-Satake algebra that is displayed. The paint group is the biggest
compact subgroup of the original isometry group that commutes with the
noncompact Cartan generators of the generating solvable algebra. It has
been explained at the end of section \ref{ss:classhomquatK} how these
tables are truncated in special cases with $q<0$.
 \par
We think that the above results are valuable to provide new insights and
tools in order to solve the following outstanding three problems:
\begin{itemize}
\item completion of the smooth cosmic billiard program by
extending, if possible, the complete integration formula found, in
\cite{Fre:2005bs}, for maximally split manifolds (that are related to
maximal supersymmetry) to the non-maximally split cases (related to lower
supersymmetry).
\item clarification of the microscopic string interpretation of
homogeneous geometries in orbifold compactifications. In particular, one
is attracted by the idea of extending the Tits-Satake projection from the
massless sector to the entire massive spectrum of superstrings. A first
example in this direction is shortly discussed in appendix~\ref{d3d7}.
\item affine and hyperbolic extension of homogeneous geometries,
upon  dimensional reduction, respectively, to $D=2$ and $D=1$ dimensions
following the procedure introduced in \cite{Fre':2005si}.
\end{itemize}
\medskip
\section*{Acknowledgments.}

\noindent We thank M. Esole for useful comments. This work is supported
in part by the European Community's Human Potential Programme under
contract MRTN-CT-2004-005104 `Constituents, fundamental forces and
symmetries of the universe'. The work of J.R. and A.V.P. is supported in
part by the FWO - Vlaanderen, project G.0235.05 and by the Federal Office
for Scientific, Technical and Cultural Affairs through the
"Interuniversity Attraction Poles Programme -- Belgian Science Policy"
P5/27.

\landscape

\begin{table}[!htb]
  \caption{\it The Solvable algebras and Tits-Satake projections of homogeneous
  special K{\"a}hler and quaternionic-K{\"a}hler manifolds that are not very special. In this table
  $P$ denotes positive definite integers. After the column
  denoting the series are mentioned the columns of generators in table \ref{tbl:genV}
  that are present in this case. The manifolds are always the coset of $G$ with its
  maximal compact subgroup, and the described algebras are the
  solvable parts of the Iwasawa decomposition of the algebras that are mentioned.}\label{tbl:SATS1}
\begin{center}
$
\begin{array}{|cc|ccc|ccc|}\hline\hline
 \mbox{name} & \mbox{gen.} &\multicolumn{3}{c|}{\mbox{Special K{\"a}hler}} & \multicolumn{3}{c|}{\mbox{Quaternionic-K{\"a}hler}} \\
   &   & G & G_{\rm TS} & G_{\rm sub-TS} & G & G_{\rm TS} & G_{\rm sub-TS}\\
   \hline\hline
 L(-3,0) &   &   &   &   & \usp(2,2) & \su(1,1) & \so(1,1)\\ \hline
 L(-3,P) &   &   &   &   & \usp(2P+2,2) & bc_1 & \so(1,1)\\ \hline
 \mathrm{SG}_4 & (0) & \mathrm{SG} & - & - & \su(2,1) & \begin{array}{c}
   bc_1=  \\
   \pmatrix{q_0\\ g_0\\ h_0}
 \end{array}    & \begin{array}{c}
   \su(1,1)= \\
   \pmatrix{g_0\\ h_0}
 \end{array}   \\ \hline
 L(-2,0) & (01)& \su(1,1) & \begin{array}{c}
   \su(1,1)= \\
    \pmatrix{g_1\\ h_1}
 \end{array}  & \begin{array}{c}
   \su(1,1)= \\
    \pmatrix{g_1\\ h_1}
 \end{array}  & \su(2,2) & \begin{array}{c}
   \so(3,2)= \\
   \pmatrix{q_0&p_1\\ g_0&g_1\\ h_0&h_1}
 \end{array}    & \begin{array}{c}
   \so(2,2)= \\
    \pmatrix{g_0&g_1\\ h_0&h_1}
 \end{array}   \\ \hline
 L(-2,P) & (01Y)& \su(P+1,1) & \begin{array}{c}
    bc_1= \\
   \pmatrix{Y^-\\g_1\\ h_1}
 \end{array}  & \begin{array}{c}
   \su(1,1)= \\
    \pmatrix{g_1\\ h_1}
 \end{array} & \su(P+2,2) &\begin{array}{c}
    bc_2= \\
   \pmatrix{\tilde Y^-&Y^-\\q_0&p_1\\ g_0&g_1\\ h_0&h_1}
 \end{array}    & \begin{array}{c}
   \so(2,2)= \\
   \pmatrix{g_0&g_1\\ h_0&h_1}
 \end{array}    \\
 \hline\hline
\end{array}
$
\end{center}
\end{table}

\begin{table}[!htb]
  \caption{\it The algebras of the very special manifolds. In this table
  $q$, $m$, $P$ and $\dot P$ are always integers $\geq 1$. After the column
  denoting the series are mentioned the columns of generators in table \ref{tbl:genV}
  that are present in this case ($+$ denotes the sum of columns 2 and 3).
  We mention which element of the series gives the Tits-Satake and
  sub-Tits-Satake algebra, and in case the corresponding manifold is
  symmetric, we mention also the isometry group $G$.
  }\label{tbl:SATS2}
\begin{center}$
\begin{array}{|lc|l|l|l|c|}
\hline\hline
 \mbox{name} & \mbox{gen.} & \mbox{Real} & \mbox{K{\"a}hler} & \mbox{quaternionic-K{\"a}hler}& \\
   &   & G_{\rm TS} & G_{\rm TS} & G_{\rm TS} &   \mbox{ \raisebox{1.5ex}[0cm][0cm]{$ G_{\rm sub-TS}$}} \\
\hline\hline
  \mathrm{SG}_5 & (01) & -  & \mathrm{SG}_5\ :\ \su(1,1) &  \mathrm{SG}_5\ :\ \mathfrak{g}_{2(2)} & \mathrm{SG}_5
  \\ \hline
 L(-1,0) & (01+) & L(-1,0)\ :\ \so(1,1)  & L(-1,0)\ :\ \su(1,1)^2 & L(-1,0)\ :\ \so(3,4) &
 \\ \cline{1-5}
 L(-1,P) & (01+Y) & L(-1,1)\ :\ \so(2,1) & L(-1,1) & L(-1,1) &  \mbox{ \raisebox{1.5ex}[0cm][0cm]{$ L(-1,0)$}} \\
 \hline
 L(0,0) & (0123) & L(0,0)\ :\ \so(1,1)\oplus \so(1,1) & L(0,0)\ :\ \su(1,1)^3 & L(0,0)\ :\ \so(4,4) &   \\
\cline{1-5}
 L(0,P) & (0123Y) & L(0,1)\ :\ \so(1,1)\oplus \so(2,1) & L(0,1)\ :\ \su(1,1)\oplus \so(3,2) & L(0,1)\ :\ \so(5,4) &   \\
\cline{1-5}
L(0,P,\dot P)  & (0123YZ) & L(0,1,1) & L(0,1,1) & L(0,1,1) &  \mbox{ \raisebox{0.5ex}[0cm][0cm]{$ L(0,0)$}} \\
\cline{1-5}
 \hspace{-2mm} \begin{array}{l}
   L(q,P) \\
   L(4m,P,\dot P)
 \end{array}   & (0123XYZ)  & L(1,1)\ :\ \mathfrak{s\ell }(3,\mathbb{R}) & L(1,1)\ :\ \mathfrak{sp}(6) & L(1,1)\ :\ \mathfrak{f_{4(4)}} &  \\
 \hline\hline
\end{array}$
\end{center}
\end{table}

\endlandscape

%%%%%%%%%%%%%%%%%%%%%%%%%%%%%%%%%%%%%%%%%
% Appendices %%%%%%%%%%%%%%%%%%%%%%%%%%%
%%%%%%%%%%%%%%%%%%%%%%%%%%%%%%%%%%%%%%%%
\appendix
\section{The $D3/D7$--brane system: an application}
\label{d3d7} Recently, \cite{Angelantonj:2003zx,D'Auria:2004qv}
considered the low energy dynamics of Type IIB theory compactified to
four dimensions on a $K3\times T^2/\mathbb{Z}_2$--orientifold
\cite{Tripathy:2002qw,Andrianopoli:2003jf}, in the presence of $n_3$ D3
and $n_7$ D7 space--filling branes and fluxes. It was found that the
scalar manifold of the four dimensional $N=2$ supergravity, describing
the vector multiplet sector, had to be the homogeneous non--symmetric
space $L_{SK}(0,n_3,n_7)$. This was the first application of this Special
K{\"a}hler space to the description of a specific microscopic setting.\par
Let us consider Type IIB superstring compactified on a $K3\times
T^2/\mathbb{Z}_2$--orientifold, in the presence of a stack of $n_3$
parallel $D3$--branes and a stack of $n_7$ parallel $D7$--branes. The
branes are all space--filling, the $D7$--branes being wrapped on $K3$.
The resulting low--energy four dimensional theory has $N=2$
supersymmetry. We suppose the branes to be in the Coulomb phase and we
integrate out the massive modes (the existence of supersymmetric vacua in
the low--energy theory would justify a posteriori this assumption). In
this phase the scalar fields describing the positions of the $D3$ branes
along $K3$, which, in the language of the four dimensional $N=2$ theory,
would belong to the hypermultiplet sector, are massive and thus do not
enter our effective description. In fact the quaternionic sector in the
model under consideration describes just the $K3$ moduli and contains no
open string excitation. \par The vector multiplet sector of the
low--energy supergravity contains three complex scalars $s,t,u$
originating from the bulk sector and spanning a manifold of the form:
\begin{equation}
\label{stu}
    \left(\frac{{\rm SU} (1,1)}{{\rm U}(1)}\right)_s\times
\left(\frac{{\rm SU}(1,1)}{{\rm U}(1)}\right)_t\times
    \left(\frac{{\rm SU}(1,1)}{{\rm U}(1)}\right)_u\,.
\end{equation}
The $s$ scalar describes the $K3$--volume and the R--R four--form
$C_{(4)}$ on $K3$, $t$ the $T^2$--complex structure and $u$ the IIB
axion--dilaton system:
\begin{eqnarray}
s&=& C_{(4)} -{\rm i}\, {\rm Vol} (K_3),\,
\nonumber\\
t&=& \frac{g_{12}}{g_{22}} -{\rm i}\,\frac{\sqrt{{\det} g}}{g_{22}}\,,
\nonumber\\
u&=& C_{(0)} -{\rm i}\, e^{\varphi}\,,
\end{eqnarray}
where the matrix $g$ denotes the metric on $T^2$. The corresponding three
vector fields $A^1_\mu,\,A^2_\mu,\,A^3_\mu$, together with the vector
contained in the graviton multiplet $A^0_\mu$, originate from the
components $B_{\mu a},\,C_{\mu a}$ of the ten dimensional NS--NS and R--R
2--forms, $a=1,2$ labelling the directions of $T^2$, and transform in the
${\bf (2,2)}$ of ${\rm SU}(1,1)_u\times {\rm SU}(1,1)_t$.\par
 The open string
excitations describing the positions of the $D3$--branes and of the $n_7$
$D7$--branes along $T^2$ are described by $n_3$ complex scalars $y^r$ and
$n_7$ complex scalars $z^k$ ($r=1,\dots, n_3$; $k=1,\dots, n_7$). These
moduli enlarge the scalar manifold (\ref{stu}) to the homogeneous,
non-symmetric space $L_{SK}(0,n_3,n_7)$. The coordinates $y^r$ and $z^k$
are in the same supermultiplets as the $D3$ and the $D7$--brane vector
fields $A^r_\mu$ and $A^k_\mu$ respectively. The prepotential ${ F}$ in
the special coordinates $s,t,u,y^r,z^k$ is expressed by the following
cubic polynomial:
\begin{equation}\label{prepot}
    {F}(s,t,u,x^k,y^r)\,=\, stu-\frac{1}{2}\,s \,z^k
    z^k-\frac{1}{2}\,u\,
    y^r y^r\,,
\end{equation}
Considerations based on the analysis of the microscopic system imply that
${\rm SU}(1,1)_s$ acts as an electric--magnetic duality transformation
\cite{Gaillard:1981rj} both on the bulk and D7--brane vector
field--strengths, while the ${\rm SU} (1,1)_u$ acts as an
electric--magnetic duality transformation on the D3--vector
field--strengths. Likewise the bulk vectors transform perturbatively
under ${\rm SU}(1,1)_u \times {\rm SU}(1,1)_t$ while the D3--brane
vectors do not transform under ${\rm SU}(1,1)_s \times {\rm SU}(1,1)_t$
and the D7--brane vectors do not transform under ${\rm SU}(1,1)_u \times
{\rm SU}(1,1)_t$. However in the special coordinate basis all the three
${\rm SU}(1,1)$ duality groups have a non--perturbative action on the
vector fields and thus a symplectic rotation is needed to define the
right symplectic frame yielding the correct interactions between vector
fields and scalar fields. This new frame does not admit a
prepotential.\par Below we give the description of the
$L_{SK}(0,n_3,n_7)$ manifold in terms of Alekseevsky's coordinates and
then define the precise correspondence between these coordinates and the
complex fields $s,t,u,y^r,z^k$ defined above.
 Our identification of the scalar
fields with solvable parameters is described by the following expression
for a generic solvable Lie algebra element:
\begin{eqnarray}
\rm{Solv} &=&\{\sum_{\alpha=1}^3\varphi^\alpha h_\alpha +\hat{\theta}_1
g_1+\theta_2 g_2+\theta_3 g_3+y^{r\pm }Y^{\pm}_r+z^{k\pm
}Z^{\pm}_k\}\,,\nonumber\\
&&\phantom{aaaaaa}\hat{\theta}_1 = \theta_1+y^{r+}\, y^{r-}+z^{k+}\,
z^{k-}\,,
\end{eqnarray}
where $(y^{r+},y^{r-})$ and $(z^{k+},z^{k-})$ are related to the real and
imaginary parts of the $D3$ and $D7$--branes complex coordinates along
$T^2$. The non trivial commutation relations between the above solvable
generators are:
\begin{eqnarray}
[h_1,Y^\pm]&=&
\frac{1}{2}\,Y^\pm\,\,\,;\,\,\,\,\,[h_1,Z^\pm]=\frac{1}{2}\,
Z^\pm\,,\nonumber\\
\left[ h_3,Y^\pm \right] &=&\pm\frac{1}{2}\,Y^\pm\,\,\,;\,
\,\,\,\,\left[h_2,Z^\pm\right]=\pm\frac{1}{2}\,
Z^\pm\,,\nonumber\\
\left[g_3,Y^-\right]&=&Y^+\,\,\,;\,\,\,\,\,\left[g_2,Z^-\right]=
Z^+\,,\nonumber\\
\left[Y^+_r,Y^-_s\right]&=&\delta_{rs}\,g_1\,\,\,;\,\,\,\,\,\left[Z^+_k,Z^-_\ell\right]=
\delta_{k\ell}\,
g_1\,\,;\,\,\,r,s=1,\dots,n_3\,\,k,\ell=1,\dots,n_7\,,\nonumber\\
\left[h_\alpha, g_\alpha\right]&=&g_\alpha\,\,;\,\,\,\alpha=1,2,3\,.
\label{comkal}
\end{eqnarray}
We exponentiate the solvable algebra using the following
coset-representative:
\begin{eqnarray}
L&=&\rme^ {\theta_3 g_3}\,\rme^ {y^{r-} Y^-_r}\,\rme^{y^{r+}
Y^+_r}\,\rme^{\theta_2 g_2}\,\rme^{z^{k-} Z^-_k}\,\rme^{z^{k+} Z^+_k}\,
\rme^{ \hat{\theta}_1\,g_1}\,\rme^{\varphi^\alpha h_\alpha}\,.
\end{eqnarray}
The order of the exponentials in the coset representative and the
particular parameter $\hat{\theta}_1$ used for $g_1$, have been chosen in
such a way that the axions $\theta_1,
\,\theta_2,\,\theta_3,\,y^{r+},\,z^{k+}$ appear in the resulting metric
only covered by derivatives. The metric reads:
\begin{eqnarray}
\rmd s^2&=&
(\rmd\varphi_\alpha)^2+\rme^{-2\varphi_1}\,\left(\rmd\theta_1+\frac{1}{2}\,\rmd\theta_2
(z^{-})^2+\frac{1}{2}\,\rmd\theta_3 (y^{-})^2+z^{k-}\,
\rmd z^{k+}+y^{r-}\, \rmd y^{r+} \right)^2+\nonumber\\
&&
\rme^{-2\varphi_2}\,\rmd\theta_2^2+\rme^{-2\varphi_3}\,\rmd\theta_3^2+\rme^{-\varphi_1-\varphi_2}\,
(\rmd z^{k+}+\rmd\theta_2\,z^{k-})^2+\rme^{-\varphi_1+\varphi_2}\, (\rmd
z^{k-})^2+\nonumber\\&& \rme^{-\varphi_1-\varphi_3}\, (\rmd
y^{r+}+\rmd\theta_3\,y^{r-})^2+\rme^{-\varphi_1+\varphi_3}\, (\rmd
y^{r-})^2 \nonumber\\&&(z^{-})^2 \equiv \sum_{k=1}^{n_7}(z^{k-})^2\,;
\,\,(y^{-})^2\equiv\sum_{r=1}^{n_3}(y^{r-})^2\,.\label{met}
\end{eqnarray}
The above metric defines the $\sigma$--model action in the vector
multiplet sector and is manifestly invariant under global ${\rm
SO}(n_3)\times {\rm SO}(n_7)$. Identifying the axionic coordinates
$\theta_1,\,\theta_2,\,\theta_3,\,y^{r+},\,z^{i+}$ with the real parts of
the special coordinates $s,\,t,\,u,\,y^r,\,x^i$, and comparing the
corresponding components of the metric one easily obtains the following
relations between the solvable coordinates and the special coordinates:
\begin{eqnarray}
s&=&\theta_3-\frac{i}{2}\,\rme^{\varphi_3}\,\,;\,\,\,u=\theta_2-\frac{i}{2}\,
\rme^{\varphi_2}\,,\nonumber\\
t&=&\theta_1-\frac{i}{2}\,\left(\rme^{\varphi_1}+\frac{1}{2}
\,\rme^{\varphi_2}\,(z^{-})^2+\frac{1}{2}\,\rme^{\varphi_3}\,
(y^{-})^2\right)\,,\nonumber\\
z^k&=&z^{k+}+\frac{i}{2}\, \rme^{\varphi_2}\,
z^{k-}\,\,;\,\,\,\,y^r=y^{r+}+\frac{i}{2}\,\rme^{\varphi_3}\, y^{r-}\,.
\end{eqnarray}
After having described the local geometry of the $L_{SK}(0,n_3,n_7)$
space and the corresponding $\sigma$--model action (\ref{met}), let us
discuss the TS projection and its microscopic meaning. The TS projection
yields the space $L_{SK}(0,1,1)$, which describes the situation in which
we have 1 $D3$ and 1 $D7$--brane. The fact that the position of the
billiard walls does not depend of the number of branes stems from the
fact that the dilatonic scalars (e.g., in the four-dimensional case, the
imaginary parts of the $s,t,u$ scalars), being bulk fields, couple to
colour--singlet combinations of open string excitations (scalars and
vector fields in four dimensions). For instance ${\rm Im}(s)$, namely the
volume of $K3$ (in the string frame), defines the effective YM coupling
constant $g_{YM}^{(7)}$ on the $D7$--brane world--volume theory
compactified on $K3$, while ${\rm Im}(u)$, namely the ten-dimensional
dilaton, defines the YM coupling constant $g_{YM}^{(3)}$ on the
$D3$--brane world--volume theory. These scalars will therefore couple to
the vector fields on the two kinds of branes as follows:
\begin{eqnarray}
-\left(\frac{1}{g_{YM}^{(7)}}\right)^2\,\int_{D7}{\rm
Tr}(F^2)\propto {\rm Im}(s)\,\int_{D7}{\rm Tr}(F^2)\,,\nonumber\\
-\left(\frac{1}{g_{YM}^{(3)}}\right)^2\,\int_{D3}{\rm Tr}(F^2)\propto
{\rm Im}(u)\,\int_{D3}{\rm Tr}(F^2)\,,\label{ff}
\end{eqnarray}
where the trace is defined on the generators of the ${\rm U}(n_7)$ and
${\rm U}(n_3)$ colour--groups respectively. In the Coulomb phase these
groups are spontaneously broken to their maximal torii. Upon dimensional
reduction to $D=3$, the $n_3+n_7$ boundary vectors yield $2(n_3+n_7)$
axions which, however, define through the couplings (\ref{ff}) just four
walls with a certain degeneracy (colour) each.
%%%%%%%%%%%%%%%%%%%%%%%%%%%%%%%%%%%%%%%%%%%%%%%%%%%%%%%%%%%
\section{Properties of real Clifford algebras} \label{realcliffalg}

In this section, we will recall some properties of real Clifford
algebras. Some reviews are in \cite{Okubo:1994ss,Alekseevsky:2003vw}. We
will restrict to Clifford algebras with positive signature. The
$(q+1)$-dimensional real Clifford algebra $\mathcal{ C}(q+1,0)$ is
generated by real matrices $\gamma_\mu$ ($\mu = 1,\cdots, q+1$)
satisfying
\begin{equation} \label{cliffcondapp}
\gamma_{\mu} \gamma_{\nu} + \gamma_{\nu} \gamma_{\mu} = 2 \delta_{\mu
\nu} \unity.
\end{equation}
The main properties are given in table \ref{tbl:RealCliff}, which we will
now further explain.
\begin{table}[!htb]
  \caption{\it Real Clifford algebras ${\cal C}(q+1,0)$, the dimension ${\cal D}_{q+1}$ of their irreducible
  representations, and the metric preserving group in the centralizer of the Clifford algebra in the
  $(P+\dot P){\cal D}_{q+1}$-dimensional representation. Here $\mathbf{F}(n)$ stands for $n \times n$ matrices
with entries in the field $\mathbf{F}$.}\label{tbl:RealCliff}
\begin{center}
\begin{tabular}{||c|c|c|c|l||}\hline
$q$ &$q+1$ &${\cal C}(q+1,0)$& ${\cal D}_{q+1}$&${\cal S}_q(P,\dot P)$
\\ \hline &&&&\\[-3mm]
$-1$ & 0&$\mathbf{R}$    &1         &$\SO(P)$     \\
0    & 1&$\mathbf{R}\oplus \mathbf{R} $&1&$\SO(P)\times \SO(\dot P)$ \\
1    & 2&$\mathbf{R}(2)$ &2         &$\SO(P)$     \\
2    & 3&$\mathbf{C}(2)$ &4         &$\U(P)$     \\
3    & 4&$\mathbf{H}(2)$ &8         &$\U(P,\mathbf{H})\equiv \USp(2P)$
\\ 4    & 5&$\mathbf{H} (2)\oplus \mathbf{H} (2)$&8&$\USp(2P)\times \USp(2\dot P)$\\
5    & 6&$\mathbf{H}(4)$ &16&$\U(P,\mathbf{H})\equiv \USp(2P)$   \\
6    & 7&$\mathbf{C}(8)$ &16&$\U(P)$     \\
7    & 8&$\mathbf{R}(16)$&16&$\SO(P)$     \\
$n+7$ & $n+8$ & $\mathbf{R}(16)\times{\cal C}(n,0)$&16 ${\cal D}_n$ &
as for $q+1=n$\\[1mm]
\hline
\end{tabular}
\end{center}
\end{table}

When $q+1 = 0,1,2 \ (\mathrm{mod\ 8})$, the matrices of the complex
Clifford algebra can be chosen to be real. So in these cases, the
dimension of an irreducible representation is given by the dimension of
the corresponding complex representation. If this occurs, the real
Clifford algebra is said to be of the normal type. In the other cases, it
is possible to obtain a real representation of dimension twice that of
the complex representation. Indeed, many representations contain only
purely real or purely imaginary matrices. Real matrices of double
dimension are then obtained by considering the following matrices:
\begin{equation} \label{realgammas}
\Gamma^a = \gamma^a \otimes \unity_2 \quad \mathrm{if}\ \gamma^a \
\mathrm{is\ real,} \qquad \Gamma^a = \gamma^a \otimes \sigma_2 \quad
\mathrm{if}\ \gamma^a \ \mathrm{is\ imaginary}\,.
\end{equation}
Consider now a real irreducible representation of the Clifford algebra
$\mathcal{C}(q+1,0)$, given by $\mathcal{D}_{q+1} \times
\mathcal{D}_{q+1}$-matrices $\gamma_\mu$, where $\mathcal{D}_{q+1}$ is
given in table \ref{tbl:RealCliff}. Consider a real $\mathcal{D}_{q+1}
\times \mathcal{D}_{q+1}$-matrix $S$ satisfying
\begin{equation} \label{defSrealcliff}
\lbrack S, \gamma_\mu \rbrack = 0 \,.
\end{equation}
According to Schur's lemma, matrices that commute with an irreducible
representation of the Clifford algebra must form a division algebra. This
leads to distinction in a normal, almost complex and quaternionic case.

\subsection{The normal case}
As already mentioned, this occurs when
\begin{equation}
q+1 = 0,1,2 \quad \mathrm{mod\ 8} \,.
\end{equation}
In this case the general form of the matrices $S$, commuting with all
$\gamma$-matrices, is
\begin{equation} \label{Snormal}
S = a \unity \,,
\end{equation}
where $a$ is a real constant. The dimension of the irreducible
representation is given by $\mathcal{D}_{q+1} = 2^l$, where $q+1 = 2l$ or
$2l+1$. For $q+1$ even this irreducible representation is unique (up to
similarity transformations), while for $q+1$ odd, the representations
$\gamma_\mu$ and $-\gamma_\mu$ are inequivalent and constitute the 2
possible irreducible representations one can have. In this case the
product of all $\gamma$-matrices is moreover given by plus or minus the
identity.

\subsection{The almost complex case}
\label{almostcomplexgamma}
This occurs when
\begin{equation}
q+1 = 3,7 \quad \mathrm{mod\ 8}\,.
\end{equation}
The irreducible representation is unique and has dimension
$\mathcal{D}_{q+1}=2^{l+1}$. The general form of the matrices $S$ is
given by
\begin{equation} \label{Salmostcompl}
S = a \unity + b J \,,
\end{equation}
where $a,b$ are real constants and where the real $\mathcal{D}_{q+1}
\times \mathcal{D}_{q+1}$-matrix $J$ commutes with all $\gamma$-matrices
and squares to $-\unity$. $J$ is given by
\begin{equation} \label{J}
J = \pm \gamma_1 \cdots \gamma_{q+1}.
\end{equation}

\subsection{The quaternionic case}
This occurs when
\begin{equation}
q+1 = 4,5,6 \quad \mathrm{mod\ 8}\,.
\end{equation}
The dimension of the irreducible representations is given by
$\mathcal{D}_{q+1}=2^{l+1}$. It is unique for $q+1$ even, while there
exist two inequivalent irreducible representations when $q+1$ is odd. The
two irreducible representations are again related to each other by a
minus-sign. The general form of the matrices $S$ is now given by
\begin{equation} \label{Squat}
S = a_0 \unity + \sum_{j=1}^3 a_j E_j \,,
\end{equation}
where the constants $a_0$, $a_i$ are all real. The three matrices $E_i$
commute with the $\gamma$-matrices and they satisfy a quaternion
relation:
\begin{equation} \label{propE}
E_j E_k = - \delta_{jk} \unity + \sum_{l=1}^3 \epsilon_{jkl} E_l \,.
\end{equation}
%%%%%%%%%%%%%%%%%%%%%%%%%%%%%%%%%%%%%%%%%%%%%%%
\subsection{The structure of $\mathcal{S}_q(P,\dot{P})$}
\label{sqppdot}
The representations of the real Clifford algebras we are
working with, need not be irreducible. If one has a reducible
representation, one can choose it to be of the form
\begin{eqnarray}
\gamma_\mu & = & \unity_P \otimes \gamma_\mu^{\mathrm{irr}} \quad
\mathrm{for} \ q \neq 0 \ \mathrm{mod \
4}\,, \\
\gamma_\mu & = & \eta \otimes \gamma_\mu^{\mathrm{irr}} \quad
\mathrm{for} \ q = 0 \ \mathrm{mod \ 4}\,.
\end{eqnarray}
where $\gamma_\mu^{\mathrm{irr}}$ is an irreducible representation of the
Clifford algebra, and where $\eta = \mathrm{diag}(\unity_P,
-\unity_{\dot{P}})$. The group $\mathcal{S}_q(P,\dot{P})$, appearing in
the isometry groups of homogeneous very special spaces is generated by
all antisymmetric matrices that commute with all $\gamma$-matrices. In
the normal case and when $q \neq 0 \ \mathrm{mod \ 4}$, the generators of
$\mathcal{S}_q(P,\dot{P})$ are given by:
\begin{equation}
S = A \otimes \unity \,.
\end{equation}
where $A$ is an antisymmetric $P \times P$-matrix. When $q=0 \
\mathrm{mod\ 4}$, the matrix $A$ has to be replaced by a matrix
consisting of 2 blocks : one $P \times P$ and one $\dot{P} \times
\dot{P}$ antisymmetric block. In the almost complex case, the generators
of $\mathcal{S}_q(P,\dot{P})$ are of the following form
\begin{equation}
S = A \otimes \unity \,, \qquad \mathrm{or} \quad S = B \otimes J\,,
\end{equation}
where $A, B$ are antisymmetric, respectively symmetric $P \times
P$-matrices. In the quaternionic case, when $q \neq 0 \ \mathrm{mod \
4}$, the generators of $\mathcal{S}_q(P,\dot{P})$ are
\begin{equation}
S = A \otimes \unity \,, \qquad \mathrm{or} \quad S = B \otimes
\left(\sum_{j=1}^3 a_j E_j\right)\,,
\end{equation}
where $A, B$ are antisymmetric, respectively symmetric, $P \times
P$-matrices. Again, when $q = 0$  mod 4,
one should look upon $A$ and $B$ as (anti)symmetric matrices consisting
of (anti)symmetric $P \times P$ and $\dot{P} \times \dot{P}$ blocks.

\newpage

%%%%%%% If using BibTex %%%%%%%%%%%%%%%%%%%%%%%%%%%%%
%\bibliography{refLectParis}

\begin{thebibliography}{10}

\bibitem{Damour:2002cu}
T.~Damour, M.~Henneaux  and H.~Nicolai, \emph{$E_{10}$ and a 'small
tension
  expansion' of M theory}, Phys. Rev. Lett. {\bf 89} (2002) 221601,
\href{http://www.arXiv.org/abs/hep-th/0207267}{{\tt hep-th/0207267}}
%%CITATION = HEP-TH 0207267;%%.

\bibitem{Damour:2002et}
T.~Damour, M.~Henneaux  and H.~Nicolai, \emph{Cosmological billiards},
Class.
  Quant. Grav. {\bf 20} (2003) R145--R200,
\href{http://www.arXiv.org/abs/hep-th/0212256}{{\tt hep-th/0212256}}
%%CITATION = HEP-TH 0212256;%%.

\bibitem{Fre:2003ep}
P.~Fr{\'e}, V.~Gili, F.~Gargiulo, A.~Sorin, K.~Rulik  and M.~Trigiante,
  \emph{Cosmological backgrounds of superstring theory and solvable algebras:
  Oxidation and branes}, Nucl. Phys. {\bf B685} (2004) 3--64,
\href{http://www.arXiv.org/abs/hep-th/0309237}{{\tt hep-th/0309237}}
%%CITATION = HEP-TH 0309237;%%.

\bibitem{Cremmer:1999du}
E.~Cremmer, B.~Julia, H.~L{\"u}  and C.~N. Pope, \emph{Higher-dimensional
  origin of $D = 3$ coset symmetries},
\href{http://www.arXiv.org/abs/hep-th/9909099}{{\tt hep-th/9909099}}
%%CITATION = HEP-TH 9909099;%%.

\bibitem{Keurentjes:2002xc}
A.~Keurentjes, \emph{The group theory of oxidation}, Nucl. Phys. {\bf
B658}
  (2003) 303--347,
\href{http://www.arXiv.org/abs/hep-th/0210178}{{\tt hep-th/0210178}}
%%CITATION = HEP-TH 0210178;%%.

\bibitem{deWit:1991nm}
B.~de~Wit and A.~Van~Proeyen, \emph{Special geometry, cubic polynomials
and
  homogeneous quaternionic spaces}, Commun. Math. Phys. {\bf 149} (1992)
  307--334,
\href{http://www.arXiv.org/abs/hep-th/9112027}{{\tt hep-th/9112027}}
%%CITATION = HEP-TH 9112027;%%.

\bibitem{Andrianopoli:2004xu}
L.~Andrianopoli, S.~Ferrara  and M.~A. Lled{\'o}, \emph{No-scale $D = 5$
  supergravity from Scherk-Schwarz reduction of $D = 6$ theories}, JHEP {\bf
  06} (2004) 018,
\href{http://www.arXiv.org/abs/hep-th/0406018}{{\tt hep-th/0406018}}
%%CITATION = HEP-TH 0406018;%%.


\bibitem{Gunaydin:1984bi}
M.~G{\"u}naydin, G.~Sierra  and P.~K. Townsend, \emph{The geometry of $N=2$
  Maxwell--Einstein supergravity and Jordan algebras}, Nucl. Phys. {\bf B242}
  (1984)
244
%%CITATION = NUPHA,B242,244;%%.

\bibitem{deWit:1992cr}
B.~de~Wit and A.~Van~Proeyen, \emph{Broken sigma model isometries in very
  special geometry}, Phys. Lett. {\bf B293} (1992) 94--99,
\href{http://arXiv.org/abs/hep-th/9207091}{{\tt hep-th/9207091}}
%%CITATION = HEP-TH 9207091;%%.

\bibitem{deWit:1984pk}
B.~de~Wit and A.~Van~Proeyen, \emph{Potentials and symmetries of general
gauged
  $N=2$ supergravity -- Yang-Mills models}, Nucl. Phys. {\bf B245} (1984)
89
%%CITATION = NUPHA,B245,89;%%.

\bibitem{Castellani:1990zd}
L.~Castellani, R.~D'Auria  and S.~Ferrara, \emph{Special geometry without
  special coordinates}, Class. Quant. Grav. {\bf 7} (1990)
1767--1790
%%CITATION = CQGRD,7,1767;%%.

\bibitem{D'Auria:1991fj}
R.~D'Auria, S.~Ferrara  and P.~Fr{\`e}, \emph{Special and quaternionic
  isometries: General couplings in $N=2$ supergravity and the scalar
  potential}, Nucl. Phys. {\bf B359} (1991)
705--740
%%CITATION = NUPHA,B359,705;%%.

\bibitem{Bagger:1983tt}
J.~Bagger and E.~Witten, \emph{Matter couplings in $N=2$ supergravity},
Nucl.
  Phys. {\bf B222} (1983)
1
%%CITATION = NUPHA,B222,1;%%.

\bibitem{Cecotti:1989qn}
S.~Cecotti, S.~Ferrara  and L.~Girardello, \emph{Geometry of type II
  superstrings and the moduli of superconformal field theories}, Int. J. Mod.
  Phys. {\bf A4} (1989)
2475
%%CITATION = IMPAE,A4,2475;%%.

\bibitem{deWit:1992nm}
B.~de~Wit and A.~Van~Proeyen, \emph{Special geometry, cubic polynomials
and
  homogeneous quaternionic spaces}, Commun. Math. Phys. {\bf 149} (1992)
  307--334,
\href{http://arXiv.org/abs/hep-th/9112027}{{\tt hep-th/9112027}}
%%CITATION = HEP-TH 9112027;%%.

\bibitem{Cremmer:1985hc}
E.~Cremmer and A.~Van~Proeyen, \emph{Classification of K{\"a}hler manifolds
in
  $N=2$ vector multiplet--supergravity couplings}, Class. Quant. Grav. {\bf 2}
  (1985)
445
%%CITATION = CQGRD,2,445;%%.

\bibitem{D'Auria:2004kx}
R.~D'Auria, S.~Ferrara  and M.~Trigiante, \emph{c-map,very special
quaternionic
  geometry and dual Kaehler spaces}, Phys. Lett. {\bf B587} (2004) 138--142,
\href{http://www.arXiv.org/abs/hep-th/0401161}{{\tt hep-th/0401161}}
%%CITATION = HEP-TH 0401161;%%.

\bibitem{D'Auria:2004cu}
R.~D'Auria, S.~Ferrara  and M.~Trigiante, \emph{Homogeneous special
manifolds,
  orientifolds and solvable coordinates}, Nucl. Phys. {\bf B693} (2004)
  261--280,
\href{http://www.arXiv.org/abs/hep-th/0403204}{{\tt hep-th/0403204}}
%%CITATION = HEP-TH 0403204;%%.

\bibitem{Smet:2004da}
G.~Smet and J.~Van~den Bergh, \emph{O(3)/O(7) orientifold truncations and
very
  special quaternionic-K{\"a}hler geometry}, Class. Quant. Grav. {\bf 22}
  (2005) 1--22,
\href{http://www.arXiv.org/abs/hep-th/0407233}{{\tt hep-th/0407233}}
%%CITATION = HEP-TH 0407233;%%.

\bibitem{Fre':2005sr}
P.~Fr{\'e}, F.~Gargiulo  and K.~Rulik, \emph{Cosmic billiards with painted
  walls in non-maximal supergravities: A worked out example}, Nucl. Phys. {\bf
  B737} (2006) 1--48,
\href{http://www.arXiv.org/abs/hep-th/0507256}{{\tt hep-th/0507256}}
%%CITATION = HEP-TH 0507256;%%.

\bibitem{Henneaux:2003kk}
M.~Henneaux and B.~Julia, \emph{Hyperbolic billiards of pure $D = 4$
  supergravities}, JHEP {\bf 05} (2003) 047,
\href{http://www.arXiv.org/abs/hep-th/0304233}{{\tt hep-th/0304233}}
%%CITATION = HEP-TH 0304233;%%.

\bibitem{Keurentjes:2002rc}
A.~Keurentjes, \emph{The group theory of oxidation. II: Cosets of
non-split
  groups}, Nucl. Phys. {\bf B658} (2003) 348--372,
\href{http://www.arXiv.org/abs/hep-th/0212024}{{\tt hep-th/0212024}}
%%CITATION = HEP-TH 0212024;%%.

\bibitem{Helgason}
S.~Helgason, {\em Differential geometry, Lie groups and symmetric
spaces}.
\newblock Graduate Studies in Mathematics, Vol. 34, AMS,
2001
% .

\bibitem{Andrianopoli:1996bq}
L.~Andrianopoli, R.~D'Auria, S.~Ferrara, P.~Fr{\'e}  and M.~Trigiante,
  \emph{R-R scalars, U-duality and solvable Lie algebras}, Nucl. Phys. {\bf
  B496} (1997) 617--629,
\href{http://www.arXiv.org/abs/hep-th/9611014}{{\tt hep-th/9611014}}
%%CITATION = HEP-TH 9611014;%%.

\bibitem{Andrianopoli:1996zg}
L.~Andrianopoli, , R.~D'Auria, S.~Ferrara, P.~Fr{\'e}, R.~Minasian  and
  M.~Trigiante, \emph{Solvable Lie algebras in type IIA, type IIB and M
  theories}, Nucl. Phys. {\bf B493} (1997) 249--280,
\href{http://www.arXiv.org/abs/hep-th/9612202}{{\tt hep-th/9612202}}
%%CITATION = HEP-TH 9612202;%%.

\bibitem{Trigiante:1998vu}
M.~Trigiante, \emph{Dualities in supergravity and solvable Lie algebras},
  \href{http://www.arXiv.org/abs/hep-th/9801144}{{\tt hep-th/9801144}},
Ph.D. thesis, University of Wales, Swansea
%%CITATION = HEP-TH 9801144;%%.

\bibitem{Fre:2001jd}
P.~Fr{\`e}, \emph{Gaugings and other supergravity tools of $p$-brane
physics},
  \href{http://www.arXiv.org/abs/hep-th/0102114}{{\tt hep-th/0102114}},
proceedings of the Workshop on Latest Development in M-Theory, Paris,
France,
  1-9 Feb 2001
%%CITATION = HEP-TH 0102114;%%.

\bibitem{Alekseevsky1975}
D.~V. Alekseevsky, \emph{Classification of quaternionic spaces with a
  transitive solvable group of motions}, Math.\ USSR Izvestija {\bf 9} (1975)
297--339
% .

\bibitem{Fre:2005bs}
P.~Fr{\`e} and A.~Sorin, \emph{Integrability of supergravity billiards and
the
  generalized Toda lattice equation}, Nucl. Phys. {\bf B733} (2006) 334--355,
\href{http://www.arXiv.org/abs/hep-th/0510156}{{\tt hep-th/0510156}}
%%CITATION = HEP-TH 0510156;%%.

\bibitem{BorelTits}
A.~Borel and J.~Tits, \emph{Groupes r{\'e}ductifs}, Publications
  Math{\'e}mathiques de l'IHES {\bf 27} (1965) 55--151,
Compl{\'e}ments {\`a} l'article: vol. \textbf{41} (1972), p. 253-276,
  \href{http://www.numdam.org/numdam-bin/item?id=PMIHES_1972__41__253_0}{{\tt
  http://www.numdam.org/}}
% .

\bibitem{deWit:1992wf}
B.~de~Wit, F.~Vanderseypen  and A.~Van~Proeyen, \emph{Symmetry structure
of
  special geometries}, Nucl. Phys. {\bf B400} (1993) 463--524,
\href{http://www.arXiv.org/abs/hep-th/9210068}{{\tt hep-th/9210068}}
%%CITATION = HEP-TH 9210068;%%.

\bibitem{Cortes}
V.~Cort{\'e}s, \emph{Alekseevskian spaces}, Diff. Geom. Appl. {\bf 6} (1996)
129--168
% .

\bibitem{Bergshoeff:2002qk}
E.~Bergshoeff, S.~Cucu, T.~de~Wit, J.~Gheerardyn, R.~Halbersma,
S.~Vandoren
  and A.~Van~Proeyen, \emph{Superconformal $N = 2$, $D = 5$ matter with and
  without actions}, JHEP {\bf 10} (2002) 045,
\href{http://www.arXiv.org/abs/hep-th/0205230}{{\tt hep-th/0205230}}
%%CITATION = HEP-TH 0205230;%%.

\bibitem{Cecotti:1988ad}
S.~Cecotti, \emph{Homogeneous K{\"a}hler manifolds and $T$ algebras in $N=2$
  supergravity and superstrings}, Commun. Math. Phys. {\bf 124} (1989)
23--55
%%CITATION = CMPHA,124,23;%%.

\bibitem{Riccioni:2001bg}
F.~Riccioni, \emph{All couplings of minimal six-dimensional
supergravity},
  Nucl. Phys. {\bf B605} (2001) 245--265,
\href{http://www.arXiv.org/abs/hep-th/0101074}{{\tt hep-th/0101074}}
%%CITATION = HEP-TH 0101074;%%.

\bibitem{Romans:1986er}
L.~J. Romans, \emph{Selfduality for interacting fields: covariant field
  equations for six-dimensional chiral supergravities}, Nucl. Phys. {\bf B276}
  (1986)
71
%%CITATION = NUPHA,B276,71;%%.

\bibitem{Andrianopoli:1996ve}
L.~Andrianopoli, R.~D'Auria  and S.~Ferrara, \emph{U-duality and central
  charges in various dimensions revisited}, Int. J. Mod. Phys. {\bf A13} (1998)
  431--490,
\href{http://www.arXiv.org/abs/hep-th/9612105}{{\tt hep-th/9612105}}
%%CITATION = HEP-TH 9612105;%%.

\bibitem{Nishino:1984gk}
H.~Nishino and E.~Sezgin, \emph{Matter and gauge couplings of $N=2$
  supergravity in six dimensions}, Phys. Lett. {\bf B144} (1984)
187
%%CITATION = PHLTA,B144,187;%%.

\bibitem{Bergshoeff:1986mz}
E.~Bergshoeff, E.~Sezgin  and A.~Van~Proeyen, \emph{Superconformal tensor
  calculus and matter couplings in six dimensions}, Nucl. Phys. {\bf B264}
  (1986)
653
%%CITATION = NUPHA,B264,653;%%.

\bibitem{VanProeyen:1985ib}
A.~Van~Proeyen, \emph{$N=2$ matter couplings in $d = 4$ and 6 from
  superconformal tensor calculus},
in {\it Superunification and extra dimensions, proceedings of the 1st
Torino
  meeting,} eds. R. D'Auria and P. Fr{\'e}, (World Scientific, 1986), 97-125
% .

\bibitem{Riccioni:1999xq}
F.~Riccioni, \emph{Abelian vector multiplets in six-dimensional
supergravity},
  Phys. Lett. {\bf B474} (2000) 79--84,
\href{http://www.arXiv.org/abs/hep-th/9910246}{{\tt hep-th/9910246}}
%%CITATION = HEP-TH 9910246;%%.

\bibitem{Gunaydin:1983mi}
M.~G{\"u}naydin and N.~P. Warner, \emph{The $G_2$ invariant compactifications
  in eleven-dimensional supergravity}, Nucl. Phys. {\bf B248} (1984)
685
%%CITATION = NUPHA,B248,685;%%.

\bibitem{deWit:1993wf}
B.~de~Wit, F.~Vanderseypen  and A.~Van~Proeyen, \emph{Symmetry structure
of
  special geometries}, Nucl. Phys. {\bf B400} (1993) 463--524,
\href{http://www.arXiv.org/abs/hep-th/9210068}{{\tt hep-th/9210068}}
%%CITATION = HEP-TH 9210068;%%.

\bibitem{deWit:1995tf}
B.~de~Wit and A.~Van~Proeyen, \emph{Isometries of special manifolds},
  \href{http://www.arXiv.org/abs/hep-th/9505097}{{\tt hep-th/9505097}},
in the Proceedings of the meeting on quaternionic structures in
mathematics and
  physics, Trieste, September 1994; available on
  \href{http://www.emis.de/proceedings/QSMP94/}{http://www.emis.de/proceedings%
/QSMP94/}
%%CITATION = HEP-TH 9505097;%%.

\bibitem{Fre':2005si}
P.~Fr{\'e}, F.~Gargiulo, K.~Rulik  and M.~Trigiante, \emph{The general
pattern
  of Kac Moody extensions in supergravity and the issue of cosmic billiards},
  Nucl. Phys. {\bf B741} (2006) 42--82,
\href{http://www.arXiv.org/abs/hep-th/0507249}{{\tt hep-th/0507249}}
%%CITATION = HEP-TH 0507249;%%.

\bibitem{Angelantonj:2003zx}
C.~Angelantonj, R.~D'Auria, S.~Ferrara  and M.~Trigiante, \emph{$K3
\times
  T^2/\mathbb{Z}_2$ orientifolds with fluxes, open string moduli and critical
  points}, Phys. Lett. {\bf B583} (2004) 331--337,
\href{http://www.arXiv.org/abs/hep-th/0312019}{{\tt hep-th/0312019}}
%%CITATION = HEP-TH 0312019;%%.

\bibitem{D'Auria:2004qv}
R.~D'Auria, S.~Ferrara  and M.~Trigiante, \emph{Orientifolds, brane
coordinates
  and special geometry}, \href{http://www.arXiv.org/abs/hep-th/0407138}{{\tt
  hep-th/0407138}},
in {\it Deserfest: A Celebration of the life and works of Stanley Deser},
  (World Scientific, 2006), eds. J.T. Liu et al.
%%CITATION = HEP-TH 0407138;%%.

\bibitem{Tripathy:2002qw}
P.~K. Tripathy and S.~P. Trivedi, \emph{Compactification with flux on K3
and
  tori}, JHEP {\bf 03} (2003) 028,
\href{http://www.arXiv.org/abs/hep-th/0301139}{{\tt hep-th/0301139}}
%%CITATION = HEP-TH 0301139;%%.

\bibitem{Andrianopoli:2003jf}
L.~Andrianopoli, R.~D'Auria, S.~Ferrara  and M.~A. Lled{\'o}, \emph{4-$D$
  gauged supergravity analysis of type IIB vacua on $K3 \times
  T^2/\mathbb{Z}_2$}, JHEP {\bf 03} (2003) 044,
\href{http://www.arXiv.org/abs/hep-th/0302174}{{\tt hep-th/0302174}}
%%CITATION = HEP-TH 0302174;%%.

\bibitem{Gaillard:1981rj}
M.~K. Gaillard and B.~Zumino, \emph{Duality rotations for interacting
fields},
  Nucl. Phys. {\bf B193} (1981)
221
%%CITATION = NUPHA,B193,221;%%.

\bibitem{Okubo:1994ss}
S.~Okubo, \emph{Representation of Clifford algebras and its
applications},
  Math. Jap. {\bf 41} (1995) 59--79,
\href{http://www.arXiv.org/abs/hep-th/9408165}{{\tt hep-th/9408165}}
%%CITATION = HEP-TH 9408165;%%.

\bibitem{Alekseevsky:2003vw}
D.~V. Alekseevsky, V.~Cort{\'e}s, C.~Devchand  and A.~Van~Proeyen,
  \emph{Polyvector Super-Poincar{\'e} Algebras}, Commun. Math. Phys. {\bf 253}
  (2004) 385--422,
\href{http://www.arXiv.org/abs/hep-th/0311107}{{\tt hep-th/0311107}}
%%CITATION = HEP-TH 0311107;%%.

\end{thebibliography}
%%Included for WinEdt Gather Purpose (do not remove the comment line below:
%             %input "C:\localtexmf\bibtex\bib\refLectParis.bib"
%             %input "C:\Program Files\MiKTeX\texmf\bibtex\bib\refLectParis.bib"
%\bibliographystyle{toine}
%%%%%%%%%%%%%%%%  Result of BibTeX %%%%%%%%%%%
\providecommand{\href}[2]{#2}\begingroup\raggedright\endgroup

%%%%%%%%%%%%%%%%%%%%

\end{document}